# AN ANALYSIS OF STATES IN THE PHASE SPACE: FROM QUANTUM MECHANICS TO GENERAL RELATIVITY

Sebastiano Tosto

ENEA Casaccia, via Anguillarese 301, 00123 Roma, Italy

sebastiano.tosto@enea.it
stosto@inwind.it

# ABSTRACT

The paper has heuristic character. The conceptual frame, based on the assumption of quantum uncertainty only, has been introduced in two non relativistic papers concerning simple quantum systems, many electron atoms and diatomic molecules [S. Tosto, Il Nuovo Cimento B, vol. 111, n.2, (1996) and S. Tosto, Il Nuovo Cimento D, vol. 18, n.12, (1996)]. Instead of attempting to increase the accuracy of some existing computational model through a new kind of approximation, these papers acknowledge since the beginning the lack of deterministic information about the local properties of the constituent particles, considered random, unknown and unpredictable and thus ignored in principle. The leading idea is therefore that the physical properties of quantum systems could be inferred merely considering the delocalization ranges of dynamical variables, rather than their local values. In effect, despite the agnostic character of the approach proposed, both papers show that the kind of physical information reachable reproduces exactly in all cases examined that obtained solving the pertinent wave equations. The concept of quantum uncertainty is further extended in the present paper to both space and time coordinates, considering thus a unique spacetime delocalization range and still discarding since the beginning the local values of the conjugate dynamical variables. The paper shows an unexpected wealth of information obtainable simply extending the concept of space uncertainty to that of spacetime uncertainty: the results are inherently consistent with that of the operator formalism of wave mechanics and with the basic postulates of special relativity, both inferred as corollaries. Moreover, even the gravity appears to be essentially a quantum phenomenon. The most relevant outcomes of special and general relativity are achieved as straightforward consequence of the space-time delocalization of particles using the simple quantum formalism first introduced in the early papers.

1 Introduction.

The formalism of wave mechanics describes the state of a quantum system and its possible evolution as a function of time through the appropriate wave equation, whose solution provides the probability distribution function for the configuration of the system as a function of time and space coordinates of the constituent particles. To find this solution is in general a difficult task when positions and momenta of the particles are perturbed by mutual interactions; complex systems like the many electron atoms require approximation methods to calculate eigenfunctions and eigenvalues. Regardless of the complexity of the concerned system, however, one finds that although the wave functions are normalized over all the space the electrons could be in, quantities like average radial distances from the nucleus and average mutual distances have finite values. It is therefore reasonable to regard the coordinates of each electron as quantities changing randomly within ranges whose average sizes are finite as well and distinctive of the kind of interaction, system configuration and boundary conditions like for instance minimum total energy. Consider in this respect that the classical concept of local space-time coordinate can be regarded as the limit case of a vanishingly small sized space-time range surrounding it. Thus one could suppose that to any local coordinate of a particle in a set can be in principle related an appropriate space-time range able to describe its state and its space-time evolution too, simply assuming that the range sizes are totally arbitrary; this requirement agrees with both limit concepts of exact particle coordinate on the one side and of complete delocalization even at infinity on the other side. The idea of considering ranges of values that encompass generalized coordinates rather than the coordinates themselves is therefore applicable to classical and quantum systems, which suggests introducing the phase space of each particle simply extending the reasoning to the conjugate momentum range. This way of thinking does not seem particularly significant in classical physics, as both local conjugate variables are exactly predictable; it appears instead much more relevant for particles subjected to the Heisenberg principle, as the impossibility of knowing simultaneously both conjugate dynamical variables does not hinder exploiting jointly the respective variability ranges. In the latter case, it is enough to regard the variability ranges of classical physics as uncertainty ranges. Appears therefore rational in principle and even more general the possibility of describing the quantum systems through the delocalization ranges of the constituent particles, while disregarding since the beginning the local values of their dynamical variables. The theoretical model introduced in [1,2] starts just from a critical review of the concept of local dynamical variable, in that it considers uniquely arbitrary delocalization ranges rather than coordinates. Instead of attempting to increase the accuracy of some existing computational model through a new kind of approximation or via some new hypothesis to handle the local terms, both papers have shown that the quantized angular momentum and the non-relativistic energy levels of harmonic oscillator, many electron atoms and diatomic molecules can be inferred utilizing one basic assumption only: the quantum uncertainty, introduced explicitly and since the beginning as conceptual requirement to formulate the respective physical problems. In fact, moving the physical interest from the conjugate coordinates and momenta of the particles to their respective ranges of delocalization has been proven essential to describe correctly all cases examined. To be more specific, consider for instance the radial distance $\rho$ of an electron from the nucleus defined by $0 < \rho \leq \rho_{max}$, being $\rho_{max}$ an arbitrary maximum distance in a reference frame centred somewhere in the nucleus. If $\rho$ changes randomly, then $\rho_{max}$ cannot be uniquely defined by a particular value specified "*a priori*"; yet is relevant in principle its conceptual significance: $\rho_{max}$, whatever its specific value might be, defines the range $\Delta\rho = \rho_{max} - 0$ allowed to the random variable $\rho$. Moreover also the variability range of local momentum $0 < p_\rho \leq p_{\rho\,max}$ can be likewise defined as $\Delta p_\rho = p_{\rho\,max} - 0$. Even in lack of detailed information about $\rho$ and $p_\rho$, these ranges enable the number of allowed states in the phase space for the electron radial motion to be calculated; to this purpose $\Delta\rho$ only is of interest, not any partial range $\Delta\rho^\S = \rho - 0$ defined by random values $\rho < \rho_{max}$ that would exclude radial distances in principle possible for the electron. Although being in the present particular case $\Delta\rho \equiv \rho_{max}$ and $\Delta p_\rho \equiv p_{\rho max}$, i.e. the total ranges coincide in practice with the maximum values of the respective variables, the notations $\Delta\rho$ and $\Delta p_\rho$ better emphasize their physical meaning of ranges encompassing local coordinates and momenta in principle possible for the electron, to which they reduce as a limit case for $\Delta\rho \to 0$ and $\Delta p_\rho \to 0$. This does not mean making any hypothesis on the range sizes, because both $\rho_{max}$ and $p_{\rho max}$ are actually completely arbitrary; even their infinite values cannot be excluded. Consider now the energy $E = E(\rho, p_\rho)$ of electron radial motion that reads according to the present way of thinking $E = E(0 < \rho \leq \Delta\rho, 0 < p_\rho \leq \Delta p_\rho)$ whatever the local $\rho$ and $p_\rho$ might randomly be; the previous considerations suggest regarding this energy as $E = E(\Delta\rho, \Delta p_\rho) = E(n)$. This last step is non-trivial because the unique information available is now the number $n$ of states in the phase space consistent with the ranges allowed to the dynamical variables of the system; the calculation of $n$ becomes therefore the central aim of the physical problem. These ideas clearly hold in general also for more complex systems, i.e. the distances $r_{ij}$ between i-th and j-th electrons in a many electron atom are replaced by the ranges $\Delta r_{ij}$ including all the possible $r_{ij}$. So, the quantum uncertainty is here regarded as unique basic postulate rather than as consequence of the commutation rules of operator formalism. For instance the basic reasoning to describe the electron moving radially in the field of the nucleus consists of the following points: (i) to replace $\rho$ and $p_\rho$ with the ranges $\Delta\rho$ and $\Delta p_\rho$; (ii) to regard these latter as radial uncertainty ranges of the electron randomly delocalized; (iii) to exploit the concept of uncertainty according to the ideas of quantum statistics; (iv) to find the link between numbers of quantum states and eigenvalues allowed to the system. In this conceptual frame, the local values of $\rho$ and $p_\rho$ do not longer play any role in describing the electron radial motion: considering uniquely the phase space of the system nucleus/electron, rather than describing the actual dynamics of the electron through the pertinent wave equation, it is possible to disregard since the beginning the local values of the conjugate dynamical variables considered random, unpredictable and unknown in principle and then of no physical interest. This is a conceptual requirement, not an expedient or a sort of numerical approximation to simplify some calculation. The point (iv) raises however the question about the effective importance of these numbers of states in describing the physical properties of quantum systems; so the link between $E(\Delta\rho, \Delta p_\rho)$ and the expected eigenvalue $E(n_\rho)$ must be explained along with the link between the quantum number $n_\rho$ and the number of allowed states $n$. The paper [1] shows in this respect that the only concept of quantum delocalization is essential and enough to calculate correctly "*ab initio*" and without any further hypothesis the energy levels of many electrons mutually interacting in the field of nuclear charge; this idea was proven more useful than a new numerical algorithm also to treat the diatomic molecules [2]. Despite the apparently agnostic character of such a theoretical basis that disregards "*a priori*" any kind of local information, in all the cases examined the results coincide with that of wave mechanics, thus showing that the possible degree of knowledge on quantum systems is in fact consistent with the only idea of particles randomly and unpredictably delocalized within their respective uncertainty ranges. According to the previous considerations, the basic assumption of the quoted papers and of the present paper too is summarized as follows

$$E(x, p_x, M^2) \to E(\Delta x, \Delta p_x, \Delta M^2) \to E(n,l) \qquad 1,1$$

where $x$ denotes any generalized coordinate. The logical steps 1,1 do not require any hypothesis or constraint about the motion of the concerned particle and even about its wave/particle nature. The first step simply replaces the local dynamical variables $x, p_x$ with the respective ranges, arbitrary and linked by the relationship

$$\Delta x \Delta p_x = n\hbar \qquad 1,2$$

where the number of states $n$ is in turn arbitrary itself for each freedom degree of the system defined by the couple of conjugate variables. The step 1,2 calculates the numbers of states through elementary algebraic manipulations, as shortly reported in the next section 2 to make the present exposition self-contained and clearer. The early non-relativistic approach was afterwards extended to more complex problems concerning the relativistic free particle, the many electron atoms and the thermodynamic properties of a metal lattice. In these papers however the compliance of the positions 1,1 and 1,2 with relativity was only shortly sketched. The connection between quantum mechanics and relativity is in effect a deep problem that requires a specific examination. The physics of the quantum world rests on the uncertainty principle [3], which replaces the concept of position with that of probability density, whereas in general relativity position and velocities of particles have definite values [4]; the space-time metric is defined by a set of numbers associated with a given point with respect to which is defined the distance of any other point [5]. The quantum theory is non-local [6,7]; Einstein's general relativity exploits local realism, i.e. it excludes interactions between space-like separated physical systems, while the outcomes of measurement reflect pre-existing properties of the systems. This theoretical dualism is in fact consequence of the respective basic assumptions. In relativity they concern invariant and finite light speed, equivalence of reference systems in reciprocal motion and equivalence of inertial and gravitational mass. Just this conceptual basis defines scalar curvature and tensor properties of space-time in the presence of matter [8]. Pillars of the quantum mechanics are the uncertainty, indistinguishability and exclusion principles. So in relativity strength and direction of the gravity field are definite, whereas in quantum theory the fields are subjected to the uncertainty principle, see e.g. [9] in the case of free electromagnetic field. Einstein's gravity does not obey the rules of field theory [10] and appears conceptually far from the weirdness of the quantum world, e.g. the wave/particle dualism; yet relevant coincidences with the electromagnetism have been remarked, e.g. the gravitational waves propagate at the light speed. The quantized results of wave mechanics and the outcomes of the continuous space-time of relativity seem apparently related to two different ways of thinking the reality. The modern theories of gravity regard the space-time metric as a field [11] and attempt to quantize it [12]; certainly, this approach makes the relativistic formalism closer to that of quantum mechanics, yet the quantum gravity is still today a puzzling problem. The string theory, see e.g. [13], is one of today's leading theories for its internal coherence of approach to the quantum gravity; yet it has mostly mathematical worth, being not completely clear how it relates to the standard model and to the known physical universe. While neither Maxwell's electromagnetism nor Einstein's relativity prospect in principle the necessity of extra-dimensions, visible (if any) only at very high energies not presently accessible, there is so far no decisive experimental evidence that confirms the hidden dimensions, e.g. via precision gravity tests. The loop quantum gravity predicts discreteness of spatial area and volume [14], yet it is still unclear how to observe these quantities to validate its theoretical frame. The current efforts aimed to introduce ideas and formalism of either theoretical frame into the other seem implicitly to acknowledge, and then try to overcome, the conceptual gap just sketched; so it is also legitimate to think that the difficulty of merging the two theories stems just from their conceptual roots, i.e. from the attempt of matching a set of mathematical rules and a series of physical enunciates seemingly dissimilar and early formulated for different targets. An attractive alternative to surmount this conceptual asymmetry is to seek a unique assumption that underlies both theories and thus reveals their actual connection. Such a task requires an essential principle of nature so fundamental to infer both basic assumptions of relativity and probabilistic character of quantum mechanics. This idea is effectively viable if the quantum nature of the gravity force could be demonstrated "*ab initio*" on the basis of the sought principle only: if so, the task of harmonizing the two theories would be replaced by that of developing a theoretical frame based on and consequence of the unique root common to both of them. It appears stimulating to identify the sought leading principle with the quantum uncertainty: in [1,2] the idea of regarding the uncertainty as a fundamental principle of nature rather than as a consequence of commutation rule of mathematical operators was formerly introduced to emphasize an approach to the physics of the quantum world alternative to the usual wave formalism. As a matter of fact however this idea does not simply reverse outcomes and assumptions of wave mechanics, rather prospect a more profound physical meaning; the correct results and indistinguishability of quantum particles found as a corollary of a unique physical idea disclose a heuristic path towards the sought generalization of the early non-relativistic model. In fact, this path considers prioritary the task of harmonizing the physical assumptions on which are based quantum mechanics and relativity rather than their respective mathematical formalism. In other words, the reasonable belief of the present paper is to infer the intimate link between both theories by introducing first the quoted key concept and then to exploit the mathematical approach adequate to generalize the early non-relativistic results in a self-consistent way. This idea is decisive to organize the paper, whose main goal is to demonstrate the quantum features and properties of the gravity field. The essential steps to this purpose aim to emphasize: (i) how to exploit in fact the positions 1,1 and eq 1,2, which are the only postulate of the present paper; (ii) how to introduce the time into this conceptual frame; (iii) how to link (i) and (ii) to the operator formalism of quantum mechanics; (iv) how to infer also the gravity from this conceptual frame. Four examples, three of which formerly examined in [1], are reported in the non-relativistic section 2: this section shows thus the validity and reliability of the assertions previously introduced by comparison with the well known results of elementary quantum mechanics. The appendices A and B show that the operator formalism is inferred as a consequence of the only concept of uncertainty together with the corollary of indistinguishability of identical particles, without additional hypotheses. The exclusion principle is also inferred as a corollary in section 3, concerning the special relativity and aimed to show that the examples of section 2 are susceptible of more profound generalization with the help of a further uncertainty equation involving the time. Next, once having proven that the positions 1,1 and 1,2 are compliant with the basic principles of special relativity, an analogous procedure is followed in section 4 to introduce the gravity force as a quantum property and describe the behaviour of particles in the presence of the gravity field. The section 4 also includes Dirac's cosmology and experimental validation of results calculated by the model. The section 5 extends to a spinless particle these ideas. The sections 3 and 4 are the most important ones of the paper: they extend the positions 1,1 to the special and general relativity with the help of eq 1,2 and next eq 2,10 introducing the time uncertainty. This means describing relativistic effects like light beam bending, time dilation and perihelion precession as mere quantum phenomena; the mathematical approach exploited throughout the paper is that of quantum mechanics, the same as it is sketched in section 2.

2 Simple non-relativistic quantum systems.
*2.1 Angular momentum.*
Let $p=|\mathbf{p}|$ and $\rho=|\boldsymbol{\rho}|$ be the moduli of the random momentum and radial distance of one electron from the nucleus of charge $-Ze$. The steps 1,1 require that only $\Delta\rho$ and $\Delta p$ must be considered to describe the system nucleus+electron. No hypothesis is necessary about $\Delta\rho$ and $\Delta p$ to infer the non-relativistic quantum angular momentum and one of its components $\mathrm{M}_w=(\boldsymbol{\rho}\times\mathbf{p})\cdot\mathbf{w}$, being $\mathbf{w}$ an arbitrary unit vector; any detail about the actual electron motion is unessential. As discussed in section 1, the first step 1,1 calculates the number of states allowed for the electron angular motion through the positions $\boldsymbol{\rho}\equiv\Delta\boldsymbol{\rho}$ and $\mathbf{p}\equiv\Delta\mathbf{p}$; putting $\Delta\mathrm{M}_w=(\Delta\boldsymbol{\rho}\times\Delta\mathbf{p})\cdot\mathbf{w}=(\mathbf{w}\times\Delta\boldsymbol{\rho})\cdot\Delta\mathbf{p}$ one finds $\Delta\mathrm{M}_w=\Delta\boldsymbol{\chi}\cdot\Delta\mathbf{p}$, where $\Delta\boldsymbol{\chi}=\mathbf{w}\times\Delta\boldsymbol{\rho}$. If $\Delta\mathbf{p}$ and $\Delta\boldsymbol{\chi}$ are orthogonal $\Delta\mathrm{M}_w=0$, i.e. $\mathrm{M}_w=0$; else, writing $\Delta\boldsymbol{\chi}\cdot\Delta\mathbf{p}$ as $(\Delta\mathbf{p}\cdot\Delta\boldsymbol{\chi}/\Delta\chi)\Delta\chi$ with $\Delta\chi=|\Delta\boldsymbol{\chi}|$, the component $\pm\Delta p_\chi=\Delta\mathbf{p}\cdot\Delta\boldsymbol{\chi}/\Delta\chi$ of $\Delta\mathbf{p}$ along $\Delta\chi$ yields $\Delta M_w = \pm\Delta\chi\Delta p_\chi$. In turn this latter equation yields $M_w = \pm l\hbar$ where $l = 1, 2\cdots$ according to eq 1,2. In conclusion $M_w = \pm l\hbar$, with $l = 0, 1, 2,\cdots$: as expected, $M_w$ is not defined by a single value function because of the angular uncertainty of the electron resulting in turn from the uncertainties initially postulated for $\rho$ and $p$. Being $\Delta\chi$ and $\Delta p_\chi$ arbitrary, the corresponding range of values of $l$ is arbitrary as well; for this reason the notation $\Delta M_w$ is not longer necessary for the quantum result. It appears that $l$ is the number of states related to the electron orbital motion rather than a quantum number, i.e. a mathematical property of the solution of the pertinent wave equation. The quantization of classical values appears merely introducing the delocalization ranges into the classical expression of $M_w$ and then exploiting eq 1,2; the physical definition of angular momentum is enough to find quantum results completely analogous to that of wave mechanics, even disregarding any local detail about the electron motion around the nucleus. The quantity of physical interest to infer $M^2$ is then $l$, since only one component of $\mathbf{M}$ can be actually known: indeed, repeating the procedure for other components of angular momentum would trivially mean changing $\mathbf{w}$. Yet, just this consideration suggests that the average values of the components of angular momentum should be equal, i.e. $\langle M_x^2\rangle = \langle M_y^2\rangle = \langle M_z^2\rangle$. Each term is averaged on the number of states summing $l^2\hbar^2$ from $-L$ to $L$, being $L$ an arbitrary maximum value of $l$; then $\langle M_i^2\rangle = \hbar^2 \sum_{l_i=-L}^{l_i=L} l_i^2/(2L+1)$ yields $M^2 = \sum_{i=1}^{3}\langle M_i^2\rangle = L(L+1)\hbar^2$. Clearly these results do not need any assumption on the specific nature of the electron and have therefore general character and validity for any particle; in effect, after the first step 1,1, the unique information available comes from the very general eq 1,2 no longer involving local coordinates and momenta of a specific kind of particle. In this first example $\Delta\rho$ was in fact coincident with the maximum value $\rho_{max}$ once having defined the random variable $\rho$ in the range of values $0 < \rho \leq \rho_{max}$. More in general, however, the radial uncertainty range could be rewritten as $\Delta\rho' = \rho'_{max} - \rho_o$ without changing the result; $\rho_o$ is the coordinate that defines the origin of $\Delta\rho'$. This is self-evident because neither $\rho_o$ nor $\rho_{max}$ need to be specified in advance and do not appear in the final quantized result. In other words, the quantum expression of $M_w$ does not change whatever in general $\rho_o \neq 0$ might be, since turning $\Delta\rho$ into $\Delta\rho'$ means trivially defining the radial coordinates in a different reference system: yet the considerations about $M_w$ hold identically even in this new reference system, since the key idea of quantum delocalization and the physical meaning of the steps 1,1 remain conceptually identical. In effect $M'_w = (\Delta\boldsymbol{\rho}'\times\Delta\mathbf{p})\cdot\mathbf{w}$ yields $M'_w = \pm\Delta\chi'\Delta p_\chi$; yet this equation provides the same result previously found, because the postulated arbitrariness of the ranges in eq 1,2 entails again arbitrary values of $l'$. In other words, regarding $l$ or $l'$ is trivially the same because both symbolize sets of arbitrary integers rather than specific values; so any particular value changed in either of them replicates identically some value allowed in the other set. Of course the same holds for the radial momentum range and, with analogous reasoning, also for any other uncertainty range; this reasoning will be profitably exploited again in section 3. On the one side $\Delta\rho$ does not compel specifying where is actually located the origin $\rho_o$ of the radial distance range, e.g. somewhere within the nucleus or in the centre of mass of the system or elsewhere; being by definition $\rho = \rho_o + \Delta\rho$, any local coordinate is the limit case of $\Delta\rho \to 0$ through the arbitrary value of $\rho_o$. On the other side this property of the ranges bypasses puzzling problems like how to define the actual distance $\rho$ between electron and nucleus; in lack of any hypothesis about the local coordinates this distance could be even comparable with the finite sizes of these latter, which however are not explicitly concerned. Then, as reasonably expected because of eq 1,2, the conclusion is that the number of allowed states depends upon the range widths only, regardless of the reference systems where these latter are defined; hence the results inferred here hold for any reference system simply by virtue of the first step 1,1. This statement, sensible in non relativistic physics, becomes crucial in relativity for reasons shown in the next sections 3 and 4. In this respect is interesting, in particular for the purposes of the next section 3, a further comment about the limit case where the angular momentum tends to the classical function; for $l \gg 1$ the quantization is not longer apparent and both $M^2$ and $M_w$ are approximately regarded as functions of the continuous variable $l$. This limit case suggests considering the classical modulus of $\Delta \mathbf{M} = \Delta \mathbf{r} \times \Delta \mathbf{p}$, which reads $|\Delta \mathbf{M}| = \Delta r \Delta p \sin \vartheta$, being $\vartheta$ the angle between $\Delta \mathbf{r}$ and $\Delta \mathbf{p}$. This way of regarding $\Delta \mathbf{M}$ is consistent with the quantized result and emphasizes that $|\Delta \mathbf{M}|$ is still due to the range widths $\Delta r$ and $\Delta p$ determining $l$. It has been also shown that even considering different $\Delta \mathbf{r}'$ and $\Delta \mathbf{p}'$ the quantized result is conceptually analogous, while being now $|\Delta \mathbf{M}'| = \Delta r' \Delta p' \sin \vartheta'$. Both expressions must be therefore also equivalent in the classical limit case; hence $\Delta r^2 \Delta p^2 \sin \vartheta^2 = \Delta r'^2 \Delta p'^2 \sin \vartheta'^2$ whatever $\Delta \mathbf{r}'$ and $\Delta \mathbf{p}'$ might be. Although in general $\Delta r^2 \neq \Delta r'^2$ and $\Delta p^2 \neq \Delta p'^2$, because the vectors defining $\Delta \mathbf{M}$ and $\Delta \mathbf{M}'$ are arbitrary, the classical equivalence is certainly ensured by a proper choice of $\vartheta$ and $\vartheta'$. The last equation, expressed as a function of the local dynamical variables included within the respective uncertainty ranges, reads $(rp \sin \vartheta)^2 = (r'p' \sin \vartheta')^2$ and yields the conservation law of angular momentum of an isolated system in agreement with the result already found $M_w = M'_w$. This result holds regardless of the analytical form of $p$. In general, the reasoning above is summarized by

$$\frac{\Delta M'^2}{\Delta M^2} = \frac{\Delta r'^2}{\Delta r^2} \qquad \Delta p^2 \sin \vartheta^2 = \Delta p'^2 \sin \vartheta'^2 \qquad \Delta p^2 \neq \Delta p'^2 \qquad \vartheta \neq \vartheta'$$

The first equation is fulfilled by arbitrary $\Delta p^2$ and $\Delta p'^2$, as it must be, and could be rewritten with the ratio $\Delta p'^2 / \Delta p^2$ at the right hand side; in this case $\Delta r^2$ and $\Delta r'^2$ would appear in the second equation. If both $\Delta M^2$ and $\Delta M'^2$ are calculated with equal ranges of values of $l$ and $l'$, then $\Delta M'^2 / \Delta M^2 = 1$, which also entails $\Delta r^2 = \Delta r'^2$. Since one component only of angular momentum can be defined in addition to the angular momentum itself, taking advantage of the fact that $M'^2 \geq M'^2_w$ and $M^2 \geq M^2_w$ the first equation is rewritten as follows

$$\frac{M'^2 - M^2_w}{M^2 - M^2_w} = \frac{\Delta r'^2}{\Delta r^2}$$

One would have expected $M'^2_w$ at numerator of this equation; yet the position $M'^2_w = M^2_w$ fulfils the limit condition $M'^2 \to M^2$ for $\Delta r'^2 \to \Delta r^2$. Hence

$$M'^2 = M^2 - (M^2 - M^2_w)(1 - \Delta r'^2 / \Delta r^2) \qquad \qquad 2,1$$

In the non-relativistic case $\Delta r^2$ and $\Delta r'^2$ are merely two different ranges by definition arbitrary. It will be shown in section 3 that eq 2,1 is also consistent with the Lorentz transformation of the angular momentum. The same reasoning and formal approach just described hold to calculate the non-relativistic electron energy levels of hydrogenlike atoms and harmonic oscillators.

*2.2 Hydrogenlike atoms.*

The starting function is the classical Hamiltonian of electron energy in the field of the nucleus, which reads in the reference system fixed on the centre of mass

$$E = E_{cm} + \frac{p_\rho^2}{2\mu} + \frac{M^2}{2\mu\rho^2} - \frac{Ze^2}{\rho} \qquad E = E(\rho, p_\rho, M^2) \qquad U = -\frac{Ze^2}{\rho}$$

Being $\mu$ the electron reduced mass and $E_{cm}$ the centre of mass kinetic energy of the atom regarded as a whole. Also now $E(n,l)$ is obtained replacing the dynamical variables, unknown in principle, with the respective uncertainty ranges. In agreement with the previous discussion, also the uncertainty on $U$, due to the random radial distances allowed to the electron, concurs to define the numbers $n$ and $l$ of quantum states unequivocally defined and necessarily consistent with the radial ranges $\Delta\rho$ and $\Delta p_\rho$. This is in effect the physical meaning of the positions $U(\rho) \Rightarrow U(\Delta\rho)$ and $p_\rho^2 + M^2/\rho^2 \Rightarrow \Delta p_\rho^2 + M^2/\Delta\rho^2$: putting $p_\rho \equiv \Delta p_\rho$ and $\rho \equiv \Delta\rho$, the number of states allowed to the electron motion in the field of nucleus are calculated in agreement with the given form of the potential and kinetic energies. The energy equation turns then into the following form

$$E^* = E_{cm} + \frac{\Delta p_\rho^2}{2\mu} + \frac{\Delta M^2}{2\mu\Delta\rho^2} - \frac{Ze^2}{\Delta\rho} \qquad\qquad E^* = E^*\left(\rho \equiv \Delta\rho, p_\rho \equiv \Delta p_\rho\right)$$

The uncertainty on $M^2$ is taken into account by the range of arbitrary values allowed to $l$, whereas a further arbitrary value $n$ is to be introduced through $n = \left(2\Delta\rho\Delta p_\rho / \hbar\right)/2$ because of eq 1,2. The factor 2 within parenthesis accounts for the possible states of spin of the electron, which necessarily appears as "ad hoc" hypothesis in the present non-relativistic example. The factor ½ is due to the fact that really $p_\rho^2$ is consistent with two possible values $\pm p_\rho$ of the radial component of the momentum corresponding to the inwards and outwards motion of the electron with respect to the nucleus; by consequence, being the uncertainty range $\Delta p_\rho$ clearly the same in both cases, the calculation of $n$ simply as $2\Delta\rho\Delta p_\rho / \hbar$ would mean counting separately two different situations both certainly possible for the electron but actually corresponding to the same quantum state. These situations are in fact physically undistinguishable because of the total uncertainty assumed "a priori" about the central motion of the electron; then the factor ½ avoids counting twice a given quantum state. In conclusion, the only information available in the energy equation concerns $n$ and $l$ consistent with the radial and angular motion of the electron; they take in principle any integer values because the uncertainty ranges $\Delta\rho$ and $\Delta p_\rho$ include arbitrary values of $\rho$ and $p_\rho$ and then are arbitrary themselves. Replacing $\Delta p_\rho$ with $n\hbar/\Delta\rho$ and $M^2$ with $(l+1)l\hbar^2$ in $E^*$, the result is

$$E^* = E_{cm} + \frac{n^2\hbar^2}{2\mu\Delta\rho^2} + \frac{l(l+1)\hbar^2}{2\mu\Delta\rho^2} - \frac{Ze^2}{\Delta\rho}$$

Trivial manipulations of this equation yield

$$E^* = E_{cm} + \frac{1}{2\mu}\left(\frac{n\hbar}{\Delta\rho} - \frac{Ze^2\mu}{n\hbar}\right)^2 + \frac{(l+1)l\hbar^2}{2\mu\Delta\rho^2} - \frac{Z^2e^4\mu}{2n^2\hbar^2}$$

$E^*$ is minimized putting equal to zero the quadratic term within parenthesis, certainly positive; being $E = \min\left(E^*\right)$ the result is

$$\Delta\rho_{\min} = \frac{n^2\hbar^2}{Ze^2\mu} \qquad \Delta p_{\rho\min} = \frac{n\hbar}{\Delta\rho_{\min}} = \frac{Ze^2\mu}{n\hbar} \qquad E = E_{cm} + \frac{(l+1)l\hbar^2}{2\mu\Delta\rho_{\min}^2} - \frac{Z^2e^4\mu}{2n^2\hbar^2} \qquad 2,2$$

Then the total quantum energy $E(n,l)$ of the hydrogenlike atom results as a sum of three terms: (i) the kinetic energy $E_{cm}$ of the centre of mass of the atom considered as a whole, (ii) the quantum rotational energy of the system consisting of a reduced mass $\mu$ moving within a distance $\Delta\rho_{\min}$ from the nucleus and (iii) a negative term necessarily identified as the non-relativistic binding energy $\varepsilon_{el}$ of the electron. The values allowed to $l$ must fulfil the condition $l \leq n-1$. So, rewriting $E$ in a reference system with the centre of mass at rest, $E_{cm} = 0$ and utilizing $\Delta\rho_{\min}$, the result is

$$\varepsilon_{el} = \left[\frac{(l+1)l}{n^2} - 1\right]\frac{Z^2e^4\mu}{2n^2\hbar^2} \qquad\qquad l \leq n-1$$

If $l \geq n$ then the total energy $\varepsilon$ would result $\geq 0$, i.e. the hydrogenlike atom would not entail an electron bound state. Since the stability condition requires the upper value $n-1$ for $l$, it is possible to write $n = n_o + l + 1$, where $n_o$ is still an integer. Hence

$$\varepsilon_{el} = -\frac{Z^2 e^4 \mu}{2\hbar^2 (n_o + l + 1)^2} \qquad 2,3$$

In conclusion, all the possible terms expected for the non-relativistic energy are found in a straightforward and elementary way, without hypotheses on the ranges and without solving any wave equation: trivial algebraic manipulations replace the solution of the appropriate wave equation. It is worth emphasizing that the correct result needs introducing the concept of electron spin to count appropriately the number of allowed states, whereas the non-relativistic wave equation solution skips such a requirement. The present approach requires therefore necessarily the concept of spin, although without justifying it; this problem will tackled in the next section 3. As concerns the positions 1,1, it is also worth noticing that only the first step $E(x, p_x, M^2) \rightarrow E(\Delta x, \Delta p_x, \Delta M^2)$ concerns the particles, whereas the second step $E(\Delta x, \Delta p_x, \Delta M^2) \rightarrow E(n,l)$ concerns in fact their phase space; indeed $E(n,l)$ is a function of the number of quantum states, which are properties of the phase space like the pertinent ranges. This is especially important when considering many electron atoms: the fact that any specific reference to the electrons is lost entails as a corollary the concept of indistinguishability; $n_i$ and $l_i$ of the $i$-th electron are actually numbers of states pertinent to delocalisation ranges where *any* electron could be found, instead of quantum numbers of a specified electron. The energy levels of many electron atoms and ions have been then inferred without possibility and necessity of specifying which electron in particular occupies a given state; in effect, the electrons cannot be identified if nothing is known about each one of them. The paper [2] shows that the same ideas hold also to calculate the binding energy of diatomic molecules. The lack of local information inherent the assumptions 1,1 and 1,2 entails then in general the indistinguishability of identical particles. A closing remark concerns the correspondence principle. The quantized angular momentum and electron energy levels approximate reasonably well the continuous behaviour of the corresponding classical quantities for $l \gg 1$ and $n \gg 1$. This result is particularly significant in the present theoretical frame based on the unique assumption of quantum uncertainty, whose formulation in eq 1,2 however never allows both $\Delta x \rightarrow 0$ and $\Delta p_x \rightarrow 0$ because of the integer values of $n$. As mentioned in the introduction, any range size tending to zero turns into a local value exactly defined; so this agrees with the lack of deterministic knowledge in the quantum world. Yet this holds for any $n$, whereas one would have expected that for large $n$ eq 1,2 should admit itself the classical limit with both conjugate dynamical variables exactly predictable. The failure of this requirement suggests that the conceptual link between quantized and classical dynamical variables is more complex than the mere choice of $n$; the question raises about why the outcomes of eq 1,2 fulfil the correspondence principle, whereas eq 1,2 itself does not. A possible answer is that the eigenvalues of quantum systems do not depend on the range sizes, which appear in effect arbitrary and indeterminate in the previous examples and in the next ones; so inquiring into their limit behaviour, classical or not, could seem superfluous or out of place. Remains however important in principle the problem of understanding how to include the concept of classical dynamical variables as a limit case of the present theoretical frame. The first key idea in this respect concerns the arbitrariness of the ranges: describing a quantum system through eq 1,2 or through any other ranges $\Delta x' \Delta p_x' = n'\hbar$ is exactly the same provided that $n'$ be still arbitrary integer; if so, this last equation is actually eq 1,2 simply rewritten with different notation in a different reference system. This appears considering that by definition $n$ does not represent a set of values assigned or somehow identifiable, rather it just symbolizes abstractly any integer value; so it is meaningless to regard in a different way $n$ and $n'$ once recognizing that any value allowed to the former is also allowed to the latter by definition. The second key idea concerns the fact that there is no reason to expect the number $n^{cl}$ of states of classical system necessarily equal to $n$ of quantum system; rather it seems sensible the exact contrary, as the ways of counting quantum and classical states are reasonably different. If so, the failure of eq 1,2 in representing the classical limit could be due just to the conceptual discrepancy between $n$ and $n^{cl}$. To explain this point consider eq 1,2 together with the classical expression from it obtained replacing $n$ with $n^{cl}$, i.e. $\Delta x^{cl} \Delta p_x^{cl} = n^{cl} \hbar$; of course $\Delta x^{cl}$ and $\Delta p_x^{cl}$ are classical ranges that define $n^{cl}$. As already emphasized in the introduction, nothing hinders in principle to introduce even in classical physics coordinate and momentum ranges including random values of the respective variables; in effect the related relationship is certainly fulfilled for any $n^{cl}$ fixing arbitrarily $\Delta x^{cl}$ and then finding the corresponding $\Delta p_x^{cl}$, just as it would happen for eq 1,2. Yet, if coordinates and momenta are both exactly known, the respective classical ranges have known sizes as well; it does not hold instead in eq 1,2. Just this is the crucial difference between the classical and quantum ways of thinking. Comparing eq 1,2 with its classical formulation aims therefore to highlight the peculiar physical meaning of the respective products of ranges and explain the divergent consequences arising despite their formal analogy. Combining the equations yields $(\Delta x^{cl}/\Delta x)(\Delta p_x^{cl}/\Delta p_x) = n^{cl}/n$. This result suggests two cases of special interest: (i) for $n^{cl} \ll n$ the left hand side tends to zero, so the chance of both $\Delta x^{cl} \ll \Delta x$ and $\Delta p_x^{cl} \ll \Delta p_x$ agrees with $\Delta x^{cl} \to 0$ and $\Delta p_x^{cl} \to 0$ simultaneously whatever $\Delta x \Delta p_x$ might be; (ii) $n^{cl} \lesssim n$ instead yields $\Delta x^{cl} \Delta p_x^{cl} \lesssim \Delta x \Delta p_x$ and thus $\delta x \delta p_x \gtrsim \hbar$ if the values of both $n$ and $n^{cl}$ are large enough, being $\delta x \delta p_x = \Delta x \Delta p_x / n^{cl}$. So the obvious conclusion is that $\Delta x$ and $\Delta p_x$ never tend both to zero because they are quantum ranges, whereas instead $\Delta x^{cl}$ and $\Delta p_x^{cl}$ do for $n \to \infty$ because they are classical ranges. As concerns the other possibilities to compare $n$ and $n^{cl}$, the inequality $n^{cl} \gtrsim n$ would yield $\Delta x^{cl} \Delta p_x^{cl} \gtrsim \Delta x \Delta p_x$ and thus $\delta x^{cl} \delta p_x^{cl} \gtrsim \hbar$ that simply replicates the previous result; eventually $n^{cl} \gg n$ would yield $\Delta x^{cl} \Delta p_x^{cl} \gg \hbar$ whatever $\Delta x \Delta p_x$ might be. This last case however does not anything of relevant physical interest. To summarize: eq 1,2 has exclusive quantum character, whereas $\delta x \delta p_x \gtrsim \hbar$ has classical character since it admits also $\delta x^{cl} \delta p_x^{cl} \gtrsim \hbar$; the former compels per se the quantization, recall indeed that $n$ is both number of allowed states and quantum number of the eigenvalues, the latter merely helps to tackle quantum problems, think for instance to the calculation of electron density in Thomas-Fermi atoms. As the case (i) is consistent with the expected classical limits of $\Delta x^{cl}$ and $\Delta p_x^{cl}$, it is interesting to justify at least in principle why the inequalities $n^{cl} \ll n$ and $n^{cl} \lesssim n$ are both reasonable and allowed to occur. Let us estimate $n^{cl}$ supposing first that both ranges $\Delta x_i^{cl}$ and $\Delta p_{xi}^{cl}$ of the $i$-th allowed state are known and distinguishable among the others; the classical case defines therefore one by one the conjugate ranges that build up the various $\Delta x_i^{cl} \Delta p_i^{cl} = \hbar$, so that $\Delta x^{cl} \Delta p_x^{cl} = \sum_i \Delta x_i^{cl} \Delta p_{xi}^{cl} = n^{cl} \hbar$ with $n^{cl} \hbar = \sum_i \hbar$. Suppose now that $\Delta x_i$ are unknown and thus indistinguishable like the respective $\Delta p_{xi}$; assigning as before specific range sizes of dynamical variables to the $i$-th state to calculate the respective $n_i$ is not longer possible. Yet, instead of the previous products, it is possible to calculate separately $\Delta x = \sum_i \Delta x_i$ and $\Delta p_x = \sum_i \Delta p_{xi}$; this yields $\Delta x \Delta p_x = \left(\sum_i \Delta x_i\right)\left(\sum_i \Delta p_{xi}\right) = n \hbar$, as the sums merged together are subjected to the uncertainty constrain. So one expects $\left(\sum_i \Delta x_i\right)\left(\sum_i \Delta p_{xi}\right) \gg \sum_i \Delta x_{xi}^{cl} \Delta p_{xi}^{cl}$ for large $n$, and thus $n \gg n^{cl}$ because of the much higher number of terms at left hand side of the inequality, which scales with $n^2$, with respect to that at right hand side, which scales with $n^{cl}$. Note however that since the inequality does not involve the same range sizes, in general also the chance $\left(\sum_i \Delta x_i\right)\left(\sum_i \Delta p_{xi}\right) \gtrsim \sum_i \Delta x_{xi}^{cl} \Delta p_{xi}^{cl}$ and thus $n^{cl} \lesssim n$ or even the other chances quoted above cannot be excluded. Nevertheless this remark does not nullify the physical importance of the previous reasoning. The main consequence of comparing the relative values of $n$ and $n^{cl}$ is that of admitting reasonably in principle, and thus legitimating, the chance $n >> n^{cl}$ thanks to which the present quantum approach allows in fact the expected classical limit of conjugate dynamical variables. It is supportive in this respect the fact that the chances alternative to $n >> n^{cl}$ cannot conflict with this conclusion as they concern different ratios $n^{cl}/n$ and thus other consequences of eq 1,2, e.g. the well known relationship $\delta x \delta p_x \gtrsim \hbar$.

*2.3 Plasma.*

Consider a non-relativistic gas of $n_e$ electrons having mass $m_e$ confined in the linear space range $\Delta \rho = |\Delta \boldsymbol{\rho}|$ at equilibrium temperature $T_e$ in the absence of applied external potential. Regard $\Delta \rho$ as the physical size of a 2,4D box where the electrons move by effect of their mutual repulsion and thermal kinetic energy; its arbitrary size corresponds thus by definition to the delocalization extent of $n_e$ electrons. The only assumption of the model is $\Delta p_\rho \approx \hbar / \Delta \rho$, where $\Delta p_\rho$ is the range including the momenta components of all the electrons along $\Delta \boldsymbol{\rho}$. No hypotheses are necessary about $\Delta \rho$ and $\Delta p_\rho$. The uncertainty principle prevents knowing local position and momentum of electrons; it is possible however to define their average distance $\overline{\Delta \rho_{ne}} = \Delta \rho / (n_e - 1)$ and also to introduce the sub-range $\delta \rho < \Delta \rho$ encompassing the random distance between any local couple of contiguous electrons. Whatever $\delta \rho$ might be, its size must be a function of time in order to contain two electrons moving away each other because of their electric repulsion. To describe the dynamics of this couple, consider first the general problem of two charges $\delta \rho$ apart and let $\delta p_\rho$ be the range including the local momenta components $p_\rho$ allowed by their electric interaction. In general and without any hypothesis $\delta p_\rho$ must have the form $p_\rho - p_{o\rho}$ or $p_{o\rho} - p_\rho$ with $p_\rho$ time dependent and $p_{o\rho}$ constant, both arbitrary; the latter is defined by the momentum reference system, the former by the interaction strength. So $\dot{p}_\rho = \delta \dot{p}_\rho \approx \pm \hbar \delta \dot{\rho} / \delta \rho^2$ is the repulsion/attraction force experienced by one charge by effect of the other. Consider now the upper sign to describe in particular the mutual repulsion between two electrons and make the expansion rate $\delta \dot{\rho}$ tending asymptotically to $c$ to ensure that the electrons cannot travel beyond $\delta \rho$ whatever their current repulsion force might be; this chance, in fact allowed by the arbitrary sizes of $\delta \rho$ and $\Delta \rho$, yields the sought repulsion force $\dot{p}_\rho \approx e^2 / (\alpha \delta \rho^2)$, being $\alpha$ the fine structure constant. Introduce now a proportionality constant, $\varepsilon'_0$, to convert the order of magnitude link provided by the uncertainty principle into an equation; merging $\alpha$ and $\varepsilon'_0$ into a unique constant, $\varepsilon_0$, the force between the electrons has the well known form $e^2 / (\varepsilon_0 \delta \rho^2)$ with $\varepsilon_0$ defined by the charge unit system. The average repulsion energy between any isolated pair of contiguous electrons at distance $\overline{\Delta \rho_{ne}}$, i.e. neglecting that of all the other electrons, is then $\overline{\eta}_{cont} = e^2 / (\varepsilon_0 \overline{\Delta \rho_{ne}})$; also, the average repulsion energy acting on one test electron by effect of all the others is $\overline{\eta}_{rep} = (n_e - 1)^{-1} \sum_{i=1}^{n_e - 1} e^2 / (\varepsilon_0 \Delta r_i)$, where $\Delta r_i$ are the distances between the $i$-th electrons having local coordinates $r_i$ and the test electron. Let us put now by definition $\overline{\Delta \rho_{ne}}^{-1} = f \sum_{i=1}^{n_e - 1} \Delta r_i^{-1}$. Formally this equation replaces the sum of all the unknown actual distances of the electrons from the test electron with the reciprocal average distance through the unique arbitrary parameter $f \neq 0$, by definition positive, describing the possible configurations of the electron system; one expects thus a simpler expression of $\overline{\eta}_{rep}$ as a function of $\overline{\Delta \rho_{ne}}$ and, through this latter,

of $\Delta\rho$ as well. The condition that all the electrons be in $\Delta\rho$ requires $\Delta r_i < \Delta\rho$, i.e. $\Delta r_i = \chi_i \Delta\rho$ with $\chi_i < 1$; so $(n_e - 1)/f = \sum_{i=1}^{n_e-1} \chi_i^{-1}$ is fulfilled with a proper choice of $f$ whatever the various $\Delta r_i$ might be. In turn $\chi_i < 1$ require $f < 1$, since for $f = 1$ all the $\Delta r_i$ should be equal to $\Delta\rho$. Hence

$$\Delta\rho^{-1} = (n_e - 1)^{-1} f \sum_{i=1}^{n_e-1} \Delta r_i^{-1} \qquad 0 < f < 1 \qquad 2,4$$

Also, expressing $\Delta r_i^{-1}$ as a function of $1/\overline{\Delta\rho_{ne}}$ as $\Delta r_i^{-1} = \xi_i / \overline{\Delta\rho_{ne}}$ through the parameters $\xi_i > 0$ one finds $\sum_{i=1}^{n_e-1} \xi_i = 1/f$; regardless of the unknown and arbitrary sizes of $\Delta\rho$ and $\Delta r_i$, this result simply requires $\chi_i \xi_i = (n_e - 1)^{-1}$. Eqs 2,4, possible in principle from a mathematical point of view, have also physical interest because they relate $\Delta\rho$ and $\overline{\Delta\rho_{ne}}$ to $\overline{\eta}_{rep}$:

$$\overline{\eta}_{rep} = \frac{e^2}{f \varepsilon_0 \Delta\rho} \qquad \overline{\eta}_{cont} = \frac{e^2}{\varepsilon_0 \overline{\Delta\rho_{ne}}}$$

In conclusion our degree of knowledge about the system is summarized by $\overline{\eta}_{rep}$ and $\overline{\eta}_{cont}$, linked by the unknown parameter $f$: the former energy concerns the average collective behaviour of all the electrons, the latter that of a couple of electrons only. On the one side, this conclusion is coherent with the general character of the present approach that disregards specific values of local dynamical variables; in effect any kind of information about $f$ would unavoidably require some hypothesis on the conjugate dynamical variables themselves, which are instead assumed completely random, unknown and unpredictable within their respective uncertainty ranges. On the other side, just the impossibility of specifying the various $r_i$, which prevents establishing preferential values of $f$, compels regarding the properties of the electron gas through its whole uncertainty range $\Delta\rho$ and the whole range of values allowed for $f$. In other words, to each value possible for $f$ corresponds a possible electron configuration of the system physically admissible. For instance, consider in particular the chances $f \to 0$ or $f \to (n_e - 1)^{-1}$ or $f \to 1$ to illustrate at increasing values of $f$ the related information about the respective electron configurations. The first chance $f \to 0$ requires at least one or several $\Delta r_i$ tending to zero, because the possibility of finite $\Delta\rho$ cannot be excluded whatever $f$ might be; to this clustering effect around the test electron corresponds thus an expected increase of $\overline{\eta}_{rep}$. Note however that even in this case, in principle possible, the average energy $\overline{\eta}_{cont}$ between any couple of electrons does not diverge being defined by $\overline{\Delta\rho_{ne}}$ only. This result alone describes of course only a partial aspect of the real plasma state; more exhaustive physical information is obtained examining the further choice of values possible for $f$. If $f \to (n_e - 1)^{-1}$ then $\overline{\eta}_{rep} \to \overline{\eta}_{cont}$, i.e. the average repulsion energy $\overline{\eta}_{rep}$ acting on the test electron tends to that of an isolated couple of contiguous electrons. Moreover, if $f \to 1$ then $\xi_i < 1$ mean $\Delta r_i > \overline{\Delta\rho_{ne}}$, i.e. the distances of the various electrons from the test one are greater than the average value; in this case $\overline{\eta}_{rep}$ tends to $\overline{\eta}_{cont}/(n_e - 1)$, i.e. it is even smaller than before. Summarizing the discussion above, the clustering of electrons around the test electron appears energetically unfavourable, whereas more likely result instead increasing values of $f$ that diminish $\overline{\eta}_{rep}$ down to $\overline{\eta}_{cont}$ or to the smaller value $\overline{\eta}_{cont}/(n_e - 1)$ tending even to zero for large $n_e$. So, without specifying any electron in particular and noting that by definition $\overline{\eta}_{cont}$ corresponds to the energy of a test charge $e$ on which act all the other $(n_e - 1)e$ charges located at the maximum possible distance $\Delta\rho$, the whole range of values allowed to $f$ reveals the preferential propensity of the system to create holes in the linear distribution of average charge around any test electron that screen the repulsive effect of other mobile electrons. Since the test electron is indistinguishable with respect to the others, this behaviour holds in fact for any electron; it is essential in this respect the random motion of mobile electrons that tend to repel each other, not $n_e$ or the local position and momentum of each electron. This picture of the system, which holds regardless of the actual size of $\Delta\rho$ and despite the lack of specific values definable for $f$, is further exploited considering that two electrons are allowed in each energy state corresponding to their possible spin states. Whatever the specific value of $f$ might be, let us examine the behaviour of one of such couples with energy $\bar{\eta}_{cont}$ formed by the test electron and one among the $n_e - 1$ residual electrons at distance $\Delta\rho$, assuming in the following $n_e >> 1$ and $T_e$ high enough to regard the electrons as a classical gas of particles characterized by random motion in agreement with the previous reasoning. Each electron of the test couple has thus average Coulomb energy $\bar{\eta}_{cont}/2 = n_e e^2 /(2\varepsilon_0 \Delta\rho)$ and thermal kinetic energy $\bar{\eta}_{th} = 3KT_e/2 + \bar{\eta}_{corr}/2$; the first addend describes the electrons of the couple as if they would be free gas particles, the second is the obvious correction due to their actual electrical interaction. It will be shown below that in fact this correction term can be neglected if the gas is hot enough, because of the propensity of the system to high values of $f$ that reduce $\bar{\eta}_{rep}$; yet we consider here for generality both terms, noting that $\bar{\eta}_{corr}$ is an unknown function of $n_e$ and $f$ (more exactly of $f^{-1} - 1$) since it depends on the shielding strength between the electrons of the couple provided by the charge holes within $\Delta\rho$. The factor 2,6, which accounts for three freedom degrees of thermal motion, does not conflict with the electron confinement within the linear delocalization range $\Delta\rho$. The previous discussion has regarded the size of uncertainty range only, rather than the vector $\Delta\boldsymbol{\rho}$ that is actually not uniquely defined: owing to the lack of hypotheses about $\Delta\rho$, any vector with equal modulus randomly oriented with respect to an arbitrary reference system is in principle consistent with the aforesaid results. So there is no reason to think $\Delta\boldsymbol{\rho}$, whatever its actual modulus might be, distinctively oriented along a prefixed direction of the space; thus cannot be excluded even the idea that the orientation of $\Delta\boldsymbol{\rho}$ changes as a function of time. In fact this conclusion suggests that, under proper boundary conditions, the whole vector $\Delta\boldsymbol{\rho}$ can be considered free to rotate randomly in the space at constant angular rate $\boldsymbol{\omega}'$. This simply means introducing into the problem the angular position uncertainty of the electrons together with their radial uncertainty, the only one so far concerned. The simultaneous angular motion of all the electrons does not change the reasoning above about the mutual repulsion energies, while any possible alteration of electron configuration in the rotating frame is still described by the parameter $f$ in its unchanged range of values: nothing was known about the possible local electron configurations before introducing the frame rotation, nothing is known even now about their possible modification. Although the space orientation of the rotation axis is clearly indefinite, introducing the angular uncertainty helps to explain the physical meaning of average energy $\bar{\eta}_{cont}$: on average, the $n_e - 1$ electrons moving randomly in radial direction are statistically distributed on the surface of a sphere of radius $\Delta\rho$ centred on the test electron. Plays a crucial role in this context the indistinguishability: this picture holds for any electron, without possibility of specifying which one, and thus in fact for all the electrons of the system. The lack of further information does not preclude however to define the total energy balance of the test couple electron delocalized within $\Delta\rho$ with velocity resulting by: (i) its momentum randomly falling within the range $\Delta p_\rho$, (ii) the angular motion of its delocalization range $\Delta\boldsymbol{\rho}$ as a whole and (iii) the thermal random contribution, whose average modulus is $v_{th}^2 = (3KT_e + \bar{\eta}_{corr})/m_e$. Of course the modulus $\omega'$ is not arbitrary: its value is in fact determined by the driving energies of the system because the angular, thermal and electric terms must fulfil the condition $\bar{\eta}_{ang} = \bar{\eta}_{th} + \bar{\eta}_{cont}/2$. In this way the energy of angular motion of the test electron described by $\bar{\eta}_{cont}$ is the same as that in the non-rotating linear range $\Delta\rho$, but simply expressed in a different form, i.e. as a function of the angular uncertainty instead of the linear uncertainty only. Moreover, whatever the random motion of each electron might be, the concurrence of radial and angular uncertainties agrees with the energy conservation expected for the whole isolated system described by average quantities only. Wherever the current position of the test electron in the space might be, the average angular motion energy reads $\bar{\eta}_{ang} = m_e \omega'^2 \Delta\rho^2 / 2$: since $\bar{\eta}_{cont}$ appearing at right hand side of the energy balance is the average energy of the concerned electron couple, the aforementioned reasoning suggests to regard the test electron at distance $\Delta\rho$ from the other $n_e - 1$ electrons around it. Then

$$\frac{m_e \omega'^2 \Delta\rho^2}{2} = \frac{3}{2} KT_e + \frac{\bar{\eta}_{corr}}{2} + \frac{n_e e^2}{2\varepsilon_0 \Delta\rho} \qquad 2,5$$

As expected, this result does not depend on some particular value of $f$ in the approximation of negligible $\bar{\eta}_{corr}$ and yields

$$\omega'^2 - \frac{\bar{\eta}_{corr}}{m_e \Delta\rho^2} = 3\frac{KT_e}{m_e \Delta\rho^2} + \frac{n_e e^2}{m_e \varepsilon_0 \Delta\rho^3}$$

Introduce now the electron number density $N_e = n_e / \Delta\rho^3$ and define the correction term according to its physical dimensions putting $\bar{\eta}_{corr}/(m_e \Delta\rho^2) = \omega_{corr}^{-2}$. Moreover exploit the fact that the test electron rotating along the circumference $2\pi\Delta\rho$ has De Broglie's wavelength $\lambda = 2\pi\Delta\rho$ and momentum $p_\lambda = \hbar k$, being $k = 2\pi/\lambda$. Here has been considered the fundamental oscillation only, omitting the shorter wavelengths described by integer multiples $n\lambda = 2\pi\Delta\rho$ of $\lambda$. One finds thus

$$\omega^2 = \frac{N_e e^2}{\varepsilon_0 m_e} + 3k^2 \frac{KT_e}{m_e} \qquad N_e = \frac{n_e}{\Delta\rho^3} \qquad \omega^2 = \omega'^2 - \omega_{corr}^2 \qquad \omega_{corr}^2 = \frac{\bar{\eta}_{corr}}{m_e \Delta\rho^2} \qquad k = \frac{1}{\Delta\rho} \qquad 2,6$$

This result assigns to the frequency $\omega$ the physical meaning of collective property of electrons, owing to the fact that it is defined through average quantities. In this result is hidden the electron characteristic length $\lambda_{eD}$ as well; the first eq 2,6 can be indeed rewritten more expressively as

$$\omega^2 = \omega_p^2 (1 + 3k^2 \lambda_{eD}^2) \qquad \omega_p^2 = \frac{N_e e^2}{\varepsilon_0 m_e} \qquad \lambda_{eD} = \sqrt{\frac{\varepsilon_0 KT_e}{e^2 N_e}}$$

This result, well known, can be further exploited considering fixed $\omega^2$ in eq 2,6. Eq 2,5 shows that for $\Delta\rho$ large enough the Coulomb term becomes negligible with respect to the thermal energy, in which case $\Delta\rho_{eD}^2 \approx 3KT_e/(m_e\omega^2)$; the subscript denotes the particular value of $\Delta\rho^2$ fulfilling this limit condition of the test electron in the gas, which justifies why $\bar{\eta}_{th}$ is in fact well approximated by the free electron energy term only, i.e. $\omega^2 \approx \omega'^2$. Replace now $m_e\omega^2$ in the first eq 2,6 regarded in particular for $\Delta\rho^2 \equiv \Delta\rho_{eD}^2$ and $\Delta p_\rho^2 \equiv \Delta p_{eD}^2$; then $\Delta\rho_{eD}^2 = (N_e e^2/(3KT_e\varepsilon_0) + \Delta p_{eD}^2/\hbar^2)^{-1}$. In general, large $\Delta\rho$ entails accordingly small $\Delta p_\rho/\hbar$. If $\Delta p_{eD}/\hbar$ is small enough and $N_e$ high enough to have $N_e e^2/\varepsilon_0 >> 3KT_e\Delta p_{eD}^2/\hbar^2$ the result is $\Delta\rho_{eD}^2 \to 3KT_e\varepsilon_0/(N_e e^2) = 3\lambda_{eD}^2$. Clearly the vanishing Coulomb term means that the screening effect due to the motion of the plasma charges is controlled by the characteristic scale length $\lambda_{eD}$. At this point it is also immediate to infer what changes if instead of an electron gas only one considers a plasma made by $n_e$ electrons at average temperature $T_e$ plus $n_i$ ions with charge $-Ze$ at average temperature $T_i$. Consider first in $\Delta\rho$ the ion gas only. Simply repeating the reasoning above, the result becomes $\Delta\rho_{iD}^2 = (N_i Z^2 e^2/(3KT_i\varepsilon_0) + \Delta p_{iD}^2/\hbar^2)^{-1}$, analogous to that obtained before; of course also now $\Delta\rho_{iD}$ is obtained through the positions $\Delta\rho^2 \equiv \Delta\rho^2_{iD}$ and $\Delta p_\rho^2 \equiv \Delta p^2_{iD}$. If ions and electrons are both confined in the same range, $\Delta\rho_{eD}$ and $\Delta\rho_{iD}$ must coincide. Observing the results just obtained, this condition appears fulfilled if $\Delta p^2_{iD}/\hbar^2 = N_e e^2/(3KT_e\varepsilon_0)$ and $\Delta p^2_{eD}/\hbar^2 = N_i Z^2 e^2/(3KT_i\varepsilon_0)$; then, the position $\Delta p^2_D = \Delta p^2_{eD} + \Delta p^2_{iD}$ yields $\Delta\rho^2_D = \hbar^2(\Delta p^2_{eD} + \Delta p^2_{iD})^{-1} = \hbar^2\Delta p^{-2}_D$, as expected. The electro-neutrality $N_p = N_e = N_i Z$ yields

$$\Delta\rho^2_D = 3\lambda^2_D \qquad \lambda^2_D = \frac{K\varepsilon_0}{e^2 N_p}\frac{1}{T_e^{-1} + ZT_i^{-1}} \qquad N_p = N_e = N_i Z \qquad \Delta\rho_D \equiv \Delta\rho_{eD} \equiv \Delta\rho_{iD} \qquad 2,7$$

Eqs 2,6 and 2,7 evidence that the expression of $\lambda_D$, the plasma frequency and the basic concepts of plasma physics are simply hidden within the quantum uncertainty, from which they can be extracted through an elementary and straightforward reasoning. For comparison purposes, it is instructive at this point to remind shortly the two key-steps through which are usually inferred the plasma properties: (i) to assume Coulomb law and Boltzmann-like number density of electrons/ions, according to the idea that a high local probability of finding a particle is related to a high local charge density; (ii) to solve the potential Poisson equation assuming $eV \ll KT_e$, being $V$ the electric potential. The present approach is of course necessarily different, because the lack of information about the distances between the electrons compels introducing the parameter $f$ that replaces the unknown quantities $\Delta r_i$ and because the plasma properties are inferred through eqs 2,4 containing average quantities only; moreover the classical assumption $\overline{\eta}_{th} \approx 3KT_e/2$ replaces $eV \ll KT_e$. Strictly speaking $f$ cannot be considered unknown, since in fact it is conceptually not definable by assigned values; its arbitrariness is nothing else but that physically inherent the uncertainty ranges $\Delta\rho$ and $\Delta p_\rho$. Yet just $\Delta\rho$ enabled the well known Debye lengths of electrons and ions to be also found through the simple energy balance of the respective test charges and the boundary condition of electro-neutrality of the plasma. Defining the characteristic lengths $\Delta\rho_{eD}$ and $\Delta\rho_{iD}$ does not mean however determining the range $\Delta\rho$, which remains arbitrary because of the presence of the terms $\Delta p^2_{eD}/\hbar^2$ and $\Delta p^2_{iD}/\hbar^2$. Note in this respect that the plasma properties require $\Delta\rho > \Delta\rho_D$, i.e. they appear when the delocalization extent of the gas of charged particles is greater than the Debye length of electrons and ions, which fix therefore the length scale above which hold the peculiar features of plasma physics. Replacing momenta and coordinates with the respective uncertainty ranges means renouncing "a priori" to any information about motion and position of particles; consequently no specific representation of the electron system could have been expected through such a conceptual background. Consider for instance the approximate solution of Poisson's equation $V \sim r^{-1}\exp(-r/\lambda_D)$, calculable as a function of $r$; here of course such an information is missing once having skipped $r$. Yet, on the one side the information inferred about the system is enough to highlight the same physical consequences, e.g. the reduced penetration depth of the electric field within the plasma; on the other side, once having found the Debye lengths that control this depth, one could easily infer the quoted form of $V$ simply repeating backwards with the help of Coulomb law and Boltzmann's distribution the well known mathematical steps. Doing so however would not add any crucial contribution to the previous considerations based on $f$ only, apart the necessity of regarding such $r$ as a parameter significant within a few Debye lengths but not exactly defined point by point like a classical coordinate. Nevertheless the present approach, apparently more agnostic, enabled also the Coulomb law to be inferred itself. Moreover further information is easily inferred from eq 2,5. Consider for simplicity an electron gas only and the energy $\hbar\omega_p$ corresponding to the plasma frequency already calculated consistently with the average energy balance $\overline{\eta}_{cont}$ of any couple; trivial manipulations yield

$$\hbar\omega_p = 2\mu_B H \qquad \mu_B = \frac{e\hbar}{2m_e} \qquad H = \sqrt{\mu_0 dc^2} \qquad \mu_0\varepsilon_0 = c^2 \qquad d = \frac{n_e m_e}{\Delta\rho^3} \qquad 2,8$$

Here $\mu_B$ is the Bohr magneton, while $dc^2$ is the electron rest mass energy density per unit volume corresponding to the number density $N_e$; elementary dimensional considerations show that the third equation defines a magnetic field. Formally the first equation 2,8 is a possible way to rewrite the acknowledged expression of $\omega_p$ through well known positions, whereas $H$ can be in principle understood because any moving charge generates a magnetic field; yet its true physical meaning would be unclear without knowing that $\omega_p$ has been introduced to describe a couple of electrons with the same average energy $\bar\eta_{cont}$, thus with anti-aligned spins, moving away each other and rotating solidly with $\Delta\boldsymbol{\rho}$ as well. In effect the test electron generates a magnetic field normal to its rotation plane; so the direction of $H$ is defined in the reference system of a single couple, whereas its macroscopic average cancels out both because motion and orientation of several mobile electron couples are random and because the rotation axis defining $\omega$ is not uniquely defined itself. Thus $-\mu_B H$ and $\mu_B H$ are the energies of magnetic dipole moment of the electrons of the couple with spin components necessarily opposite with respect to the direction of their own local field; $\hbar\omega_p$ calculates the spin flip energy gap along the local $H$, i.e. the excitation energy $\Delta\bar\eta_{cont}$ of the couple. This confirms that the plasma frequency is really a property of any local couple rotating at angular rate $\omega_p$, yet without contradicting the definition of collective property previously assigned to the plasma frequency: of course the couples are not rigidly formed by specific electrons, rather they involve any neighbours randomly approaching or moving away each other. Otherwise stated, owing to the statistical concept of average, $\bar\eta_{cont}$ represents in fact the totality of couples possible in the plasma. Thus the simple inspection of $\hbar\omega_p$ compels regarding the collective properties of plasma as due to two-body interactions between continuously exchanging electrons, whose fingerprint is just the form of eqs 2,8; remains however intriguing the fact that the local field $H$ results defined through the rest mass energy density of all the electrons. It is clear now why the global behaviour of the charges is described by the full range of values of the parameter $f$, and not by some specific values previously exemplified just to check the kind of information provided by the present reasoning: the key energy controlling the properties of plasma is $\bar\eta_{cont}$ that does not depend on $f$. So the plasma can be effectively regarded as combination of all the electron configurations physically possible whatever the respective $\bar\eta_{rep}$ might be. Also note that the first eq 2,8 should have been more properly written as $\hbar\omega_p = \pm 2\mu_B H$: the negative sign of $\sqrt{\omega_p^2}$, to be excluded in a classical context exploiting Coulomb's law and Boltzmann's statistics only, appears natural in the present quantum-mechanical context that admits a negative energy state $-\hbar\omega_p$ of plasma electrons, i.e. a positron plasma, whereas actually $\mu_B = \pm|e|\hbar/2m_e$; in effect all the reasoning so far carried out would remain unchanged considering a positron gas instead of an electron gas, with the factor 2,5 still accounting for the same energy gap of spin alignment with respect to $H$. A closing remark helps to better clarify the physical meaning of $p_\lambda$. The first eq 2,6 written in the form $\omega^2 = \omega_p^2 + k^2 V^2$, with $V^2 = v_{th}^2 - \bar\eta_{corr}/m_e = 3KT_e/m_e$, shows that $(\hbar\omega)^2$ exceeds the characteristic plasma energy $(\hbar\omega_p)^2$ by $\delta\bar\eta^2 = \hbar^2 k^2 V^2$. The fact that $k \to 0$ entails $\omega^2 \to \omega_p^2$ for $\Delta\rho \to \infty$, very large plasma size, suggests regarding $\delta\bar\eta^2$ as a local perturbation of $\omega_p^2$ in an arbitrary point encompassed by $\Delta\rho$; so the plasma oscillation deviates from $\omega_p$ only locally, of course without possibility of specifying where exactly. Thus $\delta\bar\eta$ could be a spontaneous quantum fluctuation or the consequence of a transient energy local input injected into the plasma: in any case the perturbation does not propagate to infinity because its lifetime is of the order of $\hbar/\delta\bar{\eta} = (kV)^{-1}$. This means that if for instance on the system of electrons acts a flash of thermal energy that increases the local temperature of the gas, then the electron system reacts and tends again to its natural oscillation frequency $\omega_p$, the only effect of the perturbation energy input being an increased average temperature and related Debye's length. To better explain this way of regarding eqs 2,6, rewrite identically $\delta\bar{\eta}^2 = \hbar^2 k_w^2 (wV)^2$ with $k_w = 2\pi/\lambda_w$ and $\lambda_w = w\Delta\rho$, so that the dimensionless arbitrary parameter $w$ does not affect $\omega^2$; eqs 2,6 calculated with $w=1$ and $w>1$ yield $3\omega_p^2(k^2\lambda_{eD}^2 - k_w^2\lambda_{weD}^2) = 0$, i.e. $\Delta T_e/T_e = w^2 - 1$. Then the random average velocity of the electron gas increases while the momentum $p_\lambda = \hbar k$ decreases to $p_{w\lambda} = \hbar k_w$; moreover the group velocity $V_g = kV^2/\omega$ of the circulating electron wave decreases with $k$, i.e. the perturbation energy $\delta\bar{\eta}$ is dissipated in a range of the order of $V/\omega$. So one infers: (i) the momentum decrease from $p_\lambda$ to $p_{w\lambda}$ describes an electron wave circulating along a circumference of radius $\Delta\rho$ that attenuates when the radius expands to $w\Delta\rho$ along with the related wavelength increase; (ii) the local electron wave of frequency $\omega$ is necessarily longitudinal, since propagation direction and electric field oscillation are by definition both in the radial rotation plane of $\Delta\rho$. Note that $V/\omega < V/\omega_p$ and that the right hand side ratio is nothing else but $\sqrt{3}\lambda_{eD}$; thus the perturbation wave extinguishes in a range of the order of Debye's length, which clarifies why $\omega^2 \to \omega_p^2$ for $3k^2\lambda_{eD}^2 \ll 1$. Also note that $\Delta\rho \to \infty$ concerns a longitudinal plane wave for which holds Faraday's law $\mathbf{k} \times \mathbf{E} = \omega \mathbf{H}$; recalling that in this limit $\omega \to \omega_p$ and that the local magnetic field $H$ already found is normal to the rotation plane of the test electron at distance $\Delta\rho$ for the other $n_e$ electrons, the electric field acting on the test electron calculated with the help of eqs 2,6 and 2,8 has the sensible form $E = n_e e/(\varepsilon_0 \varphi \Delta\rho^2)$, where $\varphi = \sin(\widehat{\mathbf{kE}})$. In principle therefore $E$ depends on how are mutually oriented $\mathbf{k}$ and $\mathbf{E}$, because $\varphi \to 1$ if the vectors tend to $\mathbf{k} \perp \mathbf{E}$ or $\varphi \to 0$ if the vectors tend to $\mathbf{k} \parallel \mathbf{E}$: in the former case $E$ tends surely to zero, in the latter case the limit $\varphi\Delta\rho^2$ for $\Delta\rho \to \infty$ is undetermined, i.e. $E$ could be zero or infinite depending on the rate with which $\varphi \to 0$ for $\Delta\rho \to \infty$. These limits are particularly interesting as they link this result involving directly $E$ to what we have already discussed about the energy of the test electron via $\bar{\eta}_{rep}$ as a function of the values allowed to $f$: (i) the divergent values of $\bar{\eta}_{rep}$ for $f \to 0$ correspond to the chance $\varphi\Delta\rho^2 \to 0$, (ii) the result $\bar{\eta}_{rep} \to \bar{\eta}_{cont}$ for $f \to (n_e - 1)^{-1}$ corresponds to $\varphi \to 1$ in which case $E$ tends to the expected form $n_e e/(\varepsilon_0 \Delta\rho^2)$ acting on the test electron in the field of the other $n_e$ all at distance $\Delta\rho$, (iii) $f \to 1$ corresponds to $E \to 0$, i.e. $\varphi\Delta\rho^2 \to \infty$ whatever $\varphi$ might be. Although obtained in the particular case of a plane electron wave, this result suggests a conceptual link between $f$ and $\varphi$: the physical meaning of $f$ is thus related to the coupling strength between electric field within $\Delta\rho$ and wave vector of the electron circulating along the circumference $2\pi\Delta\rho$. The possible alignments of $\mathbf{k}$ and $\mathbf{E}$ are very easily explained considering that while $\Delta\rho$ rotates the electrons move randomly within $\Delta\rho$; so the combination of radial and angular motion does not produce in general a circular path, which would require instead an electron position fixed somewhere within $\Delta\rho$. This reasoning does not contradict the positions of eqs 2,5, which concern average quantities only; here instead we are attempting to describe through $f$ or $\varphi$ the local electron configuration, whose detailed knowledge is however forbidden by the quantum uncertainty. Otherwise stated, the radial and angular uncertainties that prevent knowing how are specifically oriented **k** and **E** also prevent knowing how change $f$ and $\varphi$. Since both these parameters are unpredictable and random, one cannot expect a functional relation between them; their link is merely conceptual, i.e. both express the lack of local information about position and momentum of the electrons. Nevertheless, just the fact that eqs 2,5 agree with the experience, confirms that considering average quantities only yields correct results even disregarding since the beginning any local information. So far $w$ has been not yet specified. Consider the particular value such that $w^2(v_{th}^2 - \bar{\eta}_{corr}/m_e) = c^2$; hence $\omega^2 = \omega_p^2 + k^2 c^2$ whatever $v_{th}$ and $\bar{\eta}_{corr}$ might be. Replacing $v_{th}^2 - \bar{\eta}_{corr}/m_e$ with $c^2$ means that the dispersion relation concerns now the propagation of a transverse electromagnetic wave of frequency $\omega$ travelling in the plasma rather than a matter wave; this holds however for k real, i.e. $\omega^2 > \omega_p^2$, otherwise the wave is attenuated. Hence transverse, longitudinal or mixed waves can propagate in the plasma. The overall conclusion is at this point that it is not necessary to introduce positions and momenta of each electron to infer the basic physical properties of plasma; any local information can be disregarded conceptually since the beginning, i.e. not as a sort of approximation to simplify some calculation. If properly exploited, the lack of knowledge inherent the quantum delocalization is actually valuable source of information, in fact the only one physically allowed by the quantum mechanics. Just because consequence of the uncertainty only, the above way to infer some basic concepts of plasma physics is not trivial duplicate of other well known procedures.

*2.4 Harmonic oscillator.*

This case is particularly interesting for the purposes of the present paper and simple enough to be also reported here. With the positions 1,1, the classical energy equation $p_x^2/2m + k(x-x_o)^2/2$ becomes $\Delta p_x^2/2m + k\Delta x^2/2$; then, thanks to eq 1,2, one finds $\Delta \varepsilon = \Delta p_x^2/2m + \omega^2 m n^2 \hbar^2/2\Delta p_x^2$ with $\omega^2 = k/m$. This equation has a minimum as a function of $\Delta p_x$; one finds $\Delta p_x^{(min)} = \sqrt{mn\hbar\omega}$ and thus $\Delta \varepsilon^{(min)} = n\hbar\omega$, being $n$ the number of vibrational states. For $n=0$ there are no vibrational states; however $\Delta p_x \neq 0$ compels also $\varepsilon_0 = \Delta p_0^2/2m \neq 0$. Therefore $\Delta p_0^{(min)} = \Delta p_x^{(min)}(n=1)$ defines $\varepsilon_0^{(min)} = \left(\Delta p_0^{(min)}\right)^2/2m = \hbar\omega/2$, with $\Delta p_0^{(min)} = \sqrt{m\hbar\omega}$. Being $\Delta \varepsilon^{(min)} = \varepsilon^{(min)} - \hbar\omega/2$, the result

$$\varepsilon^{(min)} = n\hbar\omega + \hbar\omega/2 \qquad 2,9$$

is obtained considering uncertainty ranges only, once again without any hypothesis on these ranges. Note that $\Delta p_x^2/2m = \omega^2 m n^2 \hbar^2/2\Delta p_x^2 = n\hbar\omega/2$ with $\Delta p_x = \Delta p_x^{(min)}$, in agreement with the virial theorem; $\varepsilon^{(min)}$ is given by the sum of kinetic and potential terms, whereas the zero point term has kinetic character only. Also note in this respect that $\Delta p_x^{(min)}$ and $\Delta p_0^{(min)}$ are merely particular ranges, among all the ones still possible in principle, fulfilling the condition of minimum $\varepsilon$ and $\varepsilon_0$; analogous reasoning holds also for $\Delta \rho_{min}$ and $\Delta p_{\rho\, min}$ of eqs 2,2. These results do not contradict the complete arbitrariness of the uncertainty ranges, since in principle there is no compelling reason to regard these particular ranges in a different way with respect to all the other ones; yet, the comparison with the experimental data merely shows preferential propensity of nature for the states of minimum energy. In effect, it is not surprising that the energy calculated with extremal values of dynamical variables does not coincide, in general, with the most probable energy. In conclusion, these examples highlight that the physical properties of quantum systems are inferred simply replacing the random, unknown and unpredictable local dynamical variables with the respective quantum uncertainty ranges: the key problem becomes then that of counting correctly case by case the appropriate number of allowed states. Consider now that a further uncertainty equation conceptually equivalent to eq 1,2 is inferred introducing the time range $\Delta t$ necessary for a particle having finite velocity **v** to travel $\Delta x$; defining formally $\Delta t = \Delta x / v_x$ and then $\Delta \varepsilon = \Delta p_x v_x$, eq 1,2 takes a different form where the new dynamical variables $t$ and $\varepsilon$ fulfil the same $n\hbar$ that reads

$$\Delta \varepsilon \Delta t = n\hbar \qquad \Delta t = t - t_o \qquad 2,10$$

Eq 2,10 is not a trivial copy of eq 1,2, even if $n$ is unchanged: it introduces new information through **v** and shows that during successive time steps $\Delta t$ the energy ranges $\Delta \varepsilon$ change randomly and unpredictably depending on $n$. Merging eqs 1,2 and 2,10 via same $n$, whatever it might arbitrarily be, means in fact merging space and time coordinates. To clarify this assertion, consider that $1/\Delta t$ has physical dimensions of frequency; then the general eq 2,10 can be rewritten as $\Delta \varepsilon_n = n\hbar \omega^\S$, being $\omega^\S$ a function somehow related to any frequency $\omega$. If in particular $\omega^\S$ is specified to be just an arbitrary frequency $\omega$, eq 2,10 reads in this case

$$\Delta \varepsilon_n = n\hbar \omega \qquad 2,11$$

Thus $\omega^\S \equiv \omega$, i.e. $\nu = \Delta t^{-1}$, enables an immediate conceptual link with eq 2,9; having found that $n$ is according to eq 1,2 the number of vibrational states of harmonic oscillator and $n\hbar\omega$ their energy levels, then without need of minimizing anything one infers that $\Delta \varepsilon_n = \varepsilon^{(min)} - \hbar\omega/2$ is the energy gap between the $n$-th excited state of the harmonic oscillator and its ground state of zero point energy; the condition of minimum and $\Delta p_x^{(min)}$ are now replaced by the specific meaning of $\Delta t$. This conclusion shows that a particular property of the system is correlated to a particular property of the uncertainty ranges, thus confirming the actual physical meaning of these latter. In this case the random, unpredictable and unknown $\varepsilon_n$ falling within $\Delta \varepsilon_n$ are necessarily the classical energies of harmonic oscillator whose quantization leads to $\varepsilon^{(min)}$. Note that $\omega$ was previously defined through the formal position $\omega^2 = k/m$; now eq 2,10 shows its explicit link with the time uncertainty $\Delta t$.

*2.5 Quantum fluctuations*

The results so far exposed, together with those of the papers [1,2], strongly suggest the existence of a link between the wave character of quantum mechanics and the positions 1,1, here raised to the rank of fundamental principle of nature; hence it seems reasonable to expect that the eqs 1,2 and 2,10 should somehow incorporate the operator formalism of quantum mechanics. The appendixes A and B concern just this topic: the former infers the momentum operator as a consequence of the position-momentum uncertainty equation, the latter infers likewise the energy operator as a consequence of the time-energy uncertainty equation. The uncertainty principle has been formerly found examining the commutation rules of operators postulated "a priori"; the appendixes A and B show that the reverse logical path is also possible, i.e. postulating the uncertainty entails by consequence the operator formalism: the bi-directional correspondence, non-trivial although reasonably expected, explains why in effect eqs 1,2 and 2,10 infer results consistent with the solutions of appropriate wave equations. Yet, in doing so, these equations entail also the corollaries of indistinguishability of identical particles, already emphasized, and exclusion principle, to be shown in the next section. The appendixes also highlight the particle/wave dualism. Once having pointed out the sought link between the present theoretical frame and the wave mechanics, this subsection on quantum fluctuations could seem superfluous, being this topic well known and widely concerned in literature. It appears however useful to confirm the appropriateness of the present way of thinking and validate an interesting result found in the appendixes through a few short remarks. Let us relate the quantum fluctuations of a single atom to that of a system of atoms with the help of eq B4 of appendix B. Replacing $\hbar/\Delta t$ of eq 2,10 into B4, one finds $\Delta \varepsilon^2/\varepsilon^2 = n$; this result has general character, i.e. it does not concern any specific kind of system in particular, but holds however for a stationary system as it in fact eliminates the time from the problem. Let us specify therefore the system as a set of $N$ identical atoms. If so, nothing refers yet $\Delta \varepsilon^2/\varepsilon^2$ to a single atom or to a whole set of atoms; since neither possibility can be excluded, it is reasonable to think both chances in fact admissible. Write thus $\Delta \varepsilon^2/\varepsilon^2 = n_\varepsilon$ and $\Delta E^2/E^2 = n_E$ respectively for the single atom and for the whole set, regardless of whether it is in solid or liquid or gas phase; the notation emphasizes that the number of states is reasonably different in either case. Let us average both equations to calculate $\overline{\Delta\varepsilon^2/\varepsilon^2} = \overline{n_\varepsilon}$ and $\overline{\Delta E^2/E^2} = \overline{n_E}$ assigning proper values to $\varepsilon$ and $E$; for instance these latter could be regarded as minimum values $\varepsilon_{min}$ and $E_{min}$, if any, or as average values $\bar\varepsilon$ and $\bar E$. The second option is more attracting, as it allows to exploit the reasonable link $\bar E = N\bar\varepsilon$. Let us multiply both sides of this equation by $\overline{n_E}$, noting that by definition this latter must have the form $a/N$, where $a$ is a constant that defines uniquely the number of states $\overline{n_E}$ averaged on the number of atoms. Then we obtain $\overline{n_E}\,\bar E = a\,\bar\varepsilon$, which must hold of course for any $N$; the boundary condition that this result holds in particular for $N=1$, i.e. $\overline{n_\varepsilon}\,\bar\varepsilon = a\,\bar\varepsilon$, requires $a = \overline{n_\varepsilon}$. So the couple of equations $\bar E = N\bar\varepsilon$ and $\overline{n_E}\,\bar E = \overline{n_\varepsilon}\,\bar\varepsilon$ yields immediately

$$\overline{\left(\frac{E-\bar E}{\bar E}\right)^2} = \frac{1}{N}\overline{\left(\frac{\varepsilon-\bar\varepsilon}{\bar\varepsilon}\right)^2}$$

The fact that $\overline{n_E}\,\bar E$ holds for any number of atoms, thus including even $N=1$, is nothing else but the statement of indistinguishability of identical atoms. This well known result confirms the validity of the relationship $\Delta\varepsilon^2/\varepsilon^2 = n$ found in appendix B. Another interesting result inferred through this equation will be given in the next section 3. It is remarkable that the reasoning did not require any hypothesis about the number of states of the single atom in the whole set, in agreement with the fact that $n$ is actually not definable by any specific value.

*2.6 Towards the special relativity.*

The positions 1,1 seem general and reliable enough to demonstrate the conceptual self-consistency of the approach based on the uncertainty only. In this respect are significant some preliminary comments on the energy uncertainty $\Delta\varepsilon = p_2 v - p_1 v$ underlying eq 2,10 (the subscript $x$ is omitted for brevity). Like $\Delta x$ and $\Delta p$, also $\Delta\varepsilon$ and $\Delta t$ are arbitrary and range in general from zero to infinity. Yet, once having linked eq 1,2 with eq 2,10, it follows that v must be upper bound by a well defined value hereafter called $c$. To show this point, consider any finite $\Delta x$ and $\Delta p$ to which correspond finite values of $n$: if $v \to \infty$ then in eq 2,10 $\Delta t = 0$ and $\Delta\varepsilon = \infty$. These limits could be in principle simultaneously allowed regarding eqs 1,2 and 2,10 separately, i.e. with different values of $n$ whatever its specific value might be in either case; the limits are however not jointly consistent once assuming the same $n$ for both equations as done here because, according to the respective range sizes, a free particle could have momentum $p$ necessarily finite and energy $\varepsilon$ even infinite. So the condition $v \le c$ is consequence of having merged together both uncertainty equations 1,2 and 2,10, whereas it would be instead unrequired regarding separately time and space coordinates. Eq 2,10 for a free particle reads, whatever v might be,

$$\Delta\varepsilon^{(v)}\Delta t^{(v)} = n\hbar \qquad \text{in } R \qquad 2{,}12a$$

The notation at left hand side emphasizes the actual velocity of the particle delocalized in $\Delta x$. Being v arbitrary eq 2,12a is also consistent with any other $v' \ne v$, in which case it reads $\Delta\varepsilon^{(v')}\Delta t^{(v')} = n\hbar$ with the same number of states because likewise inferred from eq 1,2; the related momentum change in $\Delta x$ is still incuded within the same $\Delta p$ because the sizes of both these latter are arbitrary and thus definable consistently with possible momentum changes of interest here. If so, however, it is impossible to establish if these equations really regard two different velocities $v'$ and v equally allowed for the particle in $R$ or the motion of the particle in two different reference systems $R'$ and $R$ reciprocally sliding at constant rate; it is possible in effect to introduce an inertial reference system $R'$ with respect to which the particle has velocity $v'$. Eq 2,12a rewritten in $R'$ as a function of $v'$ reads therefore

$$\Delta\varepsilon^{(v')}\Delta t^{(v')} = n\hbar \qquad \Delta\varepsilon^{(v')} \neq \Delta\varepsilon^{(v)} \qquad \Delta t^{(v')} \neq \Delta t^{(v)} \qquad \text{in } R' \qquad 2,12b$$

The expected notation $n'$ for the number of states is actually unnecessary because neither $n$ nor $n'$ are specifically defined by assigned numerical values: thus, whatever for either observer $n$ might be, its possible change to $n'$ means transforming an arbitrary integer undetermined for the first observer into any other arbitrary integer undetermined as well for the second observer. The conclusion is that $n$ and $n'$, even though changing, trivially duplicate from $R$ to $R'$ all the possible numbers of states allowed for the given system; despite their different notation, the sets of arbitrary numbers $n$ and $n'$ are in fact indistinguishable. Clearly there is no way to distinguish either situation 2,12a or 2,12b; so, as expected from the results of section 2, the reference systems $R$ and $R'$ are indistinguishable as well, i.e. they must be in fact equivalent to describe the physical properties of the particle. Of course this must hold also for the particular and well defined value $c$ allowed to $v$; so, it is still possible to write in particular

$$\Delta\varepsilon_{\min}^{(c)}\Delta t_{\min}^{(c)} = n\hbar \qquad \text{any } R \qquad 2,13$$

provided that $c$ be equal in any reference system; this property of $c$ allows in fact the equivalence of any $R$ despite its value is not arbitrary and unknown like that of $v$. The position $\Delta t_{\min}^{(c)} = \Delta x/c$ defines $\Delta\varepsilon_{\min}^{(c)} = p_2 c - p_1 c$ as energy uncertainty of the particle having $v = c$. Then $\varepsilon^{(c)}$ ranging within $\Delta\varepsilon_{\min}^{(c)}$ must have the form $\varepsilon^{(c)} = p^{(c)} c$, where $p^{(c)}$ is any value between $p_1$ and $p_2$. Instead, for $v < c$ the same $n$ in eqs 2,12a and 2,13 requires $\Delta t^{(v)}$ a factor $c/v$ longer than $\Delta t_{\min}^{(c)}$ and so the corresponding $\Delta\varepsilon^{(v)}$ a factor $v/c$ smaller than $\Delta\varepsilon_{\min}^{(c)}$; being $c$ constant, this requires that $p^{(c)}$ scales by $v/c$ to $p^{(v)}$; this latter must have thus the form $p^{(v)} = (v/c) p^{(c)} = \varepsilon^{(c)} v/c^2$, being $\varepsilon^{(c)}$ and $p^{(v)}$ local values, random and unknown, within the respective $\Delta\varepsilon_{\min}^{(c)}$ and $\Delta p^{(v)}$. The superscripts merely recall the uncertainty ranges that define the momentum and energy local variables; yet, since the range sizes are anyway arbitrary and irrelevant as concerns the eigenvalues of quantum systems, the same equation holds in general for any $p$ and $\varepsilon$ falling within the respective $\Delta p$ and $\Delta\varepsilon$. Thus

$$p = \varepsilon v/c^2 \qquad \text{any } R \qquad 2,14$$

Note that this result and the following ones do not depend on a particular choice of reference system. Consider now the particular reference system where the particle is at rest and note that $v = 0$ yields $p = 0$; yet nothing compels thinking that in this reference system $\varepsilon$ vanishes as well. If so, regard in general finite the ratio $p/v$. Defining thus

$$m = \lim_{v \to 0} p/v \qquad \text{any } R \qquad 2,15$$

eq 2,14 yields $m = \varepsilon_{rest}/c^2$ being obviously $\varepsilon_{rest} = \lim_{v \to 0}\varepsilon$ by definition. Thus $m$ and $\varepsilon_{rest}$ are intrinsic properties of the particle, not due to its motion. Eq 2,14 is well known; yet it is non-trivial noticing that the concept of mass introduced by the consequent eq 2,15 is inferred from the that of uncertainty only. Exploit now $\delta p = \varepsilon \delta v/c^2 + v \delta\varepsilon/c^2$ showing how $p$ changes as a function of $\delta\varepsilon$ and $\delta v$: multiplying both sides by $\hbar c^2/\delta x$ and recalling eqs A11 and B4 of appendixes A and B one finds $(pc)^2 = \varepsilon^2 + \varepsilon'\varepsilon$, with $\varepsilon' = \hbar \delta v/\delta x$ and $\delta x = v \delta t$. The limit of $(pc)^2/c^4$ for $v \to 0$ reads $0 = m^2 + \lim_{v \to 0}\varepsilon'\varepsilon/c^4$, which yields $\lim_{v \to 0}\varepsilon'\varepsilon = -(mc^2)^2$. For a free particle $\varepsilon = const$; so, because of eq 2,14, this result agrees with

$$\varepsilon^2 = (pc)^2 + (mc^2)^2 \qquad p = mv/(1-v^2/c^2)^{1/2} \qquad \varepsilon = mc^2/(1-v^2/c^2)^{1/2} \qquad \text{any } R \qquad 2,16$$

Both $\varepsilon_{rest}$ and $\varepsilon'_{rest} = -mc^2$ confirm the existence of states of negative energy found in appendix B. It is obvious that the kinetic energy of the particle can be nothing else but $\varepsilon - \varepsilon_{rest}$, which yields for $v \ll c$ just the energy used for the Hamiltonians of the hydrogenlike atom and harmonic oscillator, assumed known in section 2 for graduality of exposition only. Although the position/momentum uncertainty equation contains conceptually also the energy/time uncertainty equation, the only fact of having linked explicitly the latter with the former entails that even the previous results of subsections *2.1* to *2.5* are fully self-contained through the positions 1,1. In other words, eqs 1,2 and 2,10 were regarded as abstract relationships involving momentum and energy, whose analytical form is however not specified "a priori"; yet the considerations inherent the steps 2,12 to 2,13 lead to eqs 2,14 to 2,16 where the concepts of momentum and energy find eventually their analytical expression established with the help of velocity only. Mostly important, it holds in particular for the mass $m$. In the subsection *2.2* the mass appeared in the Hamiltonian of the hydrogenlike atom via momentum and kinetic energy of the electron in the field of the nucleus; after having established the analytical form of the energy equation, only ranges of momentum and coordinates have been considered. The result in eq 2,15 changes however drastically this way of thinking; for clarity it is worth summarizing the reasoning just carried out, whose basic ingredients are space and time coordinates and velocity. Consider the last equations 2,12 to 2,16 and the way to infer them from eqs 1,2 and 2,10 only; these uncertainty equations with the same $n$ are in fact an identity regardless of any non-relativistic or relativistic concept. Let us reason as if physical meaning and analytical form of $p_x$ and $\varepsilon$ would be unknown; this means relating time and space uncertainties to the uncertainties $\Delta p_x$ and $\Delta \varepsilon$ of two physical entities defined uniquely by appropriate physical dimensions of the constant $\hbar$. Simply regarding space and time coordinates together, the physical entity $p_x$ is related to the physical entity $\varepsilon$; so, following the reasoning introduced by eqs 2,12 and 2,13 that exploit the key definition $\Delta \varepsilon = \Delta p_x v_x$, the uncertainty of space and time coordinates requires the relationship $p_x = \varepsilon v_x / c^2$, whereas it is also required that $c$ is an invariant maximum value of the arbitrary velocity $v_x$: the latter is defined by its own reference system, the former does not. After that, in the limit $v_x \to 0$ appears by consequence a further physical entity denoted as $m$. Eventually, following the reasoning from eqs 2,15 to 2,16, one finds the explicit analytical form of $p_x$ and $\varepsilon$ whose physical meaning is uniquely determined by the physical dimensions of $\hbar$. The connection with the results of subsection *2.2* is evident noting that for a free particle the limit expressions of $p_x$ and $\varepsilon$ for $v_x \ll c$ are $mv_x$ and $mv_x^2/2$. Note also at this point that the Coulomb law has been inferred from eqs 1,2 and 2,10 in subsection *2.3* as well. In conclusion, the examples carried out in section 2 are in fact fully self-consistent, i.e. based on the assumption of uncertainty only. No classical or relativistic concept is required to conclude that what we call mass is a limit property of a particle at rest, i.e. in any reference system solidal with the particle and regardless of any other information or hypothesis. Owing to the extended concept of space-time uncertainty, the mass appears itself not longer as a concept familiar because of the reality around us and thus evident "a priori", but rather as a consequence of the same principle, the agnostic positions 1,1, that governs and describes the quantum world. It appears since now the intimate connection between quantum world and gravity: the principles of indistinguishability and exclusion are related through the unique concept of uncertainty to the invariance of $c$ and to the notion of rest mass, whose unambiguous definition excludes since now any possible distinction between inertial and gravitational behaviour. On the other hand, the appendixes A and B and the next section 5 highlight a further consequence of the positions 1,1, i.e. the wave/corpuscle duality. These short notes have been purposely included at the end of this preliminary section to show that simply considering together time and space ranges to define the uncertainty, the present approach contains inherently conceptual prerequisites that include as corollaries the postulates of relativity and quantum mechanics; this is in fact possible without reference to either kind of specific postulates and even without exploiting the appropriate 4-vector algebra. This evidence prospects thus a wider and more fundamental generalization of the previous results. It is useful to close this non-relativistic section remarking that in the present approach the equation describing quantum systems is formally obtained replacing the dynamical variables of the appropriate classical equation with the respective uncertainty ranges, e.g. $p_x \to \Delta p_x$, whereas instead the non-relativistic wave equation is formally obtained replacing the dynamical variables with the respective operators, e.g. $p_x \to -(\hbar/i)\partial_x$. The question raises whether or not these positions are still adequate in relativity. The latter position does not: once having shown that the relativistic and classical expressions of momentum are substantially different for profound reasons rooted on the previous corollaries introducing the special relativity, there is no reason to assume that the naive position just quoted is applicable to both cases. As regards the positions 1,1 instead? The fact of having inferred through them eqs 2,14 to 2,16 does not exclude a more positive expectation, simply assuming that the space uncertainty ranges are actually space-time uncertainty ranges. In this way the same quantum approach can be extended to the special and general relativity. This is the aim of the next sections 3 and 4 that disregard once more the operator formalism and exploit uniquely the concepts previously introduced, with trust on the prospective effectiveness of the character absolutely general of eqs 1,2 and 2,10.

3 Special relativity.
The section 2 has shown that the concept of delocalization is the "added value" necessary and enough to plug the classical physics into the quantum world. A few comments highlight further this point and show that really the special relativity is straightforward generalization of the results therein shortly sketched. Although the approach based uniquely on eqs 1,2 and 2,10 is apparently more agnostic than that based on the formalism of wave mechanics, the physical information about the quantum systems examined in [1,2] is in fact completely analogous. Actually this consistency could be expected because the local coordinates and conjugate momenta appear explicitly only in the wave equations providing all the possible information on any physical system, not longer in their eigenvalues; on the one side the local dynamical variables result physically worthless in determining the allowed states of the system, on the other side neglecting them in principle merely avoids handling variables conceptually unessential as concerns the eigenvalues and simplifies considerably the way to infer the allowed physical information: the present approach, indeed, does not need solving any wave equation just because the local dynamical variables are disregarded since the beginning. The examples of section 2 highlight the reasons of it: the quantum properties are controlled by the uncertainty equations 1,2 and 2,10, whereas the quantum numbers are actually numbers of allowed states introduced since the beginning into the problem through these equations. Calculating the numbers of allowed states means working directly with the eigenvalues instead of extracting them from the eigenfunctions; the uncertainty is not mere restriction of knowledge but rather a sort of essential information, actually the only one available. It is worth noticing that just the abstract concept of quantum delocalisation is significant in principle because, as shown in the previous examples, size and origin of the uncertainty ranges are actually never specified. In the hydrogenlike atom $\Delta\rho$ is mere notation to indicate a space range around the nucleus, whose actual size remains however undetermined; the only essential idea is that it conceptually exists and encompasses the possible radial coordinate expected for the electron in any given physical situation, this coordinate being however completely arbitrary in principle. For this reason even an infinite size is allowed to the ranges in agreement with their complete arbitrariness, without divergence troubles: there is no reason why an infinite range should necessarily entails local values infinite themselves, although this possibility cannot be excluded in principle just by their random character. Since $M_w$, $M^2$, $\varepsilon^{(min)}$ and $\varepsilon_{el}$ do not depend explicitly on these range sizes, in fact the physical properties of matter do not diverge and admit even the infinite limit of allowed quantum states, in which case they tend to the respective classical quantities. Just because lacking any specific value, $\Delta\rho$ plays a role similar to that of the dynamical coordinate $\rho$ it replaces: e.g. an increasing radial range is consistent in principle with larger electron/nucleus distances, whereas the limit case of infinite distance from the nucleus is described by $\Delta\rho \to \infty$. Also, local derivatives like $\partial_\rho$ are replaced systematically by $\partial_{\Delta\rho}$; this is evident as $\Delta\rho = \rho - \rho_o$ is defined by a fixed value $\rho_o$ and a variable value $\rho$ allowed to change when the radial coordinate of the particle is perturbed by any physical reason. These considerations, merely reasonable in a non-relativistic frame, have central importance to introduce the basic principles of special relativity. The leading idea is that the uncertainty ranges and their space derivatives are not subjected to any hypothesis about their sizes, analytical form and coordinates $x_o$ and $\rho_o$ defining their origin; recall that for instance any local coordinate $\rho$ is effectively replaced by $\rho_o + \Delta\rho$ regardless of $\rho_o$ and $\Delta\rho$. Since however $x_o$ and $\rho_o$ are defined by their own reference system only, it means that in fact the present physical description of quantum systems does not specify in particular any reference system of space coordinates. The present approach excludes therefore "a priori" the existence of a preferential reference system to describe the physics of the quantum world, i.e. all the reference systems are physically equivalent. So any result here inferred holds by definition in any reference system; this explains why the relativistic equations 2,14 to 2,16 have been found without introducing explicitly the concept of invariance of physical laws. This conclusion holds identically also for the time; as shown in the case of harmonic oscillator, the consistency of eqs 2,9 and 2,11 does not require any hypothesis about $\Delta t$ and even $t_o$ does not need to be specified whatever $\omega$ might be. Since $t_o$ can be defined only introducing a time reference system, one concludes that there is in fact no preferential time reference system. This analogy between time and space coordinates, already emphasized also in appendix B to infer the energy operator, is too strong and significant to be merely accidental, rather it suggests the physical concurrence of both in describing the quantum properties. This assertion is clearly confirmed recalling that eqs 1,2 and 2,10 are direct consequences one of the other for a given number $n$ of quantum states. Being physically meaningless to ask which equation is "more fundamental", the angular momentum or the energy levels of hydrogenlike atoms formerly inferred from eq 1,2 could be identically regarded as a consequence of eq 2,10 rewritten in the form 1,2, as explicitly done in the case of harmonic oscillator. Yet, since the concurrence of time and space ranges entails physical equivalence of the respective random coordinates as well, the link between eqs 1,2 and 2,10 appears actually more profound than in non-relativistic physics and suggests that actually the positions 1,1 concern the more general idea of a unique space-time range $\Delta x_{xt}$ that combines together space and time uncertainties $\Delta x$ and $\Delta t$, regarded separately in section 2 for simplicity and graduality of exposition only. Consider any range $\Delta x$ with $x$ regarded as generalized coordinate, e.g. radial or Cartesian or curvilinear: in principle there is no reason nor necessity to conceive this range as a function of space coordinates only, as nothing has been specified about it; the previous considerations on angular momentum or hydrogenlike energy levels could have been identically inferred considering instead of $\Delta x$ the more general range $\Delta x_{xt}$ linear or non-linear combination of $\pm|\mathbf{v}|\Delta t$ and $|\mathbf{x}-\mathbf{x}_o|$, for instance

$$\Delta x_{xt} = |\mathbf{x}-\mathbf{x}_o| \pm |\mathbf{v}|(t-t_o) \qquad 3,1$$

As concerns $\Delta p_{xt}$ nothing changes aside from its conceptual meaning, now consistent with the space-time definition of uncertainty. The combination of space and time coordinates, possible in principle because sizes and analytical form of the ranges are completely arbitrary and not explicitly determined in eqs 1,2 and 2,10, also suggests merging together these equations with the same $n$

$$\Delta x_{xt} \Delta p_{xt} = \Delta t \Delta \varepsilon = n\hbar \qquad 3,2$$

Regarding together time and space coordinates of a system arises the question about why the results of section 2 even ignoring the time are so reasonable. Without considering explicitly the time evolution of the respective systems, the delocalization of each particle has been described via a unique uncertainty space range large enough to encompass any position allowed by its random and unknown motion; being the range arbitrary by definition, there is no reason to exclude in principle this way of thinking, which in effect yields sensible results. A possible way to link the idea of time dependent uncertainty with the results of section 2 is $\Delta x = x - x_o \pm v \Delta t$ that reads $\Delta x' = x - x'_o$ simply admitting the position $x'_o = x_o \mp v \Delta t$, i.e. changing reference system; so, having shown the physical equivalence of any $\Delta x$ and $\Delta x'$ as concerns the eigenvalues, $\Delta t$ does not longer affect explicitly the uncertainty of the system only because it becomes hidden in the arbitrary choice of the reference frame. Yet this explanation although reasonable is not exhaustive: it would be difficult to justify in a coherent way and within a unique conceptual frame why the time does not appear explicitly in the steady cases of section 2 while it should control the time evolution of unstable systems, e.g. subjected to an external perturbation. The previous definition of $x'_o$ cannot distinguish itself either case, rather it holds only if the time range is really ineffective in describing the total uncertainty of the system and can be therefore removed from the problem. In principle the link between time and uncertainty is evident recalling that in the present context the quantum numbers defining the eigenvalues of the system are actually random numbers of states allowed to the system; so, transitions between ground and excited states of a quantum system are natural consequence of the randomness inherent the positions 1,1. As these transitions are reasonably expected to alter the configuration of the particles constituting the system as a function of time, the aforesaid link relates configuration energy change $\Delta \varepsilon^\S$ and energy transient lifetime $\Delta t^\S$ necessary for it to occur. In this respect eqs 3,1 and 3,2 constructively modify the non-relativistic point of view, just because the time explicitly concurs together with the space coordinates to define the uncertainty and thus the eigenvalues of the system: although the previous definition of $x'_o$ and the results of section 2 remain certainly possible, the more general concept of time dependent uncertainty allows also the chance of describing the time evolution of quantum systems. To this purpose the positions 1,1 and eqs 1,2 and 2,10 suggest considering just time length and energy gap of the excitation transient rather than the particular nature of the perturbation and the specific interaction mechanism it triggers. Let the energy of a system deviate from the minimum value by an arbitrary range $\Delta \varepsilon^\S$ because of the unpredictability and randomness of the numbers of allowed states; e.g. $n$ and $l$ of eq 2,3 yield $\Delta \varepsilon^\S = (Z\alpha)^2 g^\S \mu c^2 / 2$ for $1 < n \leq n^\S$, with $g^\S = 1 - n^{\S-2}$ and $\alpha$ fine structure constant. Then the minimum lifetime $\Delta t^\S = \hbar / \Delta \varepsilon^\S$ of the excited state suggests that $\Delta t \geq \Delta t^\S$ is necessary condition for the system to return again in the ground state, being trivially evident that steady eigenvalues are not definable while the transient $\Delta \varepsilon^\S$ is still in progress. The time inequality involves $\Delta \varepsilon^\S$ and $\Delta \varepsilon$ as well and then, via eq 3,1, the corresponding coordinates and momenta: so the characteristic lifetime $\Delta t^\S$ of the excited state controls the time scale $\Delta t$ during which the energy and configuration changes $\Delta \varepsilon^\S$ and $\Delta x^\S_{xt}$ are allowed by $n \to n^\S \to n$. This conclusion is reasonable: the energy levels of eq 2,3 have been inferred considering uniquely the electron delocalization in the field around the nucleus; it appears natural that the link between the eigenvalues of eq 2,3 and the general eq 3,1 concerns just the uncertainty of particles. These considerations also suggest comparing $\Delta t$ and $\Delta t^\S$ to infer if $\Delta \varepsilon^\S$ is mere random fluctuation that admits time average statistically stationary or if it concerns a real time evolution of the system. To explain this point consider for instance an isolated non-relativistic hydrogenlike atom: for $\Delta t < \Delta t^\S$ has physical meaning the mean energy of the transient averaged between $n = 1$ and $n^\S$, being all the intermediate states in fact accessible; for $\Delta t \geq \Delta t^\S$ the system appears instead statistically stationary because the time range is long enough to complete any possible deviation from and return to the ground state. Being in general constant the kinetic energy $\varepsilon_{kin}$ of an isolated particle delocalized in the arbitrary space-time range $\Delta x$, the uncertainty $\Delta \varepsilon$ of eq 3,1 admits for the atom in the ground state the limit $\Delta \varepsilon \to 0$ that, as explained in the introduction, represents by definition just $\varepsilon_{kin}$; the related $\Delta t$ of eq 3,1 tends then to infinity, which means that the required time inequality $\Delta t \geq \Delta t^\S$ is certainly fulfilled for any $n^\S$. The expected conclusion is that an isolated atom is a system statistically stationary on time scale long enough to conclude any possible random transient above the ground state of minimum energy. To extend this kind of reasoning to the case where instead the system in fact evolves as a function of time, consider a real gas of hydrogenlike atoms interacting via anelastic collisions in the presence of a local temperature gradient. The energy fluctuations allowed to each atom are still described by $\Delta \varepsilon^\S$, yet now $\varepsilon_{kin}$ differs from point to point in gas volumes containing unperturbed and perturbed atoms, i.e. $\varepsilon_{kin}$ is not longer constant. So in eq 3,1 the range $\Delta \varepsilon$ of kinetic energy pertinent to atoms diffusing between zones of different density is not uniquely defined for the whole system, rather it depends on the choice of the local gas volume. Try again to relate each local $\Delta \varepsilon$ to $\Delta \varepsilon^\S$ recalling that the latter decreases with $n^\S$, whose values are arbitrary; in principle it can happen that atoms with high values of $n^\S$ verify the inequality $\Delta \varepsilon^\S < \Delta \varepsilon$, whereas atoms with low values of $n^\S$ verify $\Delta \varepsilon^\S > \Delta \varepsilon$; it can also happen that $\Delta \varepsilon$ is so large that locally holds the former inequality only. A unique statement about the corresponding time range inequalities is not longer possible: a given $\Delta t$, finite because not longer related to $\Delta \varepsilon \to 0$, could fulfil the stationary condition in some point of gas volume but not necessarily in a neighbouring point. The conclusion is that an external perturbation, coinciding in the present example with the energy input that activates the local temperature gradient, produces in general a time dependent transient wherever $\Delta t < \Delta t^\S$: the stationary state previously described for the isolated atom is not longer ensured throughout the gas volume, the time dependence of the allowed states of the system of atoms appears thus explicitly. The obvious consequence is that any non-equilibrium system is unstable and evolves locally as a function of time; the non-trivial remark is however that such a conclusion is inferred considering uniquely arbitrary uncertainty ranges that link the time scale of the excitation process to the configuration and energy changes occurring in the system. Of course the volume of gas can be also regarded as a whole isolated system for which hold the considerations of the former example, i.e. the existence of a unique $\Delta \varepsilon$ large enough to include any local energy ranges; this latter point of view concerns thus a macroscopic system where random density fluctuations are locally allowed to occur. In non-relativistic physics introducing $\Delta x_{xt}$ alternative to $\Delta x$ appears relevant to discriminate the transient or steady character of a given problem, as previously emphasized. The generalization of the concept of uncertainty and the conceptual equivalence of eq 2,10 with eq 1,2, the only equations so far considered, require that also $\Delta x_{xt}$ must have the features evidenced by the examples of section 2 whatever its analytical form might be. This means that: (i) even $\Delta x'_{xt} = \sqrt{\left|\Delta x^2 \pm v^2 \Delta t^2\right|}$, for instance, could replace legitimately $\Delta x$ without changing any step of the approach leading to eqs 2,3 or 2,9 or 2,11; (ii) even the space-time ranges must fulfil the same condition of indistinguishability, i.e. equivalence, of reference systems already emphasized in subsection *2.5*. Clearly still holds the idea that the arbitrariness of local space and time coordinates compels that of space-time uncertainty ranges, which is in turn closely linked to the equivalence of any time-space reference system. So no distinctive property makes in principle $\Delta x'_{xt}$ more or less appropriate than $\Delta x_{xt}$ of eq 3,1 in describing non-relativistic quantum systems. At this point, one more step allows to generalize farther on the previous considerations to the relativistic point of view. The uncertainty ranges of section 2, despite their ability to describe a wide variety of phenomena, are mere space-like or time-like particular cases of $\Delta x_{xt}$, which holds in general and regardless of the particular aim of the specific problem; yet, when merging together the space and time uncertainties, the expressions 2,14 and 2,16 are inferred and involve the constant parameter $c$ even if the actual velocity of the particle is $|\mathbf{v}|$. Of course $|\mathbf{v}|$ has a value arbitrary and dependent on a specific reference system; instead $c$, as emphasized in section 2, has a well defined and thus invariant value to ensure the indistinguishability of any space-time range and respective reference system. These considerations suggest that, in alternative to $\Delta x$ and $\Delta t$, should have relevant importance in relativity ranges like $\Delta x_c = |\mathbf{x} - \mathbf{x}_o| \pm c(t - t_o)$ or $\Delta x'_c = \sqrt{|\Delta x^2 \pm c^2 \Delta t^2|}$, where the space and time contributions to the uncertainty are linked by $c$ instead of $|\mathbf{v}|$. This entails a further consequence: indistinguishable reference systems require ranges invariant themselves, which just for this reason have distinctive importance to infer general physical laws. So the existence of an invariant time-space interval is not only a property but mostly a necessary conceptual requirement once merging space and time coordinates into a unique four dimensional space. Clearly these remarks summarize and correlate the basic hypotheses of special relativity: existence of invariant interval, finite value and invariance of $c$, invariance of physical laws for different inertial reference systems, non-relativistic physics as a limit case. The non-trivial fact is however that the properties of the delocalization ranges introduce these basic assumptions as straightforward corollary of the quantum uncertainty inherent any physical system of particles. It is known that the invariance of interval entails in turn the Lorentz transformation of time and space in different inertial reference systems in reciprocal constant motion. Considering "a priori" only uncertainty ranges as proposed here, i.e. conceptually and not as a sort of approximation to simplify some calculation, these transformations are inherently fulfilled and appear consequences themselves of a unique principle underlying both quantum mechanics and special relativity. The following relevant example highlights this concept and extends the invariance of $c$ to the interval invariance rule too. Consider a photon travelling a range $c\Delta t = x - x_o$ such that the arbitrary coordinates $x_o$ and $t_o$ identify a well defined space-time reference system $R$. A slower massive particle would travel a smaller range $x_s - x_o$; it is possible then to write the formal identity $c\Delta t = x_s - x_o + \delta X$ with $\delta X = x - x_s$. Consider now also the range $c\Delta t' = x' - x'_o$ obtained shifting $x_o$ to $x'_o$ and $x$ to $x'$ during the time range $\Delta t$ at average rate $V = (x'_o - x_o)/\Delta t$. Yet also $x'_o$ and $t'_o$ define their own reference system $R'$ displacing with respect to the former; this is clearly the case of two identical ranges initially overlapped and then mutually displaced by sliding one along the other at constant rate. The same reasoning yields then $c\Delta t' = x'_s - x'_o + \delta X'$ with $\delta X' = x' - x'_s$. Since the range sizes are arbitrary, it is possible to put by definition $x_s - x_o = x'_s - x'_o$ whatever their boundary coordinates might be. Two consequences are therefore possible: (i) $\Delta t' = \Delta t$ that requires $\delta X' = \delta X$, i.e. identical stationary ranges or a Galilean transformation; (ii) $\Delta t' \neq \Delta t$ that requires $\delta X' \neq \delta X$. The non-trivial conclusion is that any non-Galilean transformation $\delta X' \neq \delta X$ between $R$ and $R'$ entails $\Delta t' \neq \Delta t$: an appropriate law of transformation of coordinates excludes therefore the concept of universal time equal in all reference systems. The sought law is found regarding again the same ranges yet considering $c^2 \Delta t^2 = (x - x_o)^2$ and $c^2 \Delta t'^2 = (x' - x'_o)^2$, so that $c^2 \Delta t^2 = (x_s - x_o)^2 + \Delta X^2$ and $c^2 \Delta t'^2 = (x'_s - x'_o)^2 + \Delta X'^2$ with $\Delta X^2$ and $\Delta X'^2$ properly defined by $(x_s - x_o)^2$ and $(x'_s - x'_o)^2$ of the slow particle. Putting again $(x_s - x_o)^2 = (x'_s - x'_o)^2$, the range invariant in two reference systems reciprocally moving at constant rate V has the form $\delta s^2 = c^2 \Delta t^2 - \Delta X^2$ despite $\Delta t$ and $\Delta X$ are necessarily different from $\Delta t'$ and $\Delta X'$. This form of $\delta s^2$ deserves further attention for Lorentz's transformations of the quantities appearing in eq 3,2. Rewriting $c^2 \Delta t^2 - \Delta X^2 = c^2 \Delta t'^2 - \Delta X'^2$ as $c^2 \Delta t^2 = c^2 \Delta t'^2 - \Delta X'^2$, which is in fact possible because the condition $\delta X \neq \delta X'$ does not exclude $\Delta X = 0$, yields $\Delta t = \Delta t' \sqrt{1 - V'^2/c^2}$ with $V' = \Delta X'/\Delta t'$. Replacing the time ranges of the initial positions $c\Delta t = x - x_o$ and $c\Delta t' = x' - x'_o$ into this result one finds also $x - x_o = (x' - x'_o)\sqrt{1 - V'^2/c^2}$. Make now the sizes of the ranges $x - x_o$ and $x' - x'_o$ tending to zero so that, as stated in the introduction, the ranges tend to the respective local coordinates; is so, the same holds of course for $\Delta t$ and $\Delta t'$ as well. Let therefore $x - x_o$ tend to $x$; then $x' - x'_o$ must tend to $x \pm V't$, as $x'_o$ has been displaced by $\pm V't$ with respect to $x_o$. An analogous reasoning for $t$ and $t'$ in the respective time ranges yields then the well known result

$$x' = (x - V't)/\sqrt{1 - V'^2/c^2} \qquad t' = (t - xV'/c^2)/\sqrt{1 - V'^2/c^2} \qquad 3,3$$

This conclusion is now exploited as concerns the angular momentum of the system of particles concerned in the early eq 2,1. Previously $\Delta r$ and $\Delta r'$ were introduced as independent and arbitrary ranges. If however the sizes $\Delta r$ and $\Delta r'$ are now defined in reference systems reciprocally moving, $\Delta r'^2 / \Delta r^2 = 1 - V^2/c^2$ because of mere relativistic reasons even though eq 2,1 is calculated with $l = l'$. Since in both reference systems hold the uncertainty conditions previously discussed, replacing in eq 2,1 the length ratio just inferred $M'^2$ results to be just Lorentz's transformation of $M^2$ in a reference system where the center of mass is at rest, as in fact it is well known

$$M'^2 = M^2 - (M^2 - M_w^2)V^2/c^2 \qquad 3,4$$

The reasoning sketched in detail in appendix C concerns the relativistic quantization of angular momentum when the uncertainty ranges of space-time coordinates and linear momentum $\Delta \rho$ and $\Delta p$ transform in $R$ and $R'$ according to the Lorentz transformation consequent the interval invariance rule. The local variables are again disregarded, whereas the procedure of section 2 still holds despite $\Delta \rho, \Delta p$ change to $\Delta \rho', \Delta p'$ in the respective reference systems because the range sizes are unessential for the quantum properties of particles. It is found in this way that $M_w$ is not longer given by $\pm l\hbar$ only, rather appears a further component $\pm l'\hbar/2$, with $l' = 0, 1, 2, \cdots$. This result, inferred here without any hypothesis "ad hoc" because the interval invariance rule is itself consequence of eqs 1,2 and 2,10, can be more shortly obtained from $\delta s = \sqrt{c^2\Delta t^2 - \Delta X^2}$. As $\sqrt{a^2 - b^2} = a - (b/a + (b/a)^3/4 + (b/a)^5/8 + \cdots)b/2$ while $\Delta t$ and $\Delta X$ are arbitrary, regardless of whether $c^2\Delta t^2 > \Delta X^2$ or $c^2\Delta t^2 < \Delta X^2$ the series expansion yields $\delta s = \delta \rho_c + \delta \rho_o / 2$. The former inequality corresponds to $\delta \rho_c = c\Delta t$, while $\rho_o = \Delta X \left[ \Delta X / \rho_c + (\Delta X / \rho_c)^3 / 4 + \cdots \right]$ can be expressed in principle with any number of higher order terms. Being $\delta s$ arbitrary, the reasoning of subsection *2.1* is identically replicated utilizing the invariant vector $\delta \mathbf{s} = \delta \boldsymbol{\rho}_c + \delta \boldsymbol{\rho}_o / 2$ in $M_w = (\delta \mathbf{s} \times \delta \mathbf{p}) \cdot \mathbf{w}$. Replace the invariant form of $\mathbf{p}$ with its own uncertainty range, noting once again that neither reference system nor analytical form of the relativistic momentum need to be specified: as previously shown, the local momentum is not really calculated at any position or time, rather it is simply required to randomly change within a range of values undetermined itself. One immediately infers again $M_w = \pm l\hbar \pm l'\hbar / 2$: the second addend appears because the simple range of space uncertainty is replaced by the more general range consisting of both space and time parts, which explain why the series development defines $\delta s$ as sum of two terms. As expected, considering invariant ranges of conjugate dynamical variables or the invariant range $\delta s$ since the beginning the result is the same: $M_w$ differs from that of subsection *2.1* by the presence of the term $l'\hbar/2$ related to the time component of the space-time uncertainty. In any case, this result requires

$$\mathbf{M} = \mathbf{L} + \mathbf{S} \qquad 3,5$$

In subsection *2.1* $M^2$ has been calculated summing its squared momentum components averaged between arbitrary values $-L$ and $+L$ allowed for $\pm l$, with $L$ by definition positive; the sum gave $3 < (\hbar l)^2 > = L(L+1)\hbar^2$. Follow now an identical method. Replace $\pm l$ with $\pm l \pm s$ and let likewise $j = l \pm s$ range between arbitrary $-J$ and $J$; then $M^2 = 3 < (\hbar j)^2 > = 3(2J+1)^{-1} \sum_{-J}^{J} (\hbar j)^2 = \hbar^2 J(J+1)$ with $J$ positive by definition. In effect the obvious identity $\sum_{-J}^{J} j^2 \equiv 2 \sum_{0}^{J} j^2$ confirms that $J$ consistent with $M^2$ takes all the values allowed to $|j|$ from $|l-s|$ up to $|l+s|$ with $l \leq L$ and $s \leq S$. Since no hypothesis has been made on **L** and **S**, this result yields in general the addition rule of quantum vectors. Also, holds for **S** the same reasoning carried out in section 2 for **L**, i.e. only one component of **S** is known, whereas appendix C shows that $S^2 = \hbar^2 (L'/2+1)L'/2$. To show the physical meaning of **S**, it is instructive to compare the two ways to infer eq 3,5, which is just the eq C7 found in appendix C. The appendix follows the typical way of reasoning of special relativity that concerns observers and physical quantities in two different inertial reference systems $R$ and $R'$ in reciprocal motion: so the angular momentum is the anti-symmetric 4-tensor $\mathrm{M}^{ik} = \sum (x^i p^k - x^k p^i)$ whose spatial components coincide with the components of the vector $\mathbf{M} = \mathbf{r} \times \mathbf{p}$. In the present model, eqs C1 and C2 introduce the Lorentz transformations of length and linear momentum to define in eq C3 $\mathbf{M'}$ as a function of $\mathbf{M}$ respectively in $R'$ and $R$; in turn eq C3 coincides with eq C4 obtained directly from the transformation of the 4-tensor. The next steps achieve the sought result $\mathrm{M}_w = \pm l\hbar \pm l'\hbar/2$ in eq C13 through simple manipulations of the cross products defining **L** and **S**, as done in subsection *2.1*. Yet, exploiting directly the invariance of uncertainty ranges leads to eq 3,5 in a more straightforward and easier way simply considering the sum of space and time uncertainties inherent $\delta\mathbf{s}$. The key point of the comparison is not the greater simplicity and immediacy of the last approach with respect to that of appendix C, rather the verification that the extended concept of uncertainty including also the time efficiently surrogates the explicit reasoning based on the different points of view of observers in $R$ and $R'$ and the tensor definition itself of angular momentum. This conclusion is not surprising. The first part of this section has shown that the transformation properties between different reference systems are already inherent the concept of space-time uncertainty, of which they are natural consequences; then the requirement that the ranges are arbitrary inevitably entails that any physical event is identically described by $\Delta x$, $\Delta p_{xt}$, $\Delta t$, $\Delta \varepsilon$ or by any other $\Delta x'$, $\Delta p'_{xt}$, $\Delta t'$, $\Delta \varepsilon'$. Once regarding these ranges as quantities defined in the respective inertial reference systems $R$ and $R'$ in reciprocal motion at constant velocity V, the arbitrariness of their size previously introduced appears to be nothing else but the statement of physical equivalence of $R$ and $R'$ in describing any physical event; their invariance is ensured by the compliance with the interval rule shown above. The invariance between the points of view of different observers is thus surrogated by the indistinguishability of the respective reference systems. This reasoning avoids thus considering explicitly the points of view of different observers and explains why eq 3,5 is already basically inherent the uncertainty equation 3,1. Just for this reason eq 2,1 is in fact consistent with the Lorentz transformation of **M** through an appropriate reading of the length ratio $\Delta r/\Delta r'$, even though the early approach of section 2 is unrelated to the concepts of relativity: simply including the time into the uncertainty equation 1,2, as done in eq 3,1, it follows that the ratio matches the Lorentz transformations of lengths in $R$ and $R'$. In this respect it is also significant to show that the result of eq 3,5 can be once more obtained through the linear combination of uncertainty ranges of eq 3,1; this aims to confirm that the analytical form of the ranges is unessential as concerns the quantized results. Rewrite eq 3,1 in vector form as $\Delta\mathbf{x}_{xt} = \mathbf{r} - \mathbf{r}_o + \mathbf{v}\Delta t$ and introduce the position $2\mathbf{v}\Delta t = \Delta\mathbf{r}_t = \mathbf{r}_t - \mathbf{r}_{ot}$, i.e. $\Delta\mathbf{x}_{xt} = (\mathbf{r} - \mathbf{r}_o) + (\mathbf{r}_t - \mathbf{r}_{ot})/2$; the factor 2 is explained considering that the total displacement range $\Delta\mathbf{r}_t$ of a free particle compatible with the direction defined by $\mathbf{v}$ is actually twice the path $|\mathbf{v}|\Delta t$, because the particle can move towards two directions opposite and indistinguishable with respect to the reference point $\mathbf{r}_{ot} = \mathbf{v} t_o$. Defining then $\Delta \mathbf{M} = \Delta \mathbf{x}_{xt} \times \Delta \mathbf{P}^\S$ with $\Delta \mathbf{P}^\S$ arbitrary momentum range and proceeding as usual, one finds $\Delta \mathbf{M} = \Delta \mathbf{r} \times \Delta \mathbf{P}^\S + \Delta \mathbf{r}_t \times \Delta \mathbf{P}^\S /2$ that of course yields once again $\mathrm{M}_w = \pm l\hbar \pm l'\hbar/2$: note also here the further number $l'$ of states pertinent the time component of uncertainty. In fact $l$ and $l'$ are independent because they concern two independent uncertainty equations; the former is related to the angular motion of the particle, the latter must be instead an intrinsic property of the particle, since its value is defined regardless of whether $l=0$ or $l \neq 0$. As it actually means that exist different kinds of particles characterized by their own values of $l'$, the conclusion in agreement with the considerations of appendix C is that $\mathbf{S}$ can be nothing else but what we call spin of quantum particles; this confirms the self-consistency of the theoretical model and the conceptual link between quantum mechanics and relativity. If this conclusion is correct, then the particles should behave depending on their own $l'$. Let us consider separately either possibility that $l'$ is odd or even including 0. If $l'/2$ is zero or integer, any change of the number $N$ of particles is physically indistinguishable in the phase space: are indeed indistinguishable the sums $\sum_{j=1}^{N} l_j + Nl'/2$ and $\sum_{j=1}^{N+1} l_j^* + (N+1)l'/2$ that define the total value of $\mathrm{M}_w$ before and after increasing the number of particles, as the respective $l_j$ and $l_j^*$ of the $j$-th particle are actually arbitrary. So $\mathrm{M}_w$ and then $\mathrm{M}^2$ after addition of one particle replicate any possible value allowed to the particles already present in the system simply through a different assignment of the respective $l_j$; hence, in general, a given number of allowed states determining $\mathrm{M}_w$ in not uniquely related to the number of particles. The conclusion is different if $l'$ is odd and $l'/2$ half-integer; the properties of the phase space are not longer indistinguishable with respect to the addition of particles because now $\mathrm{M}_w$ jumps from …integer, half-integer, integer... values upon addition of each new particle: any change of the number of particles necessarily yields a total component of $\mathrm{M}_w$ and then a total quantum state different from the previous one; otherwise stated any odd-$l'$ particle added to the system entails a new quantum state distinguishable from those previously existing, then necessarily different from that of the other particles. In brief: a unique quantum state is consistent with an arbitrary number of even-$l'$ particles, whereas a unique quantum state characterizes each odd-$l'$ particle. Clearly, this is nothing else but a different way to express the Pauli exclusion principle, which is thus natural corollary itself of quantum uncertainty. This reasoning is extended considering again the eq 3,2 and requiring that the link between $\Delta \mathrm{p}_{xt}$ and $\Delta \varepsilon$ be invariant. This is possible if in eq 3,2 $\Delta \mathrm{x}_{xt}/\Delta t = c$, hence $\Delta \mathrm{p}_{xt} c = \Delta \varepsilon$ is a sensible result: it means of course that any $\varepsilon$ within $\Delta \varepsilon$ must be equal to $c \mathrm{p}_{xt}$ through the corresponding $\mathrm{p}_{xt}$ within $\Delta \mathrm{p}_{xt}$. If however $\Delta \mathrm{x}_{xt}/\Delta t < c$, the fact that the arbitrary $\mathrm{v}_x$ is not longer an invariant compels putting for instance $\mathrm{v}_x^k \Delta \mathrm{x}_{xt}/\Delta t = c^{k+1}$ with $k$ arbitrary exponent; then $(\Delta \mathrm{p}_{xt} \mathrm{v}_x^{-k}) c^{k+1} = \Delta \varepsilon$ shows in general an invariant link between $\Delta \mathrm{p}_{xt} \mathrm{v}_x^{-k}$ and $\Delta \varepsilon$ through $c^{k+1}$. Since this equation must correspond to a sensible non-relativistic limit, is mostly interesting the particular case $k=1$; so $(\Delta \mathrm{p}_{xt}/\mathrm{v}_x) c^2 = \Delta \varepsilon$, which means $\mathrm{p}_{xt} = \varepsilon \mathrm{v}_x / c^2$ as well. This result contains the particular case $\mathrm{p}_{xt} c = \varepsilon$ and entails $\varepsilon/c^2 = m$ to fulfil the non-relativistic limit $\mathrm{p}_{xt}/\mathrm{v}_x \to m$, as already found in section 2. To find other known outcomes of special relativity is so trivially obvious that it does not deserve further attention here. It is worth emphasizing however that these results, usually inferred via Lorentz transformations, confirm the validity of eqs 3,2. A simple reasoning explains now why invariant results are effectively to be expected through these equations. Let eq 1,2 be defined in $R$ and let $\Delta x' \Delta p' = \Delta x \Delta p / (1 - V^2/c^2) = n'\hbar$ be its Lorentz transformation in $R'$ moving with respect to $R$ at velocity $\mathbf{V}$; for an observer in $R'$ the product of the sizes of coordinate and momentum ranges differs by the numerical factor $1 - V^2/c^2$, while the related number of states allowed to the system appears to be $n'$. Clearly still holds here the reasoning already carried out in subsection *2.6* about eqs 2,12a and b, according which $n$ and $n'$ are not specifically defined by assigned numerical values; being arbitrary V and the ranges $\Delta x \Delta p$ and $\Delta x' \Delta p'$ in $R$ and $R'$, changing the respective numbers of states means transforming an arbitrary integer $n$ into any other arbitrary integer $n'$ undetermined as well for both observers. So, despite their different notation, the sets of numbers $n$ and $n'$ remain in fact indistinguishable, i.e. eqs 3,2 hold irrespective of the particular reference system. Let us show now that in effect eq 3,2 entails Lorentz's transformation of the energy merely exploiting the results of subsection *2.1*. Recall that replacing the conjugated dynamical variables with the respective ranges, the component $M_w = (\boldsymbol{\rho} \times \mathbf{p}) \cdot \mathbf{w}$ of $\mathbf{M}$ along $\mathbf{w}$ yields $\pm \Delta \chi \Delta p_\chi$; if the arbitrary ranges $\Delta \chi$ and $\Delta p_\chi$ fulfil the Lorentz transformations, eq 3,5 yields $M^2 = M_w^2 + M_w \hbar$, being $M_w = |L_w \pm S_w|$. In principle however even any $\Delta x_{xt} \Delta p_{xt}$ of eq 3,2 could be regarded likewise $\Delta \chi \Delta p_\chi$, i.e. linked to the component $M_w$ of an appropriate $\mathbf{M}$ along $\mathbf{w}$. As this link can be reasonably expressed through a linear relationship, $\Delta x_{xt} \Delta p_{xt} = a\hbar + bM_w$, eq 3,2 reads also

$$a\hbar + bM_w = \Delta\varepsilon\Delta t = n\hbar \qquad n = a + bM_w/\hbar \qquad 3,6$$

The condition on the number of states must hold in general, thus also for a spinless particle with $L_w = 0$. So $a = n_o$, with $n_o$ arbitrary integer, whereas $bM_w/\hbar$ must be integer as well. A possible way for $b$ to fulfil this requirement is putting $bM_w = S_w + M_w = S_w + |L_w \pm S_w|$. So $n$ reads

$$n = n_o + S_w/\hbar + M_w/\hbar \qquad a = n_o \qquad b = (n - n_o)\hbar/M_w \qquad 3,7$$

Rewrite eq 3,6 as

$$b(M^2 - M_w^2)/\hbar = \Delta\varepsilon_{xt}^\S \Delta t_{xt}^\S = n^\S \hbar \qquad \Delta\varepsilon_{xt}^\S \Delta t_{xt}^\S = \Delta\varepsilon\Delta t - n_o\hbar \qquad n^\S = n - n_o \qquad 3,8$$

and note that $\Delta\varepsilon_{xt}^\S \Delta t_{xt}^\S$ is physically equivalent to $\Delta\varepsilon\Delta t$: changing range sizes is unessential as both ranges are actually arbitrary. Examine eq 3,8 considering $b(M^2 - M_w^2)$ in $R$ and $b'(M'^2 - M_w'^2)$ in $R'$ where the particle is at rest. If $\mathbf{w}$ is chosen normal to $\mathbf{V}$, which is possible because $\mathbf{w}$ and $\mathbf{M}$ are arbitrary, then $M_w'^2 = M_w^2$. So, owing to eq 2,1,

$$\frac{M^2 - M_w^2}{M'^2 - M_w^2} = \frac{\Delta r^2}{\Delta r'^2} = 1 - V^2/c^2 = \frac{\Delta\varepsilon\Delta t}{\Delta\varepsilon'\Delta t'}$$

(for brevity of notation the unessential subscripts xt and § are omitted). So $\Delta t/\Delta t' = \sqrt{1 - V^2/c^2}$ already found yields $\Delta\varepsilon/\Delta\varepsilon' = \sqrt{1 - V^2/c^2}$ that holds for any $a$, $b$ and $S_w$; in turn, repeating here the reasoning to infer eqs 3,3, one concludes with the help of eq 2,14 $\varepsilon' = (\varepsilon - pV)/\sqrt{1 - V^2/c^2}$. A further consequence of eq 3,6 is highlighted recalling again eq B4 of appendix B $\hbar\Delta\varepsilon/\Delta t = \varepsilon^2$; eqs 3,6 and 3,7 yield

$$\Delta\varepsilon^2 = \phi + \xi M_w = n\varepsilon^2 \qquad \phi = n_o\varepsilon^2 \qquad \xi = b\varepsilon^2/\hbar \qquad 3,9$$

The result $\xi M_w = (n - n_o)\varepsilon^2$ is actually an identity. Yet, owing to the first eq 3,7, more interesting is the further result $\Delta\varepsilon^2 = (n_o + S_w + M_w/\hbar)\varepsilon^2$: the fact that $\Delta\varepsilon = \Delta\varepsilon(S_w, M_w)$ suggests in turn that $\varepsilon = \varepsilon(S_w, M_w)$ as well, because $\varepsilon$ is any random value within $\Delta\varepsilon$. If so, then

$$\varepsilon^2 = a_0 + a_1 M_w + \cdots \qquad \Delta\varepsilon^2 = \alpha_0 + \alpha_1 M_w + \alpha_2 M_w^2 + \cdots \qquad 3,10$$

The position $\varepsilon = \varepsilon(S_w, M_w)$ is in fact mere hypothesis based uniquely on a formal similarity with the dependence of $\Delta\varepsilon$ on $M_w$ and $S_w$; actually nothing is known about the random variable $\varepsilon$ in its uncertainty range. It is therefore matter of experimental evidence to establish if such a position, which appears nevertheless reasonable, is true or not. So, on the one side eqs 3,10 cannot be regarded as general properties of all particles; rather, being consequence of a specific assumption, these equations could possibly concern a particular class of particles. On the other side however the general validity of eq 3,2 does not exclude the chance of real physical meaning of the position $\varepsilon = \varepsilon(S_w, M_w)$ simply because, if correct, it does not conflict with the essential postulate of uncertainty of $\varepsilon$ within $\Delta\varepsilon$. To test eqs 3,10, let us symbolize the energy range as $\Delta\varepsilon = \varepsilon_{Mw} - \varepsilon_o$, being the variable $\varepsilon_{Mw}$ the actual particle energy and $\varepsilon_o$ a reference energy; then $(\varepsilon_{Mw} - \varepsilon_o)^2$ should be function of $M_w$, approximately linear if $\alpha_2 M_w << \alpha_1$ according to the convergence of the series development, whereas $M_w$ coincides with the spin in the case of free particles, i.e. $\Delta\varepsilon = \Delta\varepsilon(S_w)$ and $\varepsilon = \varepsilon(S_w)$. Each class of such particles should be then characterized by its own value of $\varepsilon_o$. Put without loss of generality $\varepsilon = \xi \varepsilon_{Mw}$, with $\xi$ by definition such that $\varepsilon_o \leq \xi \varepsilon_{Mw} \leq \varepsilon_{Mw}$. Actually the arbitrary factor $\xi$ is unessential, because it would merely rewrite the first eq 3,10 as $\varepsilon_{Mw}^2 = a_0' + a_1' M_w + \cdots$. This suggests that $\varepsilon = \varepsilon_{Mw} = m_H c^2$, being $m_H$ the particle mass. This position holds also for $\Delta\varepsilon$, since its size is arbitrary; if so $\varepsilon_o = m_o c^2$ as well, being $m_o$ a suitable reference mass. In conclusion eqs 3,10 can be tested plotting $S_w$ vs $(m_H - m_o)^2$ and $m_H^2$. The equation linking hadron spin and mass[2] is well known; the Regge-Chew-Frautschi diagrams of experimental data for various bosons and fermions are widely reported in literature, see e.g. [15]. Fig 1 collects for clarity all data into a unique plot. Fig 2 is more interesting; it shows the plot of $S_w$ vs $(m_H - m_o)^2$ of all the quoted particles (dots) and the global regression curve. It appears that with proper values of $m_o$, characteristic of each kind of hadron, all the data calculated with the second eq 3,10 merge reasonably well into a unique trajectory. Note that actually one best fit value of $m_o$ only is required to calculate the regression coefficients reported in figure; once having calculated the coefficients for one hadron, the $m_o$ values of all the other hadrons are determined in order to fit the respective data to the first curve. It is interesting just the existence of values of $m_o$ that fit all the data into a unique Regge trajectory, well approximated by a liner trend since $\alpha_2 << \alpha_1$; it confirms that the series converges and that the second equation 3,10 involving the energy range is more general than the first one involving the energy itself. Replacing the local coordinates with the respective uncertainty ranges requires a final comment. Despite the compliance of the concepts introduced above with the special relativity, seems however problematic their extension to the general relativity, where the gravity force is explained as mass induced space-time curvature and the invariant interval is replaced by a more complex local metrics to describe this curvature. The tensor calculus expressing mathematically these concepts entails, for instance, that the local scalars $\rho^2$ and $p^2$ have the form of sums $\rho_i \rho^i$ and $p_i p^i$. On the one hand, disregarding the local terms in principle as shown in the previous section means excluding the tensor formalism of general relativity and then the wealth of information inferred by consequence. Also, the positions 1,1 bypass familiar concepts of the gravitational field theories: covariant or contravariant derivatives, necessary in curvilinear coordinate systems, are in fact useless if the local dynamical variables do not longer play "*a priori*" any physical role; in fact the concept of local distance is physically meaningless in the present theoretical context. A question raises reasonably at this point: could unexpectedly be just these mathematical features, required by a geometrical standpoint of gravity force and thus "missing" in the present theoretical frame, that make so problematic matching general relativity and quantum mechanics? In this respect it is surely relevant the fact that the basic assumptions of special relativity have been inferred as corollaries of the uncertainty; moreover the transformation properties of length and time have been also inferred without utilizing the 4-vector algebra. Appears besides encouraging the possibility to obtain in a straightforward way the most relevant results of special relativity as sketched above. So, there is no reason to exclude that also the formalism of the present model can lead to the same final results of tensor calculus, despite a mathematical approach drastically different; just the Lorentz transformation of angular momentum $M'^2 = (1 - v^2/c^2)M^2 + M_w^2 v^2/c^2$, usually obtained from its definition of momentum 4-tensor and coincident with eq 2,1 as previously shown, is an example that supports this expectation. In conclusion, the possibility of extending further to the general relativity the procedure summarized previously by the positions 1,1, i.e. considering eqs 1,2 and 2,10 as unique postulate, appears in principle admissible provided that the physical basis of the gravity force be somehow rooted itself in the concept of uncertainty; if so, despite the mathematical formalism necessarily different with respect to that of the current field theories, the reasoning followed in section 2 should include also the gravitational interactions in the present conceptual frame. As emphasized at the end of section 2, the reason of this expectation rests on the remarkable generality of the present theoretical frame with respect to that exploiting the operators formalism of wave mechanics; it is certainly relevant that the positions 1,1 entail through eqs 1,2 and 2,10 not only the corollaries of indisguishability of identical particles and exclusion principle but also the foundations of special relativity and the concept of mass, while the appendices A and B infer the operators of momentum and energy, the concept of wavefunctions and the wave/particle dualism. With this conceptual background, the next section 4 exploits again and identically the approach outlined in section 2. The aims are: (i) to show that the gravity force is effectively rooted into the concept of quantum uncertainty; (ii) to consider various isolated systems of two particles whose interaction is uniquely due to the gravitational force; (iii) to formulate for each system the classical gravitational problem; (iv) to plug into the respective problems the positions 1,1 and then exploit eqs 1,2 and 2,10; (v) to describe the behaviour of the interacting particles with the formalism of the quantum uncertainty. The belief underlying these points rests on considering even the properties of space-time, including its curvature, as mere consequences of the quantum uncertainty; in other words, if the gravity really has quantum origin there is no reason to exclude that the space-time curvature is itself a quantum phenomenon. If so, the formalism of tensor calculus is not required to exploit the points (i) to (v); rather, as in section 2, attention must be paid instead to introduce the quantum states inherent the uncertainty of any gravitational system. Instead of describing purposely how the local mass curves the space-time, one should infer the curvature as a consequence of the same quantum principles that introduce the gravity force. The simple mathematical formalism of section 2 will be again replicated in the next section to exploit the idea of space-time uncertainty; this allows regarding the gravity like any other quantum phenomenon. The section 2 has shown that "classical physics + space uncertainty = non-relativistic quantum physics"; furthermore the section 3 has shown that "classical physics + spacetime uncertainty = quantum special relativity". What about general relativity? The next section 4 will show that the concept of quantum delocalization is effectively the sought "added value" to the classical Newton physics enough to infer the most relevant results of general relativity in a surprisingly simple and straightforward way.

4. The gravity field.
The theoretical basis of the papers [1,2] rests entirely on the concepts sketched in section 1, subsequently elaborated as shown in section 2. The physical information inherent the agnostic logic of the uncertainty has shown that a unique hypothesis, the random delocalization of particles in arbitrary ranges, describes the properties of the quantum world. Opportunely the time is inherent the relativistic definition of uncertainty; according to the considerations of section 3, its role in describing the quantum systems is self-legitimated regardless of any further explanation. For instance the spin appears mere consequence of merging together time and space uncertainties. Thus, after having formerly introduced the space range derivative $\partial_{\Delta x}$, it appears natural to consider also the time range derivative $\partial_{\Delta t}$ in agreement with the idea of regarding the change of $t - t_o$ in a conceptually analogous way as that of $x - x_o$; now is $t$ the variable coordinate in an arbitrary time reference system where is defined $t_o$, i.e. $\partial_t = \partial_{\Delta t}$. The physical meaning of $\partial_{\Delta x}$ has been highlighted in section 2 when calculating the particular space ranges that correspond to the minimum energy of the system in any reference frame; let us show now that relevant consequences are also inferred from the concept of $\partial_{\Delta t}$. On the one side the physical equivalence of eqs 1,2 and 2,10 removes the necessity of an external interaction to explain the possible time evolution of the uncertainty of a system; on the other side this equivalence suggests that the relativistic point of view allows to invert the reasoning, i.e. the natural time evolution of the system uncertainty is related to a specific form of internal interaction between particles. The elusive character of such an interaction, unexpected in the conceptual frame of section 2 and then not considered in the cases therein examined, is easily understood; the fact that it explicitly involves both space and time parts of the uncertainty ranges, explains why it was inevitably missed in the approach based on eq 1,2 only. If for instance this interaction depends on $\partial \Delta x_{xt}/\partial \Delta t$, then it becomes clear why it was skipped in the steady cases of section 2. Nevertheless, once introducing also the time into the uncertainty, the effects of such an interaction can be regarded exactly as shown in section 2, i.e. simply replacing the coulombian force of the hydrogenlike atom or the harmonic spring of the oscillator with a new force, whose analytical form is however still to be inferred. In other words, it is necessary: (i) to show that effectively the space-time dependence of the uncertainty ranges defines the sought force, (ii) to infer the analytical form of this force through the positions 1,1 and (iii) to calculate results comparable with experimental observables. In principle, it is correct to say that only the experimental observation legitimates the actual existence of such a force. Yet, since in the quoted papers the gravity has been never considered, one suspects that just the gravity could be the sought kind of interaction concerned in particular by the point (i): for instance, it could be active even between electron and nucleus in the hydrogenlike atoms, although neglected in the Hamiltonian of section 2 with respect to the Coulomb interaction. The next subsection *4.1* examines the points (i) and (ii) in a merely speculative way, i.e. regardless of any preliminary information about the forces of nature and without any conceptual hint provided by known experimental evidences. The idea of time dependent uncertainty and the consequent deformation rate of the uncertainty ranges will be introduced in abstract way, i.e. simply because nothing hinders in principle its effective occurring; this idea will be legitimated "*a posteriori*" by the results of the following subsections *4.2* to *4.8* that concern the aforesaid point (iii). All of the considerations hereafter carried out are therefore developed on deductive and self-contained basis, once again starting from the classical physics implemented with the concept of quantum time-space uncertainty. The interesting fact is that the experimental verification of the results not only validates the whole theoretical model, but also highlights the hierarchical significance of the concept of quantum delocalization among the known universal principles of nature; for instance, the question raises about whether the space-time uncertainty requires as additional hypothesis the concept of space-time curvature or infers it as a consequence, the same as the indistinguishability of the inertial reference systems entails by necessity the invariance of $c$. The next subsections aim to answer this question by introducing first the gravity force in the same conceptual frame of section 1, i.e. simply describing the behaviour of quantum particles delocalized in time-space uncertainty ranges of the phase space. For simplicity the particles are assumed having zero spin and zero charge, in order to consider their gravitational interaction only; if so the quantum results describe also the behaviour of macroscopic bodies, e.g. planets, through proper mass, time and length scale factors.

*4.1. Quantum basis of the gravitational interaction.*

Les us consider first an isolated system of two non-interacting free particles constrained to move within their respective space-time uncertainty ranges, shortly denoted $\Delta x_1$ and $\Delta x_2$ from now on. Being the problem one-dimensional by definition, let $\Delta P_1$ and $\Delta P_2$ be their conjugate momentum ranges including any local values of the components $P_1$ and $P_2$ of the respective momenta $\mathbf{P}_1$ and $\mathbf{P}_2$. Before interacting, the particles are delocalized in the respective ranges independently each other; two separate uncertainty equations hold therefore for each particle

$$\Delta P_i \Delta x_i = n_i \hbar \qquad i=1,2 \qquad 4,1a$$

Moreover, let us write the corresponding time uncertainty equations as done for eq 2,10 in section 2

$$\Delta \eta_i \Delta \tau_i = n_i \hbar \qquad 4,1b$$

The notation of eqs 4,1b emphasizes the different physical meaning of $\Delta \tau_i$ now introduced with respect to $\Delta t$ of eq 3,1: the latter is the time range that defines in general the total uncertainty of any particle regardless of whether it is free or interacting, the former is specifically related to the energy ranges including the possible $\eta_i$ of the free particles considered here. To better understand this point, consider the first particle delocalized in its own $\Delta x_1 = |\mathbf{x}_1 - \mathbf{x}_o| \pm |\mathbf{v}|\Delta t$. In the reference system where is defined $\mathbf{v}$, the limit $\Delta x_1 \to \infty$ ensures everywhere and everywhen $\Delta P_1 \to 0$; yet a vanishingly small momentum range means a unique value allowed for the particle momentum, i.e. $P_1(x) = const$ at any $x$. This holds in any inertial reference system simply with a different value of $P_1$. This result is nothing else but the inertia principle that concerns a lonely particle delocalized in an infinite range. If this is the definition of free particle, then: (i) the particles must be at finite distances to interact, if they would be infinitely apart would hold for each one of them the previous considerations; (ii) the interaction must propagate at finite rate otherwise the particle would interact even if infinitely apart. Consider now again the particles described by eqs 4,1 at mutual distance random and unknown but finite in an arbitrary space-time reference system where $|\mathbf{v}| \neq 0$. The fact that now both space and time define their total uncertainty means that the sizes of $\Delta x_i$ in eqs 4,1a change as a function of time; assuming that at least one of them expands, e.g. $\Delta x_1 = |\mathbf{x}_1 - \mathbf{x}_o| + |\mathbf{v}|\Delta t$, at an appropriate $\Delta t^*$ defined in the same reference system of $|\mathbf{v}|$ the separate ranges $\Delta x_i$ merge into a unique $\Delta x$; thereafter both particles are delocalized in the same range. The involvement of time uncertainty modifies the situation described by eqs 4,1 when the particles cannot be longer regarded separately. In this respect recall once more that according to the positions 1,1 the local conjugate dynamical variables are random, unknown and unpredictable and that both uncertainties position/momentum and time/energy determine the quantum eigenvalues. It is evident that if each space time range concerns separately either particle only, ten the respective eigenvalues $\eta_i = const_i$ must be independent each other too; if however both particles are described by a unique space-time delocalization range, then the possible eigenvalues must somehow describe the system formed by both particles. The obvious conclusion is that before merging $\Delta x_i$ into $\Delta x$ the particles do not interact, yet the particles are not longer independent when they share the same range, i.e. they someway interact. This is indeed the physical meaning of $\Delta \rho$ including the possible random distances $\rho$ between electron and nucleus in the hydrogenlike atom, in this case via Coulomb's interaction. The relativistic point of view accounts therefore not only for the transition from the non-interacting to the interacting state, but also for the existence of the interaction itself; the time range necessary for $\Delta x_1$ and $\Delta x_2$ to merge together corresponds to the finite time range for the interaction field to propagate and make the particles effectively interacting. The fact that the extended concept of space-time uncertainty entails the existence of an interaction is more directly confirmed as follows: if the interaction changes the initial values of both dynamical variables of two initially independent particles, then the size of the respective uncertainty ranges must change as well to include the new values. Let us show now that in effect a force is in general originated while $\Delta x_1$ and $\Delta x_2$ of eqs 4,1a merge into a unique $\Delta x$. When $\Delta x_i$ turn the respective sizes into $\Delta x$, new $\Delta P'_i$ are also required to encompass the local values $P'_i$ modified by the interaction; from a physical point of view it means introducing the deformation rates $\Delta \dot{x}_i$, which entail by consequence the respective changes $\Delta \dot{P}_i$ of momentum uncertainty ranges as well. This reasoning holds also for the local energies $\eta_i$ of the particles initially free; if these latter are modified by the mutual interation, then their new local random values $\eta'_i$ require the change of the respective energy ranges. In other words, the interaction changes both momentum and energy uncertainty ranges as a function of time. Thus deformation rates $\Delta \dot\eta_i$ are also to be expected when the total energy of the system is changed. Eqs 4,1a yield $\Delta \dot P_i = -(n_i \hbar / \Delta x_i^2)\Delta \dot x_i$; regarding in particular this expression at the specific time $t^*$ where both $\Delta x_i$ merge into $\Delta x$, one finds

$$F_i = \Delta \dot P_i = -\frac{n_i \hbar}{\Delta x^2}\Delta \dot x_i \qquad \Delta x_i = \Delta x \qquad \Delta \dot x_1 \neq \Delta \dot x_2 \neq \Delta \dot x \qquad 4,2$$

Eqs 4,2 still express the quantum uncertainty, yet in a form involving the time derivatives of ranges of dynamical variables and related numbers of states; also now the local values of coordinate and momentum are conceptually discarded since the beginning. While showing that $F_i$ do not depend on random time changes of the local values $P_i$ and related $\dot P_i$ as well, eqs 4,2 introduce force fields spreading within $\Delta x$ rather than local forces; owing to the positions 1,1, the present approach considers space-time ranges rather than specific points of space-time. Yet the sizes of $\Delta x_i$ and $\Delta x$ can be even regarded arbitrarily small; so the physical meaning of eqs 4,2 can be extrapolated even to a region of space-time small enough where the field is effectively approximated by a well defined local value. The forces introduced by the respective $\Delta \dot P_i$ are regarded separately because in general both $\Delta x_i$ are initially arbitrary and independent each other; so the deformation rates that allow their merging into a unique common range are in turn independent as well. The second eq 4,2 emphasizes that at the time $t^*$ both $\Delta x_i$ turn into the unique $\Delta x$; the inequalities emphasize the impossibility of specifying the rate with which $\Delta x_i$ are changing at the time of their merging into a unique range, as nothing is known about the former before interaction and about $\Delta x$ after interaction. Each field is then described by its own source particle and deformation rate: eqs 4,1a define two corresponding equations for the respective interaction driven momentum changes related to $\Delta \dot x_i$. An interesting consequence of eqs 4,2 is that the force fields are in general non additive. Regard $F_i$ as components of appropriate force vectors $\mathbf{F}_i$ along an arbitrary direction; note that no information is required here about the expected four-dimensional character of $\mathbf{F}_i$ and $F_i$, taken here into account through the definition of $\Delta x_i$ and $\Delta P_i$ as a function of both space and time coordinates. Whatever $\mathbf{F}_i$ might be, the linear combination $n_1 \Delta \dot x_1 + n_2 \Delta \dot x_2 = n \Delta \dot x$, which would yield $F_1 + F_2$ equal to $F$ consistent with $\Delta x$ and related $\Delta \dot P$, is for sure unduly incomplete: the arbitrariness of $\Delta \dot x_1$, $\Delta \dot x_2$ and $\Delta \dot x$ required by the uncertainty obliges the more general form $n\Delta \dot x = \sum_j b_j (n_1 \Delta \dot x_1 + n_2 \Delta \dot x_2)^j$ with $b_j$ constant coefficients. Yet it is immediate to guess that the classical additivity holds as a first approximation only; the Newtonian vector sum of forces corresponds to the first order term of the series development. This also entails that the force and acceleration vectors are parallel in the same approximation as well: $n\Delta \dot x = n_1 \Delta \dot x_1 + n_2 \Delta \dot x_2$ not only corresponds to $F = F_1 + F_2$ but also entails $n\Delta \ddot x = n_1 \Delta \ddot x_1 + n_2 \Delta \ddot x_2$ that relates to $F$ and to each $F_i$ a corresponding acceleration. Noting that $n$, $n_1$ and $n_2$ are arbitrary and thus independent each other, multiply both sides by an elementary mass $m_o$ so small that $nm_o = m$, $n_1 m_o = m_1$ and $n_2 m_o = m_2$ describe reasonably well any macroscopic mass of experimental interest; the expected result is then $m\Delta \ddot x = m_1 \Delta \ddot x_1 + m_2 \Delta \ddot x_2$. This shows that gravity force and acceleration are parallel vectors only in the approximation of neglecting the higher order terms of the series development. The link between force additivity and parallelism with the acceleration is reasonable: if the accelerations are not parallel to the forces, the vector sum of the former cannot coincide with the vector sum of the latter.

Eqs 4,2 have been obtained from eqs 1,2 and 3,1 in the most intuitive way, simply calculating $\partial(n\hbar/\Delta x_i)/\partial \Delta t$; thus, no matter how the interaction modifies the local velocities of the particles and regardless of the reference systems defined by $x_o$ and $t_o$ of space and time ranges, $F_i$ depend on $m_i$ and $\Delta \dot{x}_i$ through the respective $P_i$. The fact that the local dynamical variables are random and unknown, suggests that the source of the field can reasonably be nothing else but the mass; this conclusion justifies the conceptual link of $F_i$ just to the gravity, which is therefore a physical property of the mass. While the existence of the gravity field has been formerly introduced through the interaction between particles, it is also true that even an isolated particle generates its own gravity field within the pertinent $\Delta x$ simply because of the link between $\Delta \dot{x}_i \neq 0$ and mass, regardless of the presence of other massive particles. Also note at this point that each eq 4,2 consists of two equations $F_i = \Delta \dot{P}_i$ and $F_i = -n_i \hbar \Delta \dot{x}_i / \Delta x^2$: although linked together, the former involves explicitly the mass, the latter does not; in principle $\partial(n\hbar/\Delta x_i)/\partial \Delta t$ is a property of eq 3,1 definable regardless of any particle present in $\Delta x_i$. This suggests that the gravity force is not related necessarily to a source mass only, but even to the mere deformation rate of uncertainty ranges. To realize this point let us rewrite eq 3,1 identically as $\Delta x = x' - x'_o$, where $x' = x \pm v' \Delta t$ and $x'_o = x_o \mp v'_o \Delta t$ with $v' + v'_o = v$; all these quantities are defined in an arbitrary reference system $R_o$. No hypothesis is necessary about $v$; moreover $v'$ and $v'_o$ are arbitrary as well, so that the condition can be certainly fulfilled. It is possible to distinguish in principle the following three cases, whose physical meaning rests on the fact that the origin $x_o$ of $\Delta x$ is defined in $R_o$: (i) if $v'_o = 0$, the time dependent upper boundary $x'$ and the fixed lower boundary $x_o$ of $\Delta x$ describe merely the size change rate $\Delta \dot{x}$ of the range at rest with respect to the origin of $R_o$; (ii) if $v' = 0$, however, the deformation rate $\Delta \dot{x}$ is consequence of having fixed in $R_o$ the upper boundary $x'$, whereas $x'_o$ displaces as a function of time at speed $v'_o$ with respect to the arbitrary initial position $x_o$. The third chance with $v'_o \neq 0$ and $v' \neq 0$ is conceptually similar to (ii), but the translation rate of $x'_o$ in $R_o$ is $v'_o = v - v'$. For an observer in $x'_o$, the origin of $R_o$ is at rest in (i) but moves at rate $-v'_o$ in (ii); in the latter case $\Delta x$ shrinks while moving in $R_o$, i.e. the physical meaning of $\Delta \dot{x}$ is uniquely due to the translation rate of $\Delta x$ in $R_o$. In other words: the reasoning (ii) infers $F_i$ because of the mere time dependence of the uncertainty equation 3,1, the reasoning (i) shows that $F_i$ coincides with $\Delta \dot{P}_i$ related to the mass; the link is clearly $n\hbar \Delta \dot{x}_i / \Delta x^2$. Since the momentum change requires $\dot{v}$, it is impossible to distinguish if $F_i$ is originated in an accelerated reference system or in the presence of a massive particle in $\Delta x$; of course this holds for a range small enough to identify approximately the force field within $\Delta x$ with that just at $x'_o$. The fact that the gravity force is not necessarily linked to the concept of interaction with a field source only, being rather inherent eq 3,1, shows that it is really rooted in the concept of space-time uncertainty; the interaction of a mass delocalized in $\Delta x$ with the field triggered by the deformation rate $\Delta \dot{x}_i$ is a consequence of the link between its momentum change and size change of the space-time uncertainty $\Delta \dot{x}_i$. Follows the corollary: an accelerated reference frame is equivalent to the mass induced gravity field, with force proportional to acceleration. It is clear that the left hand side of eq 4,2 is linear function of $m_i$, which are just that introduced in eq 2,15; the same holds indeed for the range $\Delta \dot{P}_i$ and thus for $\Delta \dot{x}_i$ itself, which therefore can be defined as $m_i c^2 /(n_i p_o)$. The proportionality factor $p_o$ has physical dimension of momentum; so $p_o$ is the modulus of the gravity field momentum. Thus

$$\frac{F_i}{m_i} = \frac{\Delta \dot{P}_i}{m_i} = -\frac{\hbar c^2}{p_o}\frac{1}{\Delta x^2} \qquad\qquad \Delta \dot{x}_i = m_i \frac{c^2}{n_i p_o} \qquad\qquad 4,3a$$

As expected, the second equation confirms that $\Delta \dot{x}_i$ is in fact related just to $m_i$; moreover $F_i / m_i = \dot{v}_i$ shows that any mass behaves in the same way as a function of $\Delta x$ when subjected to this kind of force, whatever value and analytical form of $v_i$ might be. Clearly $p_o$ does not depend on $m_i$ in eqs 4,3a; it is instead peculiar feature of the interaction and must have necessarily the form $p_o = h/\lambda_o$, which associates a wavelength to the gravity field. This result highlights that in fact the only masses appearing in these equations are $m_i$ of the actual kinetic momenta $\mathbf{P}_i$; then the reasonable coincidence of gravitational and inertial mass follows by consequence, since a unique $m_i$ is source of its own gravitational field and also describes the acceleration in the field of another particle. To better understand the physical meaning of this position let us extend first eqs 4,3a to the case where one of the particles is massless. Both $F_i$ are still defined even in this case, since none of the considerations so far carried out excludes this possibility: no hypothesis has been made about the actual nature of the particles to infer eqs 4,2, whose validity is straightforward consequence of eqs 4,1 and then absolutely general. The particle 2 with $m_2 = 0$ is a photon having speed $c$ and momentum $P_2 = h/\lambda_2 = \hbar \omega_2 / c$, i.e. eqs 4,2 regard an isolated system formed by a light beam in the gravitational field of the mass $m_1$. Replacing in eq 4,3a $m_2 c^2$ with $\hbar \omega_2$, the second eq 4,2 reads

$$F_2 = \Delta \dot{P}_2 = -\frac{\hbar^2 \omega_2}{p_o}\frac{1}{\Delta x^2} \qquad\qquad \Delta \dot{x}_2 = \frac{\hbar \omega_2}{n_2 p_o} \qquad\qquad m_2 = 0 \qquad\qquad 4,3b$$

while $F_1$ is still given by the first eq 4,3a; since $F_1$ does not depend on the mass of the second particle on which it acts, the physical meaning of $\Delta \dot{P}_1$ is still that already introduced although the kind of interaction is clearly different. At this point, three preliminary considerations are useful before considering in detail eqs 4,3. The first is that $F_1$ and $F_2$ are coupled, being generated when both $\Delta x_1$ and $\Delta x_2$ merge into a unique $\Delta x$; it entails that even the light interacts with the gravity field generated by massive particles. The second is that, since we are considering an isolated system of two particles, $\Delta \dot{P}_2$ of the photon can be only explained admitting that its momentum $\mathbf{P}_2$ changes because of the interaction with $m_1$; there are then two possibilities: the gravitational field affects (a) the wavelength, (b) the propagation direction of the photon. The third is that the gravity field must propagate with light speed for the interaction with the photon be allowed to occur. These effects are completely outside of the previsions of Newton's law initially formulated merely to describe the dynamics of massive bodies attracted by Earth's gravity; yet the quantum origin of the force fields introduced above entails these effects as natural corollary. In conclusion the initial eqs 4,2 regard the gravity as mass induced space-time deformation, here expressed by $\Delta \dot{x}_i$, without renouncing however also to its classical definition of momentum change rate, here expressed by $\Delta \dot{P}_i$; then the present model is not mere geometrical model of the gravity force. Returning now to the case of two massive particles, the reasoning to infer $F_1$ and $F_2$ has evidenced that the present explanation of the gravitational interaction requires in general $\Delta \dot{x}_i < c$ according to the last eq 4,3a; indeed $\Delta \dot{x}_i$ equal to $c$ for $m_i \neq 0$ would define $p_o$ as a function of $m_i$, whereas it is instead by definition mere proportionality factor between $\Delta \dot{x}_i$ and $m_i$. Hence: (i) the gravity field propagates like a wave having frequency $v_o = c/\lambda_o$ defined by the momentum $p_o = \hbar \omega_o / c$; (ii) $p_o$ defines also the field energy $\varepsilon_o = \hbar \omega_o$; (iii) merging $\Delta x_1$ and $\Delta x_2$ into $\Delta x$ entails time dependence of $\eta_i$, whose $\dot{\eta}_i$ are described by the ranges $\Delta \dot{\eta}_i = -n_i \hbar / \Delta \tau_i^2$ since $\dot{\eta}_i = \Delta \dot{\eta}_i = \partial \eta / \partial \Delta \tau_i$ according to eqs 4,1b; (iv) the actual physical meaning of $\Delta \eta_i$ is related to that of $\Delta \tau_i$; (v) $\Delta \dot{P}_i = \partial \Delta P_i / \partial \Delta t$ and $\Delta \dot{x}_i = \partial \Delta x_i / \partial \Delta t$ according to eq 3,1. These assertions are correlated and now clarified through a relevant example. If the gravitational field generated by $m_1$ propagates at rate $c$ in $\Delta x_1$, faster than $\Delta \dot{x}_1$, then necessarily some field energy is lost outside $\Delta x_1$ during interaction: the field spreads indeed beyond the range allowed to the source particle 1. When the second particle starts interacting, its initial energy $\eta_2$ changes; $\dot{\eta}_2$ is controlled by the residual field energy $\varepsilon_o$ still in $\Delta x_1$ during its merging with $\Delta x_2$. To examine the consequences of the previous points (i) to (v) and highlight how $\varepsilon_o$ generated by one particle affects $\eta_i$ of the other particle initially free, let us consider for simplicity the case where the particles form a system orbiting circularly at constant distance $\Delta x_i$ from the gravity centre and specify the field angular frequency $\omega_o = n_i / \Delta \tau_i$ as a function of the revolution period $\Delta \tau_i$ of the orbiting $i$-th particle. So $-\dot{\eta}_i = (\hbar \omega_o) / \Delta \tau_i$ means that the energy $-\dot{\eta}_i \Delta \tau_i$ released from the system to the field is irradiated by this latter outside $\Delta x_i$ at rate $c$, and then lost by the system via pulses $\hbar \omega_o$ of gravitational waves having characteristic frequency $\omega_o$. Let $\eta'_i = (1 - \Delta \dot{x}_i / c) \eta_i$ be the new value of $\eta_i$ due to this specific loss mechanism; this position fulfils the conditions $\eta'_i \to 0$ for $\Delta \dot{x}_i \to c$ and $\eta'_i \to \eta_i$ for $\Delta \dot{x}_i \to 0$: the former limit highlights that just $\Delta \dot{x}_i < c$ allows the gravitational waves, the latter evidences that the $i$-th particle is actually still free unless $\Delta \dot{x}_i \neq 0$ triggers the gravity field. Since $\dot{\eta}_i = (\eta'_i - \eta_i) / \Delta \tau_i$ reads also $-\dot{\eta}_i = (\Delta \dot{x}_i / c) \eta_i / \Delta \tau_i$, one finds then $\varepsilon_o = \eta_i \Delta \dot{x}_i / c$. To check this result note that eq 4,3a yields $\eta_i = n_i (\hbar \omega_o)^2 / (m_i c^2)$. Simple dimensional analysis allows then rewriting $\dot{\eta}_i = 2 n_i \hbar^2 \omega_o \dot{\omega}_o / (m_i c^2)$ as $\dot{\eta}_i = 2 n_i^2 (G \hbar^2 \omega_o^2 / c^5)(\hbar \dot{\omega}_o \Delta \dot{x}_i / G m_i^2)$, where $G$ is the gravitation constant preliminarily introduced here and more thoroughly justified in the next subsection; the factors within parenthesis have physical dimensions $\hbar$ and $t^{-2}$. As expected, the form of $\dot{\eta}_i$ agrees with $\Delta \dot{\eta}_i = -n_i \hbar / \Delta \tau^2$ of eq 4,1b. It suggests that $\hbar \dot{\omega}_o \Delta \dot{x}_i / G m_i^2 \propto -(\eta_i^\S / \hbar)^2$ and then $\dot{\eta}_i \propto -G \omega_o^2 \eta_i^{\S 2} / c^5$, where $\eta_i^\S$ is reasonably the orbiting kinetic energy: it specifies that the field energy loss is just due to an orbiting system. To complete the reasoning note that $\eta_i^\S = -m_i \omega_o^2 \Delta x_i^2 / 2$ yields $\dot{\eta}_i = -w_o G m_i^2 \omega_o^6 \Delta x_i^4 / c^5$, where the constant factor $w_o$ summarizes all of the proportionality constants above introduced. This result, pertinent to the specific case described by the given definition of $\Delta \tau_i$ and $\eta_i^\S$, highlights the link between these latter and $\varepsilon_o$, thus explaining why during the change $\eta_i \to \eta_i^\S$ time instability is to be expected for the orbiting system. To infer the value of $w_o$, let us rewrite $-\partial \eta_i / \partial (w_o \Delta \tau_i) = G m_i^2 \omega_o^6 \Delta x_i^4 / c^5$: the right hand side describes the orbital motion of the particle and thus it is related to the force acting on the particle, as shown by $G$ introduced into $\dot{\eta}_i$ together with the identity $\Delta \dot{x}_i n_i \hbar \omega_o = m_i c^3$; the left hand side concerns the energy per unit time released to the field. The former depends on $\Delta t$ of eq 3,1 because of the time derivatives $\Delta \dot{x}_i$, the latter instead depends on $\Delta \tau_i$ according to eq 4,1b. Having defined $\omega_o = n_i / \Delta \tau_i$, then $\Delta \tau_i = \Delta t / 2 \pi$ and so $\dot{\eta}_i = (\partial \eta_i / \partial \Delta t)(\partial \Delta t / \partial \Delta \tau_i) = 2 \pi (\partial \eta_i / \partial \Delta t)$. The different physical meaning of $\Delta t$ and $\Delta \tau_i$ formerly emphasized in the points (iv) and (v) entails the time scale factor $2 \pi$ between $\partial \eta_i / \partial \Delta \tau_i$ and $\partial \eta_i / \partial \Delta t$. This factor, so far hidden by the notation $\dot{\eta}$, has been in fact included into the constants of proportionality introduced after $\Delta \dot{x}_i$ and summarized by the final coefficient $w_o$. It is reasonable to regard $w_o$ as the sought time scale coefficient $2\pi$ linear like $\Delta \dot{x}_i$ between $\partial \eta_i / \partial \Delta \tau_i$ and $G m_i^2 \omega_o^6 \Delta x_i^4 / c^5$. Assuming $w_o = 2\pi$, one finds in effect $\partial \eta_i / \partial (w_o \Delta \tau_i) = \partial \eta_i / \partial \Delta t$, so that both sides are expressed as a function of $\Delta t$; then

$$-\frac{\partial \eta_i}{\partial \Delta \tau_i} = 2\pi \frac{G m_i^2 \omega_o^6 \Delta x_i^4}{c^5}$$

This formula, inferred through intuitive elementary considerations, agrees substantially with that found in general relativity to describe the emission of gravitational waves in the presence of weak fields; since $2\pi$ differs from $32/5$ by less that 2%, the present result compares reasonably well with the approximate solution of Einstein's field equation [16]. Analogous result will be more rigorously obtained together with the contraction rate of the reciprocal distance of the particles in subsection *4.7* as a function of the reduced mass of the orbiting system. The reasoning has been shortly sketched here merely to emphasize that really the field momentum $p_o$ is related to the gravity interaction: regarding $\omega_o = p_o c / \hbar$ equal to the actual angular frequency of orbital motion means specifying the experimental situation that characterizes the consequent frequency of gravitational waves removing field energy from the system. It also clarifies the importance of eqs 4,1b to justify the time instability $\dot{\eta}$ and its link to the radiation of gravitational waves, whose quantum origin appears then evident like that of $F_1$ and $F_2$ themselves. The simple Newton law formulated without reference to the underlying quantum meaning cannot explain this effect, conceptually important although irrelevant from a practical point of view as concerns the consequences of the instability.

The remainder of the present paper describes in detail the consequences of eqs 4,1 to 4,3, thus showing that the most remarkable results of general relativity are inferred through very simple considerations having entirely quantum mechanical character: the fact that the origin of the gravity is deeply rooted in the concept of space-time uncertainty legitimates the possibility of regarding the gravitational forces as consequences of deformation rates of uncertainty ranges rather than through the local metrics of curvilinear coordinates of macroscopic bodies.

*4.2 Newton's law*.

Consider an isolated system formed by two particles having masses $m_1$ and $m_2$ interacting in $\Delta x$. In agreement with the idea that each force is defined by its gravitational mass only, eqs 4,3a read

$$F_1 = -\chi \frac{m_1}{\Delta x^2} \qquad F_2 = -\chi \frac{m_2}{\Delta x^2} \qquad \chi = \frac{\hbar c^2}{p_o} \qquad p_o = \frac{\hbar \omega_o}{c} \qquad p_o = p_o(\lambda_o, \Delta x) \qquad 4,4$$

The minus signs mean that the fields are attractive. The last eq 4,4 is understood noting that $\chi$ does not depend on the mass of the particles, being property of the field momentum only; yet there is no reason to exclude that it depends in general on $\Delta x$ through $p_o$, as it will be more comprehensively explained in subsection *4.6*. The effects of either force on the other particle, so far regarded separately, can be also combined into a unique general law $F$ describing their mutual attractive interaction. Clearly $F$ must be function of the product $m_1 m_2$ in order that it reduces as a particular case to $F_1$ or $F_2$ for unit $m_2$ or $m_1$ respectively. It suggests introducing a unit reference mass $m^u$ defining $F$

$$F = -\chi_G \frac{m_1 m_2}{\Delta x_{12}^2} \qquad \chi_G = \frac{\chi}{m^u} = \frac{\hbar c^2}{m^u p_o} = \frac{c^3}{m^u \omega_o} \qquad 4,5$$

Eq 4,5 highlights that also now the positions allowed to both particles are encompassed by their distance uncertainty $\Delta x_{12}$; yet, being this latter arbitray, any mutual position $x_{12}$ is actually described by the $m_1 m_2 x_{12}^{-2}$ force law. Introducing $m^u$ both $m_1$ and $m_2$ are expressed in the same mass unit system, which was instead in principle unnecessary when considering separately $F_1$ and $F_2$; in this way the numerical values of $\chi_G$ and $\chi$ coincide and depend on the measure units fixed by $\Delta x_{12}$ and $m^u$, with the time unit also fixed in agreement with $\Delta \dot{P}_i$. Note that it is not longer possible to distinguish in eq 4,5 which particle is the source of the field and which one is that interacting with the field itself; actually this equation combines together two forces $F_{12}$ and $F_{21}$ with the particles 1 and 2 formally interchangeable in the role of gravitational and inertial masses. Then the physical equivalence of these masses, already shown, requires $F_{12} + F_{21} = 0$ to exclude a net resulting force simply regarding the unique interaction of eq 4,5 from the points of view of either particle. Before concerning in more detail the physical meaning of $p_o$, see eq 4,21 of the next subsection, note that approximating this latter with the constant value $p_o^0$ yields $\chi_G = const$; in this case $F$ reduces to the approximate Newton gravity law written as a function of the range $\Delta x_{12}^2$ replacing random distances $x_{12}^2$

$$F_N = -G \frac{m_1 m_2}{\Delta x_{12}^2} \qquad G = \frac{\hbar c^2}{m^u p_o^0} = \frac{c^3}{m^u \omega_o^0} \qquad 4,6$$

Are useful some remarks about $F_N$, noting however that because of the Newtonian character of eq 4,6 the considerations exposed below are non-relativistic by definition. The potential energy $U_i$ corresponding to $F_i$ is of course $U_i = -\int F_i dx' = -\int F_i d(\Delta x')$; if as usual the integration limits are $\Delta x$ and $\infty$, where $U_i$ vanishes, one finds the expected result $U_i \approx Gm_i / \Delta x$. This result is approximate as $\chi$ of eqs 4,4 has been replaced by $G$ of eq 4,6. A further remark concerns just $G$. The quantum nature of this constant, self-evident because of the presence of $\hbar$ in its definition, appears more clearly noting that with trivial manipulations $F_N$ yields

$$F^* = \frac{F_N}{F_P} = -\frac{(m_1/m_P)(m_2/m_P)}{(\Delta x_{12}/l_P)^2} \qquad m^u p_o^0 = m_P^2 c \qquad m_P = \sqrt{\frac{\hbar c}{G}} \qquad l_P = \sqrt{\frac{\hbar G}{c^3}} \qquad F_P = \frac{c^4}{G} \qquad 4,7$$

It is crucial to observe that introducing here the Plank units is not mere dimensional exercise to combine $G$ with the fundamental constants of nature, because both $\hbar$ and $c$ are already inherent the conceptual definition of gravity constant; the fact that the dimensionless force $F^*$ has the proportionality constant normalized to 1 supports the definition of $G$ in eq 4,6 and thus the approach so far followed. The dependence of the value of $G$ on the choice of the measure units is obvious; yet, without evidencing the quantum origin of the force, the possible combinations of $\hbar$, $c$ and $G$ would have mere formal meaning. Eqs 4,6 and 4,7 show that $G$ enters into the equation of $F_N$ only because the usual choice of mass and length units does not take properly into account $\hbar$ and $c$, as instead it would be physically appropriate according to eqs 4,4. Simple manipulations of eqs 4,6 evidence an interesting analogy between the factors defining $F_N$. Since $m^u / p_o^0 c$ has physical dimensions of reciprocal squared velocity, $(v_o^u)^{-2}$, and $(m^u v_o^u \Delta x_{12})^2$ that of squared angular momentum, it is formally possible to rewrite eq 4,5 as follows

$$F_N = -\frac{\hbar c^3}{\mu_o} \frac{Mm}{\Delta M_{12}^2} \qquad (m^u v_o^u \Delta x_{12})^2 = \mu_o \Delta M_{12}^2 \qquad M = m_1 + m_2 \qquad m = m_1 m_2 / M \qquad 4,8$$

with $\mu_o$ dimensionless proportionality constant; $\Delta M_{12}^2$ is the squared angular momentum range allowed to a unit mass moving along a proper orbit at average distance $\Delta x_{12}$ from the gravity centre with velocity $v_o^u$: being $\Delta x_{12}$ arbitrary, the pertinent $v_o^u$ is arbitrary as well. Eqs 4,8 are interesting because $F_N$ does not longer contain dynamical variables of the single particles, but only properties of the system regarded as a whole: characteristic angular momentum replacing $\Delta x_{12}^2$, total mass and reduced mass. Rewriting $m_1 m_2$ as $mM$ emphasizes that a reduced mass travels around a fixed centre where is concentrated the total mass of the system; hence one would have expected $\Delta M_{12}^2$ expressed as a function of $m$. The second eq 4,8, related to the value of $v_o^u$, requires then $m^u v_o^u = m v_o$ in order to introduce in $F_N$ the expected reduced mass; this is certainly possible because, by definition, the new value $v_o$ leaves $F_N$ unchanged. Being $m$ arbitrary as well, it means in general $m v = \text{const}$. This result, momentum conservation, agrees in particular with $\dot{v} = 0$ for $F_N = 0$, inherent eqs 4,2 when no external force acts on the system; this is nothing else but the inertia principle previously inferred as a consequence of eqs 4,1. Note however that $v_o$ is arbitrary and that the momentum conservation holds regardless of its actual analytical form. Rewriting therefore $v_o = v_o^{rel}/\sqrt{1-(v_o^{rel}/c)^2}$ in agreement with eq 2,16 and repeating the reasoning above with the mere numerical requirement $v_o^{rel} \neq v_o$, the result reads $m v_o^{rel}/\sqrt{1-(v_o^{rel}/c)^2} = \text{const}$. So the momentum component conservation holds also in special relativity. Moreover eq 4,8 shows that once having fixed $M$ and $m$, a given value of $F_N$ determines uniquely $\Delta M_{12}^2$; the fact that $\Delta x_{12}$ does not longer appear explicitly in the first eq 4,8 means that if $F_N$ is constant, i.e. in the absence of external forces perturbing the system, $\Delta M_{12}^2$ is also constant. Hence, owing to the previous result,

$$(m^u v_o^u \Delta x_{12})^2 = (m v_o \Delta x_{12})^2 = (m v_o' \Delta x_{12}')^2 \qquad F_N = const$$

i.e. the angular momentum is conserved whatever $m v_o$, $m v_o'$ and $\Delta x_{12}$ might be, in agreement with the result already found in subsection *2.1* because of mere quantum reasons. It is significant that eqs 4,8 enable these results to be inferred without integrating any equation of motion and without any hypothesis *ad hoc*, but simply owing to the form of $G$ in eq 4,6. Let us return now to eq 4,8; in principle the reasoning just carried can be extended to the constant factor $\chi_G$ as well, thus obtaining

$$F = -\frac{(\hbar c)^3}{\mu_o' M_o^2} \frac{Mm}{\Delta M_{12}^2} \qquad \mu_o = \mu_o' M_o^2 / \hbar^2 \qquad 4,9$$

where $\mu_o'$ is a further dimensionless proportionality factor. The notation emphasizes that $M_o^2$ is inferred from $p_o$ only. Eq 4,9, with both factors inversely proportional to the respective squared angular momenta, evidences the formal analogy between the variable and constant factors of $F_N$, respectively describing the dynamics of the masses in the gravitational field and the gravitational field itself. Both $\mu_o$ and $\mu_o'$ justify the positions 4,8 and 4,9 on mere dimensional basis; yet the former equation is a formal position, whereas the latter introduces in eq 4,5 new physical information through the field angular momentum. Eq 4,9 will be further concerned in the next subsection *4.8* to estimate the numerical value of $G$. Usually proportionality factors very different from unity reveal that some relevant physical effect is hidden in the dimensional analysis; it will be found in subsection *4.8* that in fact $\mu_o' \approx 1$, whereas instead $\mu_o \approx (2\pi)^2$. This supports the idea that eq 4,9 is not a trivial way to rewrite eq 4,6 because it introduces the field angular momentum $\mathbf{M}_o$.

*4.3 Red shift.*

Consider an isolated system formed by a photon moving in the gravitational field of $m_1$. Let us rewrite the second eq 4,3b to define the average energy $F_2 \Delta x$ within the uncertainty range $\Delta x$

$$F_2 \Delta x = -\frac{\hbar^2 \omega_2}{p_o \Delta x} = \hbar \Delta \omega \qquad m_2 = 0 \qquad \Delta \omega < 0 \qquad 4,10$$

With the negative sign due to eq 4,3b, eq 4,10 describes loss of photon energy $\hbar\Delta\omega$ related to momentum change $\Delta P_2 = \hbar\Delta\omega/c$ according to the aforesaid effect (a) previously introduced in subsection *4.1*. Let us define an analogous energy for the particle 1

$$F_1 \Delta x = -\frac{\hbar c^2 m_1}{p_o \Delta x} \qquad 4,11$$

The fact that $\hbar/(p_o \Delta x)$ appears in both eqs 4,10 and 4,11 confirms that the photon momentum change is correlated to the force field $F_1$. Eliminating $\hbar/(p_o \Delta x)$ between these equations yields

$$F_1 \Delta x = \frac{m_1 c^2 \Delta\omega}{\omega_2} \qquad 4,12$$

The arbitrary value of $\Delta x$ determines $\Delta\omega$ as a function of the corresponding $F_1 \Delta x$, whatever the values of the former and the latter regarded separately might in principle be. Since both eqs 4,10 and 4,11 concern $\Delta x$, the propagation direction of the photon is by definition within the same range where is also located particle 1; hence the frequency change is that of a light beam moving radially with respect to the gravitational source $m_1$. Eq 4,12 rewritten as

$$\frac{\Delta\varphi}{c^2} = \frac{\Delta\omega}{\omega_2} \qquad \Delta\varphi = \frac{F_1 \Delta x}{m_1} \qquad m_2 = 0 \qquad 4,13$$

shows that the change of gravitational potential $\Delta\varphi$ within $\Delta x$ is a property of the particle 1 only. In effect $\omega_2$ does not depend on $\Delta\varphi$, being by definition the photon proper frequency. Instead is related to $\Delta\varphi$ the frequency shift $\Delta\omega$, which is to be expressed with respect to $\omega_2$ through the reasonable position $\Delta\omega = \omega - \omega_2$; indeed for $\Delta\varphi \to 0$ eqs 4,13 and 4,12 yield both $\Delta\omega \to 0$ and $\Delta\omega/(\Delta\varphi/c^2) \to \omega_2$ in any point of $\Delta x$. Moreover according to eq 4,12 $\Delta x \to \infty$ entails $\Delta\omega \to 0$ because $F_1 \Delta x \to 0$, i.e. photon and $m_1$ are infinitely apart; as expected $\Delta\varphi \to 0$ is also compatible with a vanishingly weak field within $\Delta x$, in which case the frequency tends obviously to that of a free photon exempt of gravitational effects, $\omega \to \omega_2$. Thus eq 4,13 can be also regarded in a formally different way: it calculates the frequency shift $\Delta\omega$ with respect to $\omega_2$ due to the gravitational potential $\Delta\varphi = \varphi - 0$ generated by $m_1$, which means $\varphi/c^2 = (\omega - \omega_2)/\omega_2$ with $\varphi < 0$ because $\Delta\varphi < 0$ according to eqs 4,4. In any case, $\Delta\omega < 0$ means $\omega < \omega_2$ i.e. the proper frequency $\omega_2$ is shifted by $\Delta\omega$ down to the lower value $\omega$ when the photon moves through $\Delta x$ towards a gravitational potential decreased by $\Delta\varphi$. No hypothesis is necessary about $\Delta x$, which in effect does not appear in the final result. Note that regarding eq 4,10 only, $(\hbar\Delta\omega/c)\Delta x = n_2 \hbar$, the uncertainty does not prevent in principle $\Delta x \to 0$ and then $\Delta\omega \to \infty$, which of course is nothing else but $\Delta P_2 \to \infty$ in eq 4,1a. Yet, the crucial concept that discriminates any option mathematically possible from the actual physical behaviour of the system is the coupling of $F_1$ and $F_2$, i.e. the interaction between the photon and $m_1$. Actually eqs 4,10 and 4,11 cannot be regarded separately just because of this interaction; they describe a real physical event when merged together into eq 4,12, which effectively concerns what really happens. The interaction is the boundary condition of the physical system that eliminates in fact the infinities admitted in principle by the unavoidable arbitrariness of any uncertainty range.

*4.4 Time dilation.*

Since the first eq 4,4 yields $\Delta\varphi/c^2 = -\hbar/(p_o \Delta x)$, it follows that

$$\frac{\Delta\varphi}{c^2} = -\frac{\Delta\tau}{\tau_0} \qquad \tau_0 = \frac{\Delta x}{c} \qquad \Delta\tau = \frac{\hbar}{c p_o}$$

The notation emphasizes that the time range $\Delta\tau$ is related to the gravitational potential through $p_o$, whereas $\tau_0$ does not by definition. Moreover the limit $\Delta\varphi \to 0$ requires $\Delta\tau/(\Delta\varphi/c^2) \to -\tau_0$ and $\Delta\tau \to 0$. Let us define $\tau_0$ as reference time in the absence of gravitational potential; then $\Delta\varphi = 0 - \varphi > 0$, because $\varphi < 0$. Since $\Delta\tau < 0$ if $\Delta\varphi > 0$, considerations completely analogous to those previously carried out show that the time $\tau$ deviates from $\tau_0$ by $\Delta\tau = \tau - \tau_0 < 0$ as a result of the rising of gravitational potential. In conclusion the first equation above reads

$$\tau = \tau_0 \left(1 + \frac{\varphi}{c^2}\right) \qquad 4,14$$

and shows therefore that time dilation occurs in the presence of gravitational potential.

*4.5 Light beam bending.*

The subsection *4.3* has described the frequency shift of a photon moving radially with respect to the gravitational mass $m_1$; the link between $F_1 \Delta x$ and $F_2 \Delta x$, eq 4,13, was exploited to this purpose. This result can be further extended; the system photon/$m_1$ can be also described by the force $F_N$ writing eq 4,6 in the form $-Gm_1(\hbar\omega_2/c^2)/\Delta x_{12}^2$, i.e. concerning the interaction of $m_1$ with the photon as if this latter would have virtual mass $m_2 = \hbar\omega_2/c^2$. Eq 4,13 suggests considering the ratio

$$\Delta\zeta = -\frac{F_N \Delta x_{12}}{\hbar\omega_2/c^2} = G\frac{m_1}{\Delta x_{12}} \qquad 4,15$$

conceptually analogous to $\Delta\varphi$; the sign has been chosen in order to define the range $\Delta\zeta$ positive for reasons shown below. Now the photon is not longer constrained to travel within the same range including $m_1$, as explicitly requested in the previous case by the simultaneous conditions 4,10 and 4,11. Eq 4,15 introduces the force between particles $\Delta x_{12}$ apart, yet without necessity that this latter coincides with the range $\Delta x$ where the photon is allowed to move; then, in lack of specific boundary conditions, one must assume that in general the photon moves outside $\Delta x_{12}$ where is located $m_1$. Since $\Delta\zeta$ has physical dimensions of squared velocity, it is certainly possible to write $\Delta\zeta = \xi c^2$, being $\xi$ a dimensionless proportionality factor:

$$\frac{\Delta\zeta}{c^2} = \xi \qquad 4,16$$

It is reasonable to expect that by analogy with eq 4,13 also $\Delta\zeta$ controls the photon momentum change, yet in this case the aforementioned effect (b) of subsection *4.1* is expected to occur: the fact that the uncertainty range $\Delta x$ of photon position does not include the source of gravitational potential, suggests that the deformation rate $\Delta\dot{x}$ to which is related $\Delta\dot{P}_2$ of the photon concerns now the stretching of $\Delta x$ due to its bending by effect of $F_N$; otherwise stated, the photon deflection reproduces the curvature of its allowed uncertainty range $\Delta x$ caused by the gravitational potential $\Delta\zeta/c^2$. Since any hypothesis about the uncertainty ranges is neither possible nor necessary, the curvature of $\Delta x$ does not contradict or rule out any result so far obtained: the crucial idea is instead conceptual impossibility to know where $m_1$ and the photon are exactly, irrespective of whether their position uncertainty is described by linear or curved ranges. As one sensibly expects that the local interaction increases as long as the photon approaches the gravitational mass $m_1$, the next step should seemingly be to estimate of curvature radius at any point of $\Delta x$. Yet such a calculation would be ineffective once having assumed since the beginning the quantum uncertainty, as the local position of the photon is considered "a priori" unknown, random and physically meaningless. Even discarding the local coordinates, however, it is possible to estimate the average curvature in any point $x_0$ of $\Delta x$ through the angle $\delta\phi$ between the tangents in two arbitrary points $x'$ and $x''$ around $x_0$; moreover, renouncing to describe the actual time dynamics of the progressive bending process, it is not necessary to specify these points and the actual local value of $\delta\phi$ for $x'' \to x'$. Accordingly, also the concept of distance of the photon from $m_1$ is disregarded and replaced by the uncertainty range $\Delta x_{12}$; being indefinable the local coordinates of the particles, a range of possible distances is in fact consistent with the unknown and random positions of $m_1$ and photon. Actually this is the only approach conceptually possible, regardless of any simplifying aim to bypass the mathematical difficulty of the dynamical problem. Examining the quantities that appear in eq 4,16, the only way to introduce $\delta\phi$ into eq 4,15 is through the parameter $\xi$, whose physical meaning has been not yet specified. In principle it is possible to put $\delta\phi = a + b\xi + b'b\xi^2 \cdots$, being $a$, $b$ and $b'b$ power series development coefficients; the form of the second order coefficient will be explained soon after. Replacing into eq 4,16, the result is $Gm_1/(c^2 \Delta x_{12}) = \delta\phi/b - a/b$ at the first order of approximation owing to eq 4,15. It requires $a = 0$ because if $m_1 = 0$ there is no gravitational effect and then no photon deflection. Hence

$$\delta\phi = \frac{Gm_1 b}{c^2 \Delta x_{12}} \qquad 4,17$$

Clearly $\delta\phi$ depends on the particular choice of the points $x'$ and $x''$ if the size of $\Delta x$ is finite, i.e. $\delta\phi = \delta\phi(\Delta x, \Delta x_{12})$. Consider instead the total deflection angle $\delta\phi_{tot}$ defined for an infinite range $\Delta x_\infty$ with the points $x' \to -\infty$ and $x'' \to \infty$ where the gravitational field vanishes; $\delta\phi_{tot}$ is then by definition greater than any local value $\delta\phi$ calculated between two arbitrary points of $\Delta x$ near $x_o$. If so, $\delta\phi_{tot}$ is also defined by the nearest approach distance $\Delta x_{12}^{min}$ where the gravitational effect of $m_1$ on the photon is reasonably strongest. The asymptotic tangents to the actual photon path define two infinite linear ranges $\Delta x'_{-\infty}$ and $\Delta x''_{\infty}$ crossing in a point $x_{cr}$, whereas the total deflection angle $\delta\phi_{tot}$ of the photon path is given by the relative slope between $\Delta x'_{-\infty}$ and $\Delta x''_{\infty}$. Further information comes from the quantum uncertainty: wherever the points $x'$ and $x''$ might be, the travel directions of the photon cannot be regarded separately depending on which point it moves from; whatever size and local curvature of $\Delta x$ might be, the paths from $x'$ to $x''$ or from $x''$ to $x'$ are indistinguishable. This means that $\Delta P_2$ within $\Delta x$ is given by $h/\lambda_2 - (-h/\lambda_2) = 2h/\lambda_2$; therefore $\Delta P_2 = (2/c)\hbar\omega_2$ yields $(2/c)\omega_2 \Delta x = n_2$, being $n_2$ the number of states allowed for any frequency $\omega_2$ of the quantum system photon/$m_1$. Hence for any $\omega_2$ and $n_2$ one expects that $\delta\phi$ is a function of $c/2$, being the factor ½ the fingerprint of the quantum uncertainties $\Delta P_2$ and $\Delta \dot{P}_2$ within $\Delta x$; it means that $\delta\phi = \delta\phi(\Delta x_{12}, \Delta\zeta/(c/2)^2)$. This result agrees with $\Delta \dot{P}_2 = -(2/c)\hbar\omega_2 \Delta\dot{x}/\Delta x = -\Delta P_2 \Delta\dot{x}/\Delta x$, which reads also $\Delta \dot{P}_2 = -n_2 \hbar \Delta\dot{x}/\Delta x^2$; as expected, the bending effect of the photon path is due to the attractive force that entails $\Delta\dot{x}$ and vanishes for $\Delta x \to \pm\infty$, where the gravitational field tends to zero with $\Delta x^{-2}$ law. This confirms the sign in eq 4,15. For $x'$ and $x''$ ranging from $-\infty$ to $\infty$, i.e. for $\Delta x \to \infty$, the limit of maximum deflection $\delta\phi \to \delta\phi_{tot}$ also corresponds to $\Delta x_{12} \to \Delta x_{12}^{min}$; then $\delta\phi_{tot} = Gm_1 b/(c^2 \Delta x_{12}^{min})$. In this way $\delta\phi_{tot}$ is defined uniquely by asymptotic paths of the photon and minimum distance of approach, regardless of any detail about the real curvature at any point of $\Delta x$ and actual photon position under gravitational potential progressively increasing; $\delta\phi_{tot}$ depends only on the boundary conditions of the system at infinity and is clearly a constant parameter of the present dynamical problem. Let us plug these considerations into eq 4,17: since the dimensionless constant $b$ must be equal to 4 in order that $\delta\phi_{tot}$ depends on $c/2$, the result is

$$\delta\phi_{tot} = \frac{Gm_1}{\Delta x_{12}^{min}}\left(\frac{2}{c}\right)^2 \qquad 4,18$$

Let us emphasize once again that this formula is expressed as a function of the reciprocal range $\Delta x_{12}^{min}$; of course it means that actually $\delta\phi_{tot}$ is inversely proportional to the minimum approach distance $x_{12}^{min}$ conceptually unknown and thus, whatever it might be, appearing here through its uncertainty range. In other words, eq 4,18 is the quantum expression of the well known result $-4Gm_1/(x_{12}^{min}c^2)$ found in general relativity. Analogous considerations hold for the results of the next subsection. Before closing the present problem, let us remark an interesting consequence of the series development of $\delta\phi$. Instead of truncating the series at the first order, exploit now the expression $\delta\phi = a + b\xi + bb'\xi^2$ including also the second order term with coefficient expressed for convenience as a function of $b$ through the arbitrary factor $b'$. Eqs 4,15 and 4,16 yield then

$$\frac{Gm_1}{c^2 \Delta x_{12}^{min}} = \frac{\sqrt{1 - 4b'(a - \delta\phi)/b} - 1}{2b'}$$

As this equation reduces to eq 4,17 for $b' = 0$ and $a = 0$, assume again without loss of generality $b = 4$; so $\delta\phi$ still depends at the first order of approximation on $c/2$, as found; eq 4,17 reads now

$$\rho = \frac{\sqrt{1+b'\delta\phi}-1}{b'} \qquad \rho = \frac{r_{Schw}}{\Delta x_{12}^{min}} \qquad r_{Schw} = \frac{2Gm_1}{c^2} \qquad \frac{\sqrt{1+b'\delta\phi}-1}{2} < 1$$

The last inequality expresses the condition $b'\xi < 1$ for the series to converge; this is enough for the present purpose regardless of how rapidly the series converges. As the reasoning below is trivially extended to any number of higher order terms, $b'\xi \ll 1$ is not required. The first equation relates $\delta\phi$ to the minimum distance of approach $\rho^{-1}$ of light beam to $m_1$ expressed in units $r_{Schw}$, whose physical meaning will be clear soon and further concerned in subsection 4.6. The fact that $\delta\phi \to 0$ for $\rho \to 0$ regardless of the specific value of $b'$ explains why eq 4,17, although inferred at the first order approximation of $\delta\phi$ only, describes well small light beam deflections. The following discussion concerns instead the significant case where $\delta\phi$ takes large values that require at least three terms of series development; if so the values of $\rho$ consistent with the convergence condition depend on the higher order coefficient, e.g. $\delta\phi = \pi$ yields $\Delta x_{12}^{min} = r_{Schw}$ for $b' = \pi - 2$. The condition of convergence yields

$$-\delta\phi^{-1} < b' < 8\delta\phi^{-1} \qquad \delta\phi > \rho_\pi > \delta\phi/4 \qquad r_{Schw}\delta\phi^{-1} < \Delta x_{12}^{min} < 4r_{Schw}\delta\phi^{-1}$$

With $\delta\phi = \pi n_t$ and $n_t$ arbitrary integer, the photon makes $n_t$ half circles around $m_1$; the reason why $\delta\phi$ has not been set directly equal to $2\pi n_t$ becomes clear putting first in these inequalities $\delta\phi = \pi$. So, (i) the photon coming for instance from infinity loops one half circle around $m_1$ at distance $\rho_\pi^{-1}$ and then points back to infinity along a parallel path $2\rho_\pi^{-1}$ apart. This chance is indistinguishable from the other one identically possible, i.e. (ii) the photon coming from minus infinity after one half circle points back to minus infinity. Actually, however, these paths are consequent each other: being still at distance $\rho_\pi^{-1}$ from $m_1$ after the first half turn, the photon of the step (i) is in the same situation of the step (ii); so the photon is again deflected by $\delta\phi = \pi$ regardless of whether it comes from minus infinity or not. Then it returns once more in the situation (i) after this second half turn, and so on. It means that the photon transiting at distance $\rho_\pi^{-1}$ from $m_1$ in fact "orbits" around $m_1$, i.e. the light is trapped by the gravitational field of this latter. Thus $\rho_\pi^{-1}$, whatever its value might be, represents the characteristic "black body distance" of $m_1$ in $r_{Schw}$ units. Set now $\delta\phi = 2\pi$: the photon transiting at distance $\rho_{2\pi}^{-1}$ from $m_1$ make one full orbit around $m_1$. After one complete turn, however, the photon is in the same situation as it previously was when coming from infinity; this therefore compels it making a second turn, and so on. Hence also $\rho_{2\pi}^{-1}$, whatever its value might be, represents another characteristic "black body distance" of $m_1$. i.e. the light is again trapped by the gravitational field of this latter. Of course the same holds for any $\delta\phi = n_t\pi$, so the conclusion does not change whatever $n_t$ might be. Hence the various chances $(\pi n_t)^{-1} r_{Schw} < \Delta x_{12}^{min} < 4(\pi n_t)^{-1} r_{Schw}$ with $n_t = 1, 2, ...$ agree with convergent series development of $\delta\phi$ and light beam trapping by the gravity field of $m_1$ if the transit distance included in $\Delta x_{12}^{min}$ is of the order of $0 < x_{12}^{min} < 4\pi^{-1} r_{Schw}$. Introducing explicitly since the beginning larger values of $n_t$ merely simulates photon transits at closer and closer distances from $m_1$, to which correspond stable circular orbits closer to $m_1$. For $n_t \to \infty$ both boundaries of $\Delta x_{12}^{min}$ tend asymptotically to zero likewise the distance range $\Delta x_{12}^{min}$ itself. As expected, the actual gravitational distances enabling the peculiar black body behaviour of matter are expressed through ranges of values in the same way as in eq 4,17; moreover the size of $\Delta x_{12}^{min}$ is not uniquely specified by preset boundary values, it is only known that its order of magnitude must be $r_{Schw}$ as well. Exploit now once again the basic idea of the present theoretical frame: to any space-time range can be related a corresponding momentum range. This requirement matches the reasonable assumption that in effect the initial momentum $h/\lambda$ of the free photon at infinity changes after gravitational interaction with $m_1$. If for instance the photon is confined in a range of distances of the order of $2\rho_\pi^{-1}$ from $m_1$ due to its diametric displacement, the uncertainty equation reads $2\Delta x_{12}^{min} h(\lambda^{-1} - \lambda'^{-1}) = n'\hbar$: whatever the actual value of $\Delta x_{12}^{min}$ might be, $h/\lambda'$ is by definition the photon momentum in its circular orbit around $m_1$. The chance $\lambda' < \lambda$ is unreasonable: the photon cannot gain energy falling on $m_1$ otherwise it should also accelerate beyond its own limit speed $c$, whereas instead the key element of the present discussion is that the value of $c$ remains always constant. Is instead plausible $\lambda' > \lambda$, i.e. the photon loses energy $hc(\lambda^{-1} - \lambda'^{-1})$ when it starts orbiting around $m_1$. With the position $\pi' = 2\pi(\lambda' - \lambda)/\lambda$ the uncertainty equation reads $2\pi'\Delta x_{12}^{min} = n'\lambda'$. This result suggests a well known idea: when a particle with momentum $h/\lambda'$ moves along a circular orbit of radius $\Delta x_{12}^{min}$ the circumference must be equal to an integer number of wavelengths in order to allow a stable orbit motion; yet, in general, here $\pi' \neq \pi$. More precisely, if $\lambda < \lambda' \leq (3/2)\lambda$ then $\pi' \leq \pi$, whereas instead $\pi' \geq \pi$ for $\lambda' \geq (3/2)\lambda$; the Euclidean value $\pi' = \pi$ is found by chance in the particular case $\lambda' = (3/2)\lambda$ only. The circumference/diameter ratio equal to non-Euclidean values $\pi'$ is in general fingerprint of space-time curvature around $m_1$; this confirms the considerations carried out about the previous interpretation of light beam bending effect eq 4,17.

*4.6 The Kepler problem.*

The previous cases have been discussed without specifying in detail the analytical form of the field momentum $p_o$ introduced in eqs 4,3; it has been merely emphasized that if $p_o = h/\lambda_o$ is regarded approximately as a constant $p_o^0$, then $F$ of eq 4,5 takes the form of Newton's law, eq 4,6. Yet there is neither reason nor necessity to assume $p_o$ constant; rather, any consideration about how it could possibly change as a function of a suitable parameter has been so far ignored merely because eqs 4,5 to 4,18 did not require closer insight about its actual form. To formulate correctly the present problem, however, it is necessary to write explicitly $p_o = p_o^0 + p_o'$, where $p_o' = p_o'(\Delta x)$ denotes a proper correction to $p_o^0$ function of $\Delta x$; for simplicity of notation, $\Delta x$ without subscript indicates the distance uncertainty range between the particles. The power series development of $p_o$ must have then the form $p_o^0 + \gamma'/\Delta x + \cdots$, with $\gamma'$ constant, having neglected for simplicity the higher order terms $\Delta x^{-k}$: indeed the correction is expected to vanish for weak fields, i.e. for $\Delta x \to \infty$, where the classical Newtonian mechanics is quite accurate. The dependence of $p_o$ upon $\Delta x$ is sensible. The second eq 4,3a yields $\Delta \ddot{x}_i = (\Delta \dot{x}_i / \Delta x)^2 \gamma' / p_o$ at the first order, whereas instead $\Delta \ddot{x}_i = 0$ for $\gamma' = 0$, i.e. $\Delta \dot{x}_i$ should be either zero or constant values. Since nothing in nature changes instantaneously seems more reasonable the idea of $\Delta \ddot{x}_i \neq 0$ enabling a gradual increase of $\Delta \dot{x}_i$ up to values consistent with eqs 4,2, which therefore read also $F_i = -(n_i \hbar p_o / \gamma')(\Delta \ddot{x}_i / \Delta \dot{x}_i)$. So at infinite distance, $\Delta x \to \infty$, the force vanishes because $\Delta \ddot{x}_i \to 0$; moreover the $m_i \Delta x^{-2}$ dependence of $F_i$ in eqs 4,4 is replaced by the deformation rates $\Delta \dot{x}_i$ and $\Delta \ddot{x}_i$ and by the weak dependence of $p_o$ upon $\Delta x$; the fact that mass does not longer appears explicitly in the last expression of $F_i$ confirms the idea that indeed the gravity force is nothing else but the experimental appearance of the deformation rate of the time-space uncertainty ranges. There is no reason to exclude that even the first order correction $p_o' \ll p_o^0$ is conceptually important to account for small gravitational effects unexplained by the simple Newton law. This is the classical case of the perihelion precession of orbiting bodies. Let $E < 0$ be the total energy of a system of two bodies of masses $m_1$ and $m_2$ subjected to gravitational interaction. The elementary classical mechanics shows that perihelion precession is allowed to occur if the potential energy $U$ of a reduced mass $m$ orbiting in the gravitational central field of $M$ in agreement with eq 4,8 has the form $U = U_{eff} + U'$ [17]: here $U_{eff} = -GMm/x + M^2/2mx^2$ is the effective potential and $M^2$ the squared angular momentum of the particle at distance $x = x(t)$ around the gravitational centre; $U' = U'(x)$ is a proper correction to the effective potential. The particular case $U' = \beta/x^2$, with $\beta$ arbitrary coefficient, is explicitly reported in several textbooks; the solution of the Kepler problem with

$$U = -\frac{\zeta}{x} + \frac{M^2 + 2\beta m}{2mx^2} \qquad \zeta = m_1 m_2 G = MmG \qquad 4,19$$

is summarized then by the following relevant equations [17]:

$$U_{min}^{cl} = -\frac{\zeta^2 m}{2M^2} \qquad p = \frac{M^2}{m\zeta} = (1-e^2)a = b\sqrt{1-e^2} \qquad b = \frac{\sqrt{M^2}}{\sqrt{2m|E|}} \qquad \zeta = 2a|E| \qquad 4,20$$

$$T = 2\pi a \sqrt{\frac{am}{\zeta}} \qquad \delta\varphi = -\frac{2\pi\beta}{\zeta p}$$

where $\delta\varphi$ is the precession angle, $b$ and $a$ are the minor and major semi-axes of the ellipse having eccentricity $e$ and $T$ the revolution time. Here $U'$ is arbitrarily introduced as mere additive term to $U_{eff}$, so that $\beta$ appears in the solution only as separate additive term as well. Yet the plain case where $G$ is the mere Newtonian constant inferred from the experiment does not justify terms additional to $U_{eff}$; relativistic concepts must be introduced since the beginning into the problem to explain why $\delta\varphi$ is actually observed. To tackle this point let us carry out the quantum approach of section 2; rewrite first to this aim eq 4,5 considering the most general form expected for $p_o$

$$F = -\frac{\hbar c^2}{m^u p_o^0 \sum_{k=0}^{\infty} a_k (\gamma'/\Delta x)^k} \frac{m_1 m_2}{\Delta x^2} \qquad 4,21$$

in agreement with eq 4,4; $a_k$ are proper coefficients of series development with $a_0 = 1$ in order that $p_o \simeq p_o^0 + p_o'(\Delta x)$ at the first order. Then the condition $p_o'(\Delta x) \ll p_o^0$ yields $p_o^0 + p_o' \simeq p_o^0(1 + \gamma/\Delta x)$, with $\gamma = a_1 \gamma'$, for uncertainty ranges much larger than $\gamma$. Hence, owing to eqs 4,8,

$$F = -\left(1 - \frac{\gamma}{\Delta x}\right)\frac{\zeta}{\Delta x^2} \qquad G = \frac{\hbar c^2}{m^u p_o^0} \qquad \frac{\gamma}{\Delta x} \ll 1$$

The third equation recalls that the constant value $p_o^0$ fits the experimental gravitation constant, i.e. $G$ is by definition the zero order constant term such that $F$ tends to $F_N$ of eq 4,6 for $\gamma = 0$. Significant consequences are inferred introducing $\gamma/\Delta x$ to approximate the most general form possible for $p_o$, since the present expression of $F$ justifies in principle the additional potential term $U'$ necessary to describe the precession. The potential energy of the reduced mass $m$ in the gravitational field of $M$ reads $-\zeta/\Delta x + \zeta\gamma/(2\Delta x^2)$, to which must be summed the angular orbital term $\mathrm{M}^2/(2m\Delta x^2)$ to obtain $U$. Let us follow now the same quantum formalism already shown for the hydrogenlike atom, since the distance $x = x(t)$ between the orbiting body and the gravitational centre has been already replaced by the corresponding $\Delta x = \Delta x(t)$ of eq 3,1; we are describing therefore a quantum system of two particles orbiting at random distance included within the uncertainty range $\Delta x$ by effect of their gravitational interaction only. Also in this case it is essential that $\Delta x$ be space-time function: if $x(t)$ changes with $t$, then $\Delta x$ must depend on $t$ as well in order to include any possible change of $x$ at various times. This agrees with the fact that just the time dependence of $\Delta x$ introduces the deformation rate $\Delta \dot{x}$ of the uncertainty ranges defining the gravitational force. It is easy to show that the classical eqs 4,20, obtained integrating $\dot{x}(t)$, are inferred as limit case solving the present quantum problem exactly as shown in the examples of section 2: the quantum approach, non-relativistic, takes advantage of the non-Newtonian term $-\gamma\zeta\Delta x^{-3}$ in the interaction of two particles subjected to the gravitational force only. The kinetic and total potential energies as a function of $m$ read

$$E_{kin} = \frac{\Delta P_x^2}{2m} \qquad U = -\frac{\zeta}{\Delta x} + \frac{\mathrm{M}^2 + \gamma m\zeta}{2m\Delta x^2} \qquad \delta\varphi = -\frac{\pi\gamma}{\mathrm{p}} \qquad 4{,}22$$

The last equation is found comparing $U$ of eqs 4,22 and 4,19; $\gamma\zeta/2$ corresponds to the coefficient $\beta$ defining $\delta\varphi$ in the last eq 4,20. The perihelion precession is then explained simply through the most general form possible for $p_o$, without any specific hypothesis and in agreement with the Newtonian limit 4,6; for this reason eq 4,21 has relevant theoretical interest, even considering the first order correction to Newton's law only. Let $E = E_{kin} + U < 0$ be the total energy; then $E_{kin} > 0$ and $U < 0$ define the positive term $1 - E/U > 0$. Replacing thus $\Delta x$ with $n\hbar/\Delta P_x$ in eqs 4,22 and minimizing $E = \Delta P_x^2/2m - \zeta\Delta P_x/(n\hbar) + (\mathrm{M}^2 + \gamma m\zeta)\Delta P_x^2/(2mn^2\hbar^2)$ with respect to $\Delta P_x$, one finds

$$\Delta P_{x\min} = \frac{m\zeta}{(n\hbar)^2 + \mathrm{M}^2 + \gamma m\zeta} n\hbar \qquad \Delta x_{\min} = \frac{(n\hbar)^2 + \mathrm{M}^2 + \gamma m\zeta}{m\zeta} \qquad 4{,}23\mathrm{a}$$

and then also

$$E_{\min} = -\frac{1}{2}\frac{\zeta^2 m}{(n\hbar)^2 + \mathrm{M}^2 + \gamma m\zeta} \qquad U_{\min} = -\frac{m\zeta^2}{2}\frac{2(n\hbar)^2 + \mathrm{M}^2 + \gamma m\zeta}{\left((n\hbar)^2 + \mathrm{M}^2 + \gamma m\zeta\right)^2} \qquad 4{,}23\mathrm{b}$$

It appears that

$$|E_{\min}| = \zeta/2\Delta x_{\min} \qquad 4{,}24$$

Moreover, introducing also now a suitable parameter $p$, eqs 4,23 define the following quantities

$$p = M^2/m\zeta \qquad 1 - p/\Delta x_{min} = 1 + 2E_{min}M^2/(m\zeta^2) \qquad p/\sqrt{1-e^2} = \sqrt{M^2/(2m|E_{min}|)} \qquad 4,25$$

The revolution period is calculated easily from eqs 4,23 to 4,25. In lack of local information about the coordinates of the orbit, let us introduce the average momentum $P_{av}$ of the orbiting reduced mass and define the pertinent uncertainty range $\Delta P_{av}$ including it as $\omega = \Delta P_{av}/(m\Delta x_{min})$; of course this equation is obtained from $\Delta P_{av} = P^{\S\S} - P^{\S}$, with $P^{\S\S} = m\omega x^{\S\S}$ and $P^{\S} = m\omega x^{\S}$ putting the range $x^{\S\S} - x^{\S}$ equal to $\Delta x_{min}$. Since $\Delta P_{av}$ must be consistent with the energy $E_{min}$, it must be true that $\Delta P_{av}^2/2m = |E_{min}|$; hence $\omega = \Delta x_{min}^{-1}\sqrt{2|E_{min}|/m}$. The angular frequency of orbital motion is then $\omega = 2\pi/\delta t$, being $\delta t$ the time range of one revolution. Eqs 4,24 yield $\omega = \Delta x_{min}^{-1}\sqrt{\zeta/(m\Delta x_{min})}$, i.e.

$$\delta t = 2\pi \Delta x_{min} \sqrt{\frac{\Delta x_{min} m}{\zeta}} \qquad 4,26$$

As expected, the quantities just found are expressed through uncertainty ranges within which are delocalized the quantum particles, rather than through orbit coordinates applicable to macroscopic bodies. Yet, it is interesting to compare eqs 4,24 to 4,26 with eqs 4,20; we note that

$$\Delta x_{min} \Rightarrow a \qquad \sqrt{\frac{M^2}{2m|E_{min}|}} \Rightarrow b \qquad 1 + \frac{2E_{min}M^2}{m\zeta^2} \Rightarrow e^2 \qquad \delta t \Rightarrow T \qquad E_{min} \Rightarrow E \qquad p \Rightarrow a(1-e^2)$$

The correspondence between $\Delta x_{min}$ and $a$ is evidenced comparing eqs 4,24 and 4,26 with the fourth and fifth eqs 4,20 and confirmed by the correct calculation of revolution time $\delta t$, here appearing as characteristic time range as well; also $b$ is related to the range of arbitrary values allowed to $M^2 = n_{or}(n_{or}+1)\hbar^2$, where $n_{or}$ is the quantum number of orbital angular momentum. Analogous considerations hold for the other quantities; indeed also now $(a^2 - b^2)/a^2 = e^2$. The quantum approach evidences that the uncertainty ranges have the features of the classical orbital parameters exactly defined for macroscopic massive bodies, yet without contradicting the postulated uncertainty of the quantum approach; being $n$ arbitrary, it actually means that the plane of the orbit trajectory and the local orbit distances between $m$ and $M$ remain in fact unknown. However these results are not peculiar of the quantum world only, since they do not require $\hbar \to 0$ and $\gamma \to 0$ to infer the aforementioned classical results; it follows that the gravitational behaviour of a particle, although expressed through the uncertainty of its orbit coordinates, is in principle analogous to that of a planet, apart from mass, time and length scale factors. The reduced mass $m$ moving around $m_1 + m_2$ follows an elliptic orbit. Note that the orbit parameters $a$, $b$, $T$ and related $E_{min}$, given in eqs 4,20 and summarized by the positions above, entail an expression of minimum potential energy $U_{min}$ slightly different from the classical $U_{min}^{cl}$

$$\frac{U_{min}}{U_{min}^{cl}} = \frac{\left(2(n\hbar)^2 + M^2 + \gamma m\zeta\right)M^2}{\left((n\hbar)^2 + M^2 + \gamma m\zeta\right)^2}$$

The result $U_{min} > U_{min}^{cl}$ is not surprising: $U_{min}^{cl}$ is calculated simply minimizing $U$ of eq 4,19, thus it is mere consequence of its own analytical form; here instead $U_{min}$ is by definition calculated in connection with $E_{min}$, i.e. minimizing the global energy of the system that includes also the orbital kinetic energy. The explicit expression $e^2 = \left((n\hbar)^2 + \gamma m\zeta\right)/\left((n\hbar)^2 + M^2 + \gamma m\zeta\right)$ inferred from the second eq 4,25 shows that $e^2 \to 0$ for $M^2 \gg (n\hbar)^2$, whereas $e^2 \to 1$ for $M^2 \ll (n\hbar)^2$; hence the orbit eccentricity is a quantum effect defined by the angular momentum $M^2$ and by $(n\hbar)^2$, thus controlled by the relative values of quantum numbers $n_{or}$ and $n$ in principle arbitrary. For classical bodies like planets $M^2$ is overwhelmingly much larger than $\hbar^2$, so that $M^2 \ll (n\hbar)^2$ requires $n \to \infty$ i.e. a number of states $n \gg n_{or} \gg 1$ allowed to the orbiting system; it in turn also means that $n_{or}(n_{or}+1) \to n_{or}^2$, i.e. $M^2 \to M_z^2$; the classical Newtonian orbit lies then on the plane that corresponds to $M_z^2$. The classical $U_{min}^{cl}$ of eq 4,20 coincides with $U_{min}$ putting $\gamma$ equal to zero, as it is obvious, and when $M^2 \gg (n\hbar)^2$, i.e. for circular orbits that entail in effect the minimum value of total energy; in this particular case, i.e. for $\Delta x_{min} \to M^2/(m\zeta)$, eqs 4,24 to 4,26 read

$$\Delta r^3 \omega_r^2 = \frac{\zeta}{m} = (m_1+m_2)G \qquad \Delta r = \frac{M^2}{m\zeta} \qquad e^2 = 0 \qquad 4,27$$

$$E_{min}^r = -\frac{\zeta^2 m}{2M^2} = -\frac{\zeta}{2\Delta r} = U_{min}^r \qquad \omega_r = \frac{1}{\Delta r}\sqrt{\frac{\zeta}{m\Delta r}}$$

Also the form of these equations, well known, is a further check of the present approach and will be profitably used in the next subsection. In principle eqs 4,27 require simply proper values of $n$ and $n_{or}$, which are however so large for macroscopic bodies that in practice $\gamma m\zeta$ is expected to be completely negligible. Then, eqs 4,20, 4,22 and 4,25 yield

$$\delta\varphi = -\frac{\pi\gamma\zeta m}{M^2} = \frac{\pi\gamma mMG}{2\Delta x_{min} E_{min}\, \mathrm{p}} \qquad 4,28$$

For $\gamma \neq 0$ the major axis of the ellipse rotates by an angle $\delta\varphi$ after one revolution period; then the mass $m$ not only moves along its orbit but also rotates because of the angular precession within concentric circles $2e\Delta x_{min}$ apart. From the dimensional point of view it is possible to write

$$\gamma = 2q'\Delta x_{min} \qquad m = q''E_{min}/c^2 \qquad 4,29$$

being $q'$ and $q''$ proper dimensionless coefficients: the former measures the parameter $\gamma$ in $2\Delta x_{min}$ units, the latter relates $m$ to the constant energy $E_{min}$ of orbital motion in the field of $M$ and is therefore negative by definition. Being both masses arbitrary in principle, eqs 4,29 are only a formal way to rewrite $\delta\varphi$ more compactly as

$$\delta\varphi = q\pi\frac{MG}{\mathrm{p}c^2} \qquad q = q'q'' \qquad 4,30$$

Although $q''$ is defined by the second eq 4,29, $q$ remains unknown; owing to eq 4,24 one finds indeed $q'' = -2mc^2\Delta x_{min}/\zeta$, i.e. $q' = -q\zeta/(2mc^2\Delta x_{min})$ and $\gamma = -q\zeta/(mc^2) = -qMG/c^2$. The further reasoning necessary to define $q$ exploits the fact that $\gamma$ is related to the precession angular momentum of $m$. From a quantum point of view, this situation is described introducing an angular momentum $M_{prec}^2$ additional to $M^2$ and specifically due to the precession effect, i.e. such that $\delta\varphi \propto M_{prec}^2/M^2$, through $q = M_{prec}^2/\hbar^2 = l_{prec}(l_{prec}+1)$; if so, $q' = -l_{prec}(l_{prec}+1)\zeta/(2mc^2\Delta x_{min})$ and $q > 0$. Then $\gamma$ and $\delta\varphi$ read

$$\gamma = -\frac{M_{prec}^2}{(\hbar c)^2}MG \qquad \delta\varphi = l_{prec}(l_{prec}+1)\pi\frac{MG}{\mathrm{p}c^2} \qquad l_{prec} = 0,1,2,\cdots$$

Since $\gamma < 0$, eq 4,23a can be rewritten as $\Delta x_{min} = ((n\hbar)^2+M^2)/m\zeta - |\gamma|$, i.e. the right hand side has the form $x^{\S\S} - x^{\S}$ expected for any distance uncertainty range; in other words, $|\gamma|$ is actually the lower boundary of $\Delta x_{min}$. As concerns the value of $l_{prec}$, the trivial case $l_{prec} = 0$ has been already considered: this value is acceptable as Newtonian approximation only, being however irrelevant and unphysical in the present context. Considering therefore $l_{prec} > 0$ only, one finds

$$\gamma = -r_{Schw}^M \frac{l_{prec}(l_{prec}+1)}{2} \qquad r_{Schw}^M = \frac{2MG}{c^2} \qquad 4,31$$

One expects in general the condition $\Delta x_{min} > r_{Schw}^M$, because $r_{Schw}^M$ defines the classical boundary where the escaping velocity from $M$ is $c$: necessary condition to enable the orbiting system of massive particles is therefore that $\Delta x_{min}$ including all the possible distances between $M$ and $m$ be larger than $r_{Schw}^M = r_{Schw}^{m_1} + r_{Schw}^{m_2}$. On the one hand $|\gamma|$ is the lower coordinate of the range $\Delta x_{min}$ allowed to both particles, then by definition accessible to these latter; so it follows that necessarily $|\gamma| > r_{Schw}^M$. Consider now eq 4,31: if $l_{prec} = 1$ then $|\gamma|$ would be equal to $r_{Schw}^M$, which is not acceptable; hence it must be true that $l_{prec} > 1$. On the other hand, increasing $l_{prec}$ means that both $E_{min}$ and $U_{min}$ become less negative; then the condition of minimum energy suggests

$$\delta\varphi = 6\pi \frac{MG}{\mathrm{p} c^2} \qquad l_{prec} = 2 \qquad \gamma = -3r_{Schw}^M \qquad 4,32$$

as it is well known. Here the coefficient 6 is the fingerprint of the quantum angular momentum related to the orbital precession effect. Note that from the point of view of the orbiting particle the central mass $M$ appears rotating at rate $\Omega = \delta\varphi/\delta t$, which therefore also defines an orbital momentum $\mathbf{M}_M$ related to the precession angular velocity $\mathbf{\Omega} = \Omega \mathbf{u}$ around the direction of an arbitrary unit vector $\mathbf{u}$. The Poisson relationship that links $\mathbf{M}_M$ in the perihelion precession reference system (where the central particle does not rotate) and in the reference system where the central particle rotates with angular rate $\mathbf{\Omega}$ equal to the precession rate yields then the known result

$$\frac{d\mathbf{M}_M}{dt} = \mathbf{\Omega} \times \mathbf{M}_M \qquad 4,33$$

Hence the perihelion precession of an orbiting particle entails also the existence of a drift force in the gravitational field of a rotating body.

*4.7 Gravitational waves.*

Let us return now to the field energy loss related to the emission of gravitational waves from an orbiting system. The equations found in subsection *4.1* were inferred considering explicitly that the gravitational waves remove energy through pulses $\hbar\omega_o$ propagating at rate $c > \Delta\dot{x}_i$, i.e. faster than the deformation rate originating itself the force, with $\omega_o$ related to the orbital period. The result was

$$\varepsilon_o = \eta_i \frac{\Delta\dot{x}_i}{c} = \frac{\eta_i m_i c}{n_i p_o} \qquad p_o = \frac{h}{\lambda_o} = \frac{\hbar\omega_o}{c} \qquad \dot{\eta}_i = -2\pi G m_i^2 \omega_o^6 \Delta x^4 / c^5 \qquad 4,34$$

A better calculation is now carried out starting directly from the results of the Kepler problem. The average loss of energy radiated after one revolution of the mass $m$, during which $E_{min}$ changes by $\delta E_{min}$ and $\Delta x_{min}$ by $\delta\Delta x_{min}$, reads at the first order $\delta E_{min} = (\partial E_{min}/\partial\Delta x_{min})\delta\Delta x_{min}$; eq 4,24 yields $\delta E_{min}/\delta\Delta x_{min} = -E_{min}/\Delta x_{min}$ with good approximation for a small change of $\delta\Delta x_{min}$ during $\delta t = 2\pi/\omega$. It is easy to verify that this equation is also fulfilled by $\delta E_{min} = -wE_{min}^2\omega$ and $\delta\Delta x_{min} = -w\zeta\omega/2$, where $w$ is a proportionality constant: through the factor $\omega$ the former equation calculates the energy radiated, showing reasonably that $-\delta E_{min}$ is proportional at any time to the current value of $E_{min}^2$, i.e. the loss is expressed by a negative quantity. The same holds also for $\delta\Delta x_{min}$, since the loss requires also contraction of the minimum approach distance $\Delta x_{min}$ between the particles. Hence follow the positions $-\delta E_{min}/\delta t = wE_{min}^2\omega^2/2\pi$ and $\delta\Delta x_{min}/\delta t = -w\zeta\omega^2/4\pi$, where $w$ must be proportional to $G$: in absence of gravitational field, i.e. for $G = 0$, one expects that $\delta E_{min}$ and $\delta \Delta x_{min}$ vanish. By dimensional reasons $w$ is appropriately expressed as $w = w'G/c^5$ being $w'$ a further dimensionless constant. It is easy at this point to show again the results in the particular case of circular orbit; replacing $E_{min}$, $\omega$, $\Delta x_{min}$ with $E_{min}^r$, $\omega_r$, $\Delta r$ of eqs 4,27 yields

$$-\frac{\delta E_{min}^r}{\delta t} = w'\frac{Gm^2 \Delta r^4 \omega_r^6}{8\pi c^5} \qquad -\frac{\delta \Delta r}{\delta t} = w'\frac{G\zeta \omega_r^2}{4\pi c^5} = w'\frac{G^3 m_1 m_2(m_1+m_2)}{4\pi \Delta r^3 c^5} \qquad 4,35$$

The first eq 4,35 compares well with eq 4,34 previously inferred in subsection *4.1*. The connection between these equations is clear: the former is obtained starting directly from the energy $E_{min}$ of the orbiting system, the latter was obtained instead from the definition of field energy loss and then introducing into $\dot{\eta}_i$ of eqs 4,1b the specific energy $m_i \omega_o^2 \Delta x_i^2 / 2$ of the mass $m_i$ of the $i$-th particle in circular orbit. It explains why eq 4,35 more correctly replaces $m_i$ with the reduced mass $m$ and confirms that the field frequency $\omega_o$ really corresponds to the characteristic orbiting frequency $\omega_r$ of the specific quantum system concerned in particular. It also confirms that, as expected, $\dot{\eta}_i$ inferred from eq 4,1b is just $-\delta E_{min}^r / \delta t$ of the orbiting system, in agreement with the idea that the energy lost by the orbiting system is effectively released via pulses during each revolution period of $m$. Also now appears a coefficient $w'$; the comparison between eqs 4,34 and 4,35 is legitimate to infer the numerical value of $w'$. Putting $w' = (4\pi)^2$ eqs 4,35 read

$$-\frac{\delta E_{min}^r}{\delta t} = 2\pi \frac{Gm^2 \Delta r^4 \omega_r^6}{c^5} \qquad -\frac{\delta \Delta r}{\delta t} = 4\pi \frac{G\zeta \omega_r^2}{c^5} = 4\pi \frac{G^3 m_1 m_2(m_1+m_2)}{\Delta r^3 c^5} \qquad 4,36$$

The numerical agreement between $2\pi$ and the coefficient 32/5 of general relativity has been already emphasized in subsection *4.1*; it is now significant the same agreement between $4\pi$ and the known factor 64/5 of the relativistic formula of orbit radius contraction related to the energy loss [15]. Eqs 4,36 show clearly that the present point of view does not concern the actual dynamics of radius contraction described point by point along the orbit path of $m$; instead of describing an intuitive spiral motion progressively approaching the gravity centre, eqs 4,35 only show that after a time range $\delta t$ the current orbit radius $\Delta r$ is contracted to $\Delta r - \delta \Delta r$ while a pulse of gravitational wave propagates at speed $c$ with frequency corresponding just to $1/\delta t$. This is why in effect the orbiting frequency $\omega_r$ appears in eqs 4,36. This way to regard the emission of a gravitational wave pulse prevents knowing where or when exactly takes place the emission, which in fact cannot be regarded as gradual process progressively occurring along specified points of the gravitational field; it is only possible to calculate the emission frequency. Eqs 4,36 read also

$$\frac{\delta E_{min}^c}{\delta t} = -W_P \frac{\pi}{16}\frac{f^2}{\rho^5} \qquad \frac{\delta \Delta r}{\delta t} = -\frac{\pi}{2} c \frac{f}{\rho^3} \qquad \rho = \frac{\Delta r}{r_{Schw}^M} \qquad f = \frac{m}{M} \qquad W_P = \frac{c^5}{G} \qquad 4,37$$

The average power radiated depends on the mass ratio and not on the masses themselves; hence it is equal in principle for planets and quantum particles at proper orbital distances expressed in $r_{Schw}^M$ units. The value of $W_P$ is extremely large, about $3.6 \cdot 10^{52}$ watts; if $\Delta r$ is of the order of planetary distances the factor $\rho^{-5}$ makes irrelevant $\delta E_{min}^r / \delta t$, and then $\delta \Delta r / \delta t$ as well, even for $f \approx 1$. Considering that $\delta t$ is the time range corresponding to one orbital revolution, the planet orbits are practically stable. However increasingly large powers are to be expected as long as $\Delta r \to r_{Schw}^M$. This is typically the case of quantum particles orbiting in their own gravitational field; since the values of each mass do not appear explicitly in the equation, but only their ratio, the radiation of energy is expected to increase in an orbital system of quantum particles subjected only to gravitational interaction depending on how much their mutual distances approach their own $r_{Schw}^M$ during a reasonably short time range $\delta t$. The results of subsection *4.1* show that the gravity force in a range of distances larger than or equal to $r_{Schw}$ is described by the following properties of the particle generating the field: (i) mass $M$, eq 4,31, (ii) angular momentum $M^2_{Schw}$ (i.e. $M^2_{12}$ calculated with $\Delta x_{12} \to r^M_{Schw}$) and then possible spin, (iii) possible charge. In effect there is no reason to exclude spin and charge of the particles concerned in eqs 4,3 and 4,5, although both have been so far disregarded for practical purposes only, i.e. simply to focus the discussion on the gravity force.

*4.8 The gravity constant.*

The following discussion aims to estimate the numerical value of the constant $G = \hbar c^2 / m^u p^0_o$ of eq 4,6 with the help of eq 4,9. In the former equation $F_N$ is expressed as a function of the ratio $m_i / \Delta x_{12}$, the latter equation has instead a different form because $F_N$ is expressed as a function of the angular momenta of the gravity field and system of masses; for this reason the constant factor $(\hbar c)^3 / \mu'_o M^2_o$ is in effect formally different from $G$. Yet, the fact that $\Delta M^2_{12}$ has dimensions $\hbar^2$ suggests introducing in eq 4,9 the modulus $|\mathbf{\Pi}^u|$ of the unit momentum $\mathbf{\Pi}^u$ into eq 4,9, in order that each mass is again expressed as $mass |\mathbf{\Pi}^u|/\hbar$, i.e. mass per unit length in analogy with eq 4,6; these positions identify the reference system $R^u$ where is calculated $G$. Let us rewrite identically eq 4,9 as follows

$$F = -\chi_G \frac{M^* m^*}{\Delta M^2_{12}/\hbar^2} \qquad \chi_G = \frac{1}{\mu'_o} \frac{(\hbar c)^3}{M^2_o \Pi^{u2}} \qquad M^* = M \frac{|\mathbf{\Pi}^u|}{\hbar} \qquad m^* = m \frac{|\mathbf{\Pi}^u|}{\hbar} \qquad 4,38$$

So $\chi_G$ has again the same dimensions of $G$ whereas, being $m_1$ and $m_2$ arbitrary, $m^*$ and $M^*$ can reproduce in principle any desired value definable experimentally. Yet the physical interest of these equations rests mostly on the field angular momentum $M^2_o$, which is further examined just now exploiting the fact that $M^2_o$ can take selected values only. Eqs 4,38 and 4,5 yield

$$\hbar \omega_o = \mu'_o \frac{\Pi^{u2}}{m^u} J_o(J_o+1) \qquad \mathbf{M}_o = \mathbf{L}_o + \mathbf{S}_o \qquad \frac{M^2_o}{\hbar^2} = J_o(J_o+1)$$

The second equation recalls that, once having defined $\mathbf{M}_o$, one in principle expects its orbital and spin components $\mathbf{L}_o$ and $\mathbf{S}_o$ according eq 3,5. The fact that $M^2_o$ is a property of the gravity field only requires its link with the field momentum $p_o$ introduced in eqs 4,4 together with the characteristic field wavelength $\lambda_o$; so, in lack of other quantities related to $p_o$, one can guess nothing else but $M^2_o \approx \lambda^2_o p^2_o = h^2$ and thus $J_o(J_o+1)$ of the order of $(2\pi)^2$. Some values of $J_o(J_o+1)$ of interest in the present context among those calculated with integer and half-integer trial values of $J_o$ are: ... 24.75, 30, 35.75, 42, 48.75, 56,... The value closest to $(2\pi)^2$ corresponds to $J_o = 6$ and yields $M^2_o / \hbar^2 = 42$, i.e. the gravity field has boson properties; then with $\mu'_o \approx 1$ for the reasons explained at the end of subsection 4,2, the second eq 4,38 reads

$$\chi_G \approx \frac{\hbar c^3}{42 \Pi^{u2}} \qquad 4,39$$

Being $|\mathbf{\Pi}^u| = 1$ Kg m s$^{-1}$ by definition, $\chi_G \approx 6.76 \cdot 10^{-11}$ m$^3$ Kg$^{-1}$ s$^{-2}$ agrees with the experimental value $G_{exper} = 6.67 \cdot 10^{-11}$ m$^3$ Kg$^{-1}$ s$^{-2}$. The result confirms the validity of eq 4,9, the reasoning to obtain eq 4,39 and the position $p^0_o \approx p_o$. So eq 4,6 yields

$$p^0_o \simeq 1.4 \cdot 10^{-7} \text{ Kg m/s} \qquad \lambda^0_o \simeq 4.7 \cdot 10^{-27} \text{ m} \qquad \hbar \omega^0_o = p^0_o c = 42 \text{ J} \qquad 4,40$$

Note that this estimate of $G$ does not exclude its weak dependence on time mentioned in some theories, as it will better appear in the next subsection *4.9* that shows further independent ways to calculate $G$. A short remark concerns the fact that actually $\chi_G = \chi_G(\Delta x)$. Consider eq 4,21 and the subsequent equation of force $F = -(1-\gamma/\Delta x)\zeta/\Delta x^2$, related to the approximate constant $G$ via $\zeta = mMG$ of eq 4,19. To account for the dependence of $\chi_G$ on $\Delta x$ in eq 4,5, we have explicitly introduced in the Kepler problem the first order correction term $\gamma/\Delta x$ to the Newton law. In this respect it is possible to infer an interesting result susceptible of experimental comparison. Rewrite $F$ as a function of the average value of $G$ in a proper interval by means of eqs 4,6 and 4,8

$$F = -\bar{G}\frac{mM}{\Delta x^2} \qquad \bar{G} = \bar{G}(x) = \frac{G}{x-x_0}\int_{x_0}^{x}\left(1-\frac{\gamma}{\Delta x'}\right)d\Delta x' \qquad G = \frac{\hbar c^2}{m^u p_o^0} \qquad \gamma \ll \Delta x$$

If $G$ is the value at $x_0$, then the integral calculates the mean value of $\chi_G$ between the fixed coordinate $x_0$ defining an appropriate reference system and the current coordinate $x$. Eq 4,31 yields

$$\bar{G} = \left(1 + \frac{|\gamma|}{\Delta x - \Delta x_o}\log\frac{\Delta x}{\Delta x_o}\right)G \qquad |\gamma| = \frac{6MG}{c^2} \qquad \Delta x = x - x_o \qquad \Delta x_o = x_0 - x_o \qquad |\gamma| \ll \Delta x_o$$

with $x_o$ constant. The limits $x \to \infty$ or $x_0 \to \infty$ yield the expected Newtonian value $\bar{G} \to G$; also, since the dependence of $G$ on $x$ is weak, a significant deviation $\bar{G} - G$ is expected for $x_0 \gg x$ only. In a typical laboratory test one expects the condition $x \gtrsim x_o$, which means calculating the deviation $\bar{G} - G$ in proximity of the arbitrary reference coordinate $x_o$. The condition $x_0 \gg x \gtrsim x_o$, i.e. $\Delta x \ll \Delta x_o$, describes integration limits that reasonably prospect an experimentally detectable $x$ dependence of $\bar{G} = \bar{G}(x)$; thus the formula suitable for the comparison with the experiment is

$$\bar{G} = \left(1 - \frac{6MG}{\Delta x_o c^2}\log\rho\right)G \qquad \rho = \frac{\Delta x}{\Delta x_o} \qquad \left|\frac{6MG}{c^2}\right| \ll \Delta x_o \qquad \Delta x \ll \Delta x_o$$

recalling eqs 4,32. In conclusion, the deviation of $F$ from the mere Newtonian approximation $G\Delta x^{-2}$ due to the term $\gamma\zeta\Delta x^{-3}$ defines an average value $\bar{G}$ as a function of the dimensionless parameter $\rho$ only; moreover $\Delta x$ is expressed in $\Delta x_o$ units, which determines the real length scale corresponding to the parameter $\rho$. In a laboratory experiment with a test mass, $M = m_{Earth} + m_{test}$ actually reduces to the first addend only; with the numerical values, the formula above reads

$$\bar{G} = \left(1 + \frac{0.027}{\Delta x_o}|\log\rho|\right)G \qquad \frac{6m_{Earth}G}{c^2} = 0.027 \text{ m} \qquad 4,41$$

This equation agrees with the best fit of experimental data at laboratory scale: ref [18] proposes the function $G(R) = (1 + 0.002\log R)G_N$ at distances $R$ equal to 29.9 and 4.48 cm, being $G_N$ a normalization constant. Rewriting $G(R) = [1 + 0.002 R_o^{-1}\log(R/R_o)]G(R_o)$ with $G_N = R_o^{-1}G(R_o)$, it appears that $\bar{G}$ calculated with the first order term of eq 4,21 agrees significantly with $G(R)$ from the analytical and numerical point of view: the range $\Delta x_o$ plays the role of $R_o$, the ratio $\rho$ that of $R/R_o$ and $G(R_o)$ corresponds to $G$. The numerical factor 0.002 differs from 0.027 of eq 4,41 because of the different measure units of $\Delta x_o$ and $R_o$ that fix the length scales of $\Delta x$ and $R$: the latter fits two values of $G(R)$ of the order of $10^{-1}$ m apart, the former is defined in meters by the second eq 4,41, whence the factor 10 between the dimensional coefficients of the logarithmic function. An essential remark is that, once more, in the present approach the ranges of distances $\Delta x$ and $\Delta x_o$ replace the local coordinates defining the point to point distances $R$ and $R_o$. The interesting conclusion is that the present result has been obtained through the correction factor $\gamma$ that accounts for the perihelion precession. A closing remark regards again the physical dimensions of the constant $G$. If $m_G$ and $v_G$ denote generically an arbitrary mass and frequency, then $V_G = Gm_G v_G^{-2}$ has physical dimensions of volume. Let us introduce therefore the dimensionless quantity $\xi = Gd_G v_G^{-2}$, where $d_G = m_G/V_G$ has physical dimensions of density. The choice of $\xi$ depends on how are specified case by case $d_G$ and $v_G$. Simple considerations explain the result

$$v_G = \sqrt{Gm_G(\xi V_G)^{-1}} = \sqrt{Gd_G \xi^{-1}} \qquad 4,42$$

Imagine an arbitrary region of space-time as a box of volume $V_G$ where is delocalized a free particle of mass $m_G$; for instance, this could be the case of a particle confined in an infinite potential well. Thus $v_G$ yields the frequency of shocks against the walls of the box. Note that the arbitrary factor $\xi$ could have been omitted if included in $V_G$ arbitrary as well. Yet for sake of generality it has been explicitly quoted here to emphasize that in principle a proper choice of $\xi$ makes eq 4,42 suitable to tackle various kinds of problems, for brevity however not concerned here. The more important question whether $V_G$ is mere dimensional definition or it has actual physical meaning, clearly depends on the specific way to define $m_G$ and $v_G$. A relevant example highlights this point. Define $v_o^0 = c/\lambda_o^0$ as the frequency related to $\lambda_o^0$ previously introduced in eq 4,40 and express for convenience $m_G$ as the proton rest mass $m_p$; the interest of this choice will be explained in the next subsection *4.9*. Let the factor $\xi$ specify the position $m_G = m_p$ in the particular case where $v_G$ is identified by $v_o^0$; this position yields $V_G^P = Gm_p \xi^{-1}(\lambda_o^0/c)^2$. If $\xi^{-1} = 156.3$, then eqs 4,42 yield $V_G^P = 4.22 \cdot 10^{-105}$ m$^3$: so the Plank volume defines the hypothetical particle $m_H$ of mass $156.3 m_p$. This conclusion would have an interesting physical meaning if such a particle could be experimentally observed, as it would mean that $V_G^P$ is not mere dimensional definition. Actually the link between the Plank units and $p_o^0$ is not accidental, as eq 4,6 that defines $G$ reads also $m^u = (m_P c/p_o^0)m_P$; so $m^u$ formally introduced in eq 4,5 is proportional to the Plank mass $m_P$, the proportionality factor being the ratio between the Plank momentum and the gravity field momentum. Note in this respect that owing to eqs 4,40 the pure number $m_P c/p_o^0$ is such that $2\pi(m_P c/p_o^0)^3 = 6.09 \cdot 10^{23}$, very close to the Avogadro number despite the constant $p_o^0$ is the zero order approximation of the field momentum $p_o$ only; this result, although empirical, supports the numerical value of $p_o^0$ through its relationship with a fundamental constant of the nature. The next subsection will show how to exploit further eqs 4,42 in the present conceptual context.

*4.9 Dirac cosmology and vacuum energy.*

The present subsection aims merely to show through order of magnitude estimates that even basic information on cosmology is framed in the theoretical context so far outlined. Simple considerations on eq 4,6 highlight the physical meaning of $G$ beside its numerical value; the presence of $\hbar$ and $c$ in its definition suggests the possibility of introducing the fine structure constant $\alpha$ as

$$G = \frac{\hbar c^2}{m^u p_o^0} = \frac{c}{m^u p_o^0} \frac{e^2}{\alpha} \qquad 4,43$$

The following reasoning exploits the fact that are formally related to $G$ the unit mass $m^u$ and charge $e$; being introduced without specific reference to any real particle, these latter are regarded as virtual mass and charge concurring to define physically and numerically $G$. Rewrite eq 4,43 as

$$G\Delta\tau_o^u \alpha = \frac{\varepsilon_*^u}{m^u p_o^0} l_o^u l_*^u \qquad l_o^u = c\Delta\tau_o^u \qquad \varepsilon_*^u = \frac{e^2}{l_*^u} \qquad 4,44$$

being $\Delta\tau_o^u$ an arbitrary time range and $\varepsilon_*^u$ an energy defined by the arbitrary length $l_*^u$ (here and in the following the symbols $\varepsilon$ and $\eta$ denote respectively energy and energy density; the superscript $u$ follows the notation of $m^u$). Despite this equation actually defines neither $l_*^u$ nor $l_o^u$, whose physical meaning is therefore still hidden in eqs 4,43 and 4,44, it is formally possible to write

$$G\Delta\tau_o^u \alpha \varphi^u \lambda^u = \frac{1}{d_*^u} \frac{\varepsilon_*^u}{p_o^0} \qquad d_*^u = \frac{m^u}{V_*^u} \qquad V_*^u = \zeta^u l_o^u l_*^u (\chi^u \lambda^u) \qquad \varphi^u = \chi^u \zeta^u \qquad \lambda^u = \frac{h}{m^u c} \qquad 4,45$$

where $d_*^u$ has physical dimensions of density. The numerical factor $\varphi^u$ accounts for actual size and geometrical shape of $V_*^u$: the former is expressed in general as $\chi^u \lambda^u$ via the numerical parameter $\chi^u$ and the Compton length $\lambda^u$ of the unit mass $m^u$, the latter via the geometrical coefficient $\zeta^u$. Let $V_*^u$ be the delocalization volume of $m^u$ and $e$; if so, $\Delta\tau_o^u$ is by definition the time range during which $m^u$ is delocalized within $V_*^u$, whatever its actual size and shape might be. Regarding the virtual mass $m^u$ likewise the mass of any real particle, the definition of $G$ yields $d_*^u$ and related energy density $\eta_*^u = d_*^u c^2$ per unit volume of empty space-time the physical meaning of properties of the vacuum. As $\varepsilon_*^u / p_o^0$ has physical dimensions of velocity whatever $\varepsilon_*^u$ and $p_o^0$ might be, then

$$V_*^u = \frac{\varphi^u G\Delta\tau_o^u \alpha \lambda^u m^u}{v_*^u} \qquad d_*^u = \frac{v_*^u}{\varphi^u G\Delta\tau_o^u \alpha \lambda^u} \qquad \eta_*^u = \frac{v_*^u c^2}{\varphi^u G\Delta\tau_o^u \alpha \lambda^u} \qquad v_*^u = \frac{e^2}{p_o^0 l_*^u} \qquad 4,46$$

with $v_*^u$ arbitrary velocity. All of these quantities are defined in an arbitrary reference system $R^u$ where $m^u$ is at rest. In agreement with the concept of uncertainty, no hypothesis is possible about $\Delta\tau_o^u$ and $V_*^u$; is only known the finite amount of energy $\varepsilon_*^u = \eta_*^u V_*^u = m^u c^2$ within $V_*^u$. Yet this lack of information does not prevent to emphasize a relevant property of eqs 4,46 evident for finite values of $\Delta\tau_o^u$, i.e. $d_*^u$ and $\eta_*^u$ are upper limited for $v_*^u = c$ and $\chi^u = 1$; the last statement follows the idea that the Compton length defines the minimum limit on measuring the position of a particle. So, physical reasons require the existence of a minimum value $V_{min}^u$ of $V_*^u$ defined by the lengths $\lambda^u$, $l_o^u$ and $l_{min}^u = e^2 / p_o^0 c$; in effect $V_{min}^u$ cannot be equal to zero, otherwise $m^u$ would be exactly located somewhere in the space-time. According to eq 4,44, for $v_*^u = c$ one finds

$$l_{min}^u = \frac{\hbar\alpha}{p_o^0} \qquad V_{min}^u = \frac{\zeta^u \alpha \Delta\tau_o^u G h}{c^2} \qquad d_{max}^u = \frac{c}{\zeta^u \alpha \Delta\tau_o^u G \lambda^u} \qquad \eta_{max}^u = \frac{c^3}{\zeta^u \alpha \Delta\tau_o^u G \lambda^u} \qquad 4,47$$

In an analogous way it is also possible to infer for $v_*^u < c$ the following results

$$d_{min}^u = m^u / V_{max}^u \qquad \eta_{min}^u < c^3 (\zeta^u \alpha \Delta\tau_o^u G \Lambda^u)^{-1} \qquad 4,48$$

being $\Lambda^u$ the maximum value allowed to $\chi^u \lambda^u$; wathever $\Lambda^u$ might be, the inequality introduces an appropriate volume of vacuum $V_{max}^u$ such that $\eta_{min}^u = m^u c^2 / V_{max}^u$, being $V_{max} > \zeta^u G\Delta\tau_o^u \alpha \Lambda^u m^u / c$. Also $\eta_{min}^u$ is defined in $R^u$. So $\eta_{min}^u$ and $\eta_{max}^u$ fulfil the condition $\eta_{max}^u > \eta_{min}^u \chi^u$, with $\chi^u > 1$ arbitrary. Regard $V_{max}^u$ as a cluster of several elementary volumes $V_{min}^u$ of empty space-time; the impossibility of establishing where the mass $m^u$ is delocalized corresponds to the chances for $m^u c^2$ of being in a single box $V_{min}^u$ or in several boxes up to $V_{max}^u$, whence the respective energy densities. The fact that the elementary volume element $V_{min}^u$ is a property of the vacuum defined by physical requirements and allowing in principle measurable outcomes suggests the quantization of empty space-time, as in effect it will be confirmed later. Let us follow therefore the idea of exploiting eqs 4,46 to 4,48 without hypotheses about the actual sizes of $\Delta\tau_o^u$ and $V_*^u$, regarded in the following simply as parameters unknown. After having introduced $d_*^u$ and $\eta_*^u$ regardless of the presence of any real particle, one expects that this way of describing the properties of the vacuum can be extended to include the properties of matter possibly present in any space-time uncertainty range. So, before calculating $\eta_{min}^u$ and $\eta_{max}^u$, let us show how eqs 4,46 describe also the presence of matter. Consider a real free particle delocalized in an arbitrary region of space-time previously empty; this is equivalent to say that, for any physical reasons, $m^u$ is replaced by or turns into a particle of mass $m$. The mere definition of $G$ helps to highlight in a formal way also this point, i.e. regardless of appropriate energy considerations (vacuum quantum fluctuations, see below) inherent such an event. Rewrite eq 4,42 as $m_x = V_G^{-1} G m_G^2 v_G^{-2}$, where $m_\xi = \xi m_G$; being $\xi$ arbitrary, $m_\xi$ and $m_G$ are in principle two different masses, i.e. two different particles, whose relative values depend on how are specified the frequency $v_G$ and the volume $V_G$; it suggests the possibility of relating $m^u$ to $m$ simply putting $m_G = m$ and $m_\xi = m^u$. So replacing $m^u$ with $m$ does not need any additional hypothesis, as also $m$ is introduced in the same conceptual frame of eqs 4,43 to 4,47. Specifying in particular $V_G = \zeta^u (l_{min}^u)^3$, one finds thanks to the first equation 4,47

$$m^u = (\zeta^u)^{-1}(\hbar\alpha/p_o^0)^{-3} G m^2 v_G^{-2} \qquad 4,49$$

Eq 4,49 defines $v_G$ as a function of $m$ delocalized in $V_G$. Note that the initial eq 4,43 introduces the virtual charge $e$ appearing as $e^2$ in all of the following equations 4,44 to 4,46, whereas in eq 4,49 $m$ is introduced as $m^2$ as well; regardless of the particular kind of particles $e$ and $m$ might represent, the fact that these equations are consistent with both signs of mass and charge related to the virtual mass $m^u$ agrees with the conclusions of appendixes A and B about the existence of matter and antimatter. This means that both these latter can be generated from the vacuum, for instance thanks to the energy of a quantum fluctuation of the vacuum itself. To examine how the presence of $m$ within $V_*^u$ affects the properties of the vacuum, let us rewrite eqs 4,46 as follows

$$V_* = \frac{\varphi G \Delta\tau_o \alpha\lambda m}{v_*} \qquad d_* = \frac{v_*}{\varphi G \Delta\tau_o \alpha\lambda} \qquad \varphi = \chi\zeta \qquad \lambda = \frac{h}{mc} \qquad 4,50$$

The notations emphasize that replacing $m^u$ with $m$ affects the local space-time geometry, perturbs the neighbouring vacuum virtual charges and modifies $V_*^u$ and $\Delta\tau_o^u$; moreover, the allowed states of empty space-time around $m^u$ and $m$ are also expected to be in principle different. In analogy with the reasoning about $\Delta\tau_o^u$ and $\zeta^u$, the particle of mass $m$ is assumed delocalized during the time range $\Delta\tau_o$ in a volume $V_*$ of space-time whose geometry is described by the coefficient $\zeta$; accordingly, the Compton wavelength $\lambda^u$ of the virtual mass $m^u$ has been replaced by that pertinent to $m$. Putting again $\chi = 1$ eqs 4,47 turn into

$$V_{min} = \frac{\zeta\alpha\Delta\tau_o Gh}{c^2} \qquad d_{max} = \frac{c}{\zeta G \Delta\tau_o \alpha\lambda} \qquad \eta_{max} = \frac{c^3}{\zeta G \Delta\tau_o \alpha\lambda} \qquad 4,51$$

Eqs 4,47 and 4,51 are limit cases of the respective eqs 4,46 and 4,50. So far $\lambda^u$ and $\lambda$ have been regarded as mere reference lengths to express the sizes $\chi^u\lambda^u$ and $\chi\lambda$ characterizing $V_*^u$ and $V_*$. Let us introduce also the characteristic local density

$$V_\lambda = \zeta\lambda^3 \qquad d_\lambda = \frac{m}{V_\lambda} \qquad \eta_{max}^\lambda = \frac{mc^2}{V_\lambda} \qquad 4,52$$

noting that $V_{min}$ of eq 4,51 is a consequence of the space-time uncertainty, whereas $V_\lambda$ has instead character of definition inherent the physical meaning of Compton length. Regard therefore $V_\lambda$ as a reference volume expressing the maximum density measurable for $m$, whose corresponding energy density is given by the third equation for $m$, and calculate $\Delta\tau_o$ in the particular case where $V_{min}$ coincides with $V_\lambda$; putting by consequence $\eta^\lambda_{max} = \eta_{max}$, one finds

$$G\Delta\tau_o = \frac{c\lambda^2}{m\alpha} \qquad 4,53$$

Of course eq 4,53 does not exclude the possibility that $V_{min} > V_\lambda$, being the evolutive character of $V_{min}$ in fact inherently consequent to its own space-time definition. Then eq 4,53 refers to an arbitrary time at which hold the assumption $V_{min} \gtrsim V_\lambda$; both therefore remain in fact still indeterminate, in agreement with the considerations previously introduced about the time uncertainty, but simply estimated at the particular time that defines the corresponding $\Delta\tau_o$. It is natural at this point to ask if $\Delta\tau_o$ really describes any experimental observable confirming the steps from eqs 4,43 to 4,53. The analogy between the ways of introducing $V^u_*$ and $V_*$ suggests that $\Delta\tau_o$ should regard the lifetime of $m$ within $V_*$, likewise $\Delta\tau^u_o$ describes the lifetime of $m^u$ within $V^u_*$; moreover the volume change rate $(V_* - V^u_*)/\Delta\tau_o$ around the space-time region where $m$ replaces $m^u$, evocative of the concept of mass driven $\Delta\dot{x}$ leading to eqs 4,2, should describe the rate with which massive particles change the size of the whole space-time volume where they are delocalized. This idea can be further extrapolated: if the universe effectively reproduces on cosmic scale the local quantum scale of space-time around one particle, then $\Delta\tau_o$ appears to be the kinetic parameter determining how the size of an empty volume changes along with the amount of mass in it created by an appropriate fluctuation of the vacuum energy. To verify this conclusion, let us rewrite eq 4,53 as

$$\frac{G}{H_0} \approx \frac{c\lambda^2}{m\alpha} \qquad H_0 \approx \frac{1}{\Delta\tau_o} \qquad V_{min} \approx V_\lambda \qquad 4,54$$

This equation does not concern a particular kind of particle, whose actual nature is hidden in the mass $m$. If, in particular, $m$ coincides with the rest mass $m_p$ of the proton and $\lambda$ is accordingly regarded, then eq 4,54 reads also $G/H_0 \approx 4\pi^2 \lambdabar_p^2 c / \alpha m_p$; a numerical factor of the order of $3\pi$ apart, this is just the Dirac equation known in literature as $G/H_0 \approx 1.3\pi \lambdabar_p^2 c / \alpha m_p$ [19] with $H_0$ equal to the Hubble constant. On the one side it is not surprising the existence of a time range $\Delta\tau_o$ necessary for the empty space-time range to "respond" to the presence of a real particle therein delocalized, similarly as should also do the whole universe around the matter in it created. On the other side, however, at least two questions raise: why one particle only seems representative of the behaviour of the whole universe and why just the proton? An exhaustive answer to these questions is clearly beyond the scope of the present paper; so the next part of this subsection is deliberately restricted to a minimum amount of calculations only and introduces a few notions strictly necessary to delineate the broadened landscape of conclusions achievable through the positions 1,1. For this reason the geometry factors $\zeta^u$ and $\zeta$ will be systematically put both equal to 1 in the following, yet explicitly quoted in the formulae to emphasize their prospective role in a more dedicated analysis. Here, further considerations about the Hubble constant highlight the physical meaning of eq 4,6 that defines $G$ and support the validity of eq 4,54. Regard for simplicity $H_0$ as a time constant, remarking however that actually this position is not necessarily required and regards likely an approximated mean value; eq 4,54 estimated with the proton rest mass yields $H_0 \approx 1.6 \cdot 10^{-18}$ s$^{-1}$, sensibly close to the experimental values in the acknowledged range $(1.7 \div 2.3) \cdot 10^{-18}$ s$^{-1}$. So the reasoning from eqs 4,43 to 4,54, in particular the position $V_{min} \gtrsim V_\lambda$, should be in principle correct. It is also interesting to regard again $H_0$ with the help of eqs 4,49, specifying $\nu_G$ as the frequency $v_p / \Lambda_{un}$ of a matter wave propagating through the universe at rate $v_p$; this is nothing else but a way to describe the proton delocalized everywhere in the universe having size $\Lambda_{un}$. Eqs 4,42 calculate then the density $d_{un} = (v_p / \Lambda_{un})^2 G^{-1}$. As massive particles travel with any velocity between zero and the asymptotic limit $c$, a rough guess would suggest an average value $v_p \approx c/2$. A better estimate is obtained calculating eq 4,49 with the frequency $\nu_G = 1.56 \cdot 10^{12}$ s$^{-1}$, whose physical meaning and numerical value will be justified below, see the next eq 4,61: the result is $m_G = \pm 2.4 \cdot 10^{-27}$ Kg. This value, close to the proton rest mass, suggests that $m_G = m_p / \sqrt{1 - (v_p/c)^2}$, i.e. $m_G$ could represent in fact the kinetic mass $m_{pk} = 1.44 m_p$ of a proton moving in $R^u$ at rate $v_p$; this entails $v_p / c = 0.72$. Recalculating the first eq 4,54 with $m = m_{pk}$, one finds $H_0 \approx 2.2 \cdot 10^{-18}$ s$^{-1}$; this further way to calculate $H_o$ confirms that actually the Hubble constant is not at all a constant and should be calculated revising the position $V_{min} \approx V_\lambda$ to account for the time dependence of the properties of the universe. Eqs 4,54 yield also

$$\Delta \tau_o \approx \frac{c\lambda^2}{m_{pk} G \alpha} = 4.5 \cdot 10^{17} \text{ s} \qquad \Lambda_{un} \approx 4c\Delta\tau_o = 5.4 \cdot 10^{26} \text{ m} \qquad V_{min} \approx V_\lambda \qquad 4,55$$

The second equation is explained as follows. Consider a light beam that travels during the time range $\Delta\tau_o$ a distance $c\Delta\tau_o$; as the light can propagate from an arbitrary point along two opposite directions physically indistinguishable, the time uncertainty $\Delta\tau_o$ defines a space-time range having size $\delta_o = 2c\Delta\tau_o$. This reasoning entails therefore the source point around which propagates the light beam necessarily located in the middle of $\delta_o$. Yet the idea of a fixed point exactly located somewhere within a range is clearly inconsistent with the early concept of uncertainty so far exploited; the source point must be instead randomly located anywhere within the range, even on its boundaries wherever these latter might be as a function of time. This compels regarding $c\Delta\tau_o$ as a subrange randomly delocalized in a larger range $2\delta_o$; if so, the point where is located the light source is in effect delocalized itself in $\delta_o$. Therefore $2\delta_o$ is the range size consistent with both all of the properties inferred according to the fundamental positions 1,1 and with the impossibility of localizing the light source that determines the physical validity of eqs 4,45 to 4,54. The uncertainty pushes the size limit of the universe up to $\Lambda_{un}$ of eq 4,55, still related however to the distance reachable by a light beam during the life time of the universe itself; in effect $\Lambda_{un}$ is reasonably close to the size today acknowledged. Further interesting properties are inferred from eqs 4,55 with the help of eqs 4,42; the ratio $v_p / c$ and the way of defining $d_{un}$ allow to calculate the average energy density of matter in the universe. Still admit the proton mass as representative of the total mass of matter in the universe; the total mass $M_p^{tot}$ and total number $n_p^{tot}$ of protons and their average energy density are therefore

$$d_{un} = \frac{v_p^2}{\Lambda_{un}^2 G} = 2.4 \cdot 10^{-27} \text{ Kg m}^{-3} \qquad M_p^{tot} = d_{un} \Lambda_{un}^3 = 3.7 \cdot 10^{53} \text{ Kg} \qquad n_p^{tot} = \frac{M_p^{tot}}{m_{pk}} = 2.2 \cdot 10^{80}$$

$$\eta_{min} = d_{un} c^2 = 2.2 \cdot 10^{-10} \text{ J m}^{-3} \qquad 4,56$$

These numbers are quite reasonable and explain why just the proton is the particle that fits the Dirac equation 4,54. Moreover, there is an interesting observation about the second eq 4,55 and the first eq 4,56; eliminating $\Lambda_{un}$ one finds

$$G = d_{un}\left(\frac{4c^2\lambda^2}{m_{pk}\, \mathrm{v}_p \alpha}\right)^2$$

Recall that the kinetic mass of the proton has been inferred through the frequency $v_G$ only: if the values of this latter and $v_p$ are correct, then this equation calculates $G$ merely through the early Dirac intuition. Indeed, regarding $d_{un}$ equal to $m_{pk}$ per unit volume of space-time, the result is $G = 6.60 \cdot 10^{-11}$ m$^3$ Kg$^{-1}$ s$^{-2}$. This result supports the values of eqs 4,56, in particular the way to calculate the proton kinetic mass. Note however in this respect that eqs 4,54 and 4,55 do not necessarily assign the notable numerical coincidence found by Dirac just to the proton itself; $m$, in principle not specified, could be actually regarded as mean value of different masses, i.e. particles, approximately consistent with the mass of the proton that fits surprisingly well the experimental value of Hubble constant. The coincidence is sensible because just the protons represented more than 75% about 1 second after the big-bang and have risen up to more than 85% after about 13 seconds; today mostly H and He account for the number of of atoms in the universe. Yet cosmologists also say that up to 90% of the mass in the universe is accounted for by dark matter, whose nature is still under investigation. On the one side thus the coincidence found by Dirac, although physically profound, seems accidental as concerns the real identity of the particle hidden in the numerical value of $m$: protons or neutrons or even hydrogen atoms would bring to the same order of magnitude agreement with $H_0$ as their masses are similar. Moreover, the same conclusion would hold also in the presence of large numbers of neutrinos and small amounts of heavy atoms too: a wide variety of chances, including also dark matter, is hidden in $m$ of eq 4,54; in particular, it results evident that any change of $m$ during the lifetime of the universe affects the value of $H_0$, which is therefore not necessarily constant. Nevertheless, even regarding in principle $\Delta\tau_o$ as an effective value averaged on the relative abundance of several kinds of particles, the remarkable fact justifying the present digression is that also the Hubble constant is inferred in the theoretical context so far outlined merely exploiting the analytical form of $G$ in eq 4,6. On the other side, just the definition of $G$ and the conceptual analogy between eqs 4,46 and 4,50 support the connection between eqs 4,55 and dark matter and energy, both suggested by the vacuum density $d_*^u = \eta_*^u c^{-2}$: this result requires that $\eta_*^u$ and thus also $m^u$ formally introduced in eq 4,5 have a real physical meaning. The next subsection *4.10* will also concern this point. The conclusions inferred from eqs 4,55 seem important enough to justify the physical properties of the vacuum described by eqs 4,46 and stimulate further considerations. Rewrite the third eq 4,46 as $\eta_*^u < (\zeta^u hG\alpha/c^3)^{-1}(hv_o^u)(\chi^u \lambda^u)^{-1}$ through the intuitive position $\Delta\tau_o^u = 1/v_o^u$ and note that the first factor having physical dimensions *length*$^{-2}$ can be opportunely expressed as $(\zeta^u)^{-1}(c/v_{Gh\alpha})^2$ through the frequency $v_{Gh\alpha}^2 = c^5(Gh\alpha)^{-1}$; expressing analogously the third factor as $c/v^u$, one finds $\eta_*^u < (\zeta^u)^{-1}hv_o^u v_{Gh\alpha}^2 v^u c^{-3}$. Being $v^u$ arbitrary and the variable parameter $\chi^u$ not yet specified, it is possible to write $\eta_*^u < (\zeta^u)^{-1}hv_c^4 c^{-3}$ with $v_c = (v_o^u v_{Gh\alpha}^2 v^u)^{1/4}$. The geometrical coefficient $\zeta^u$ could have been included into the resulting value $v_c$, yet it is preferable to let it appearing explicitly in the expression of $\zeta^u$ for reasons that will be clear soon. As an appropriate interval of values of $v_c$ must correspond to $\eta_*^u$ ranging between the minimum value $\eta_{\min}^u$ and the maximum value $\eta_{\max}^u$, it is convenient to introduce a ground frequency $v_{oc}$ such that

$$\eta^u_{min} = (\zeta^u)^{-1} h\nu^4_{oc} c^{-3} \qquad \eta^u_{max} = (\zeta^u)^{-1} h\nu_{oc} \nu^3_c c^{-3} \qquad \nu_c > \nu_{oc} \qquad 4,57$$

Previously the energy densities $\eta^u_{min}$ and $\eta^u_{max}$ were defined by a unique energy, $\varepsilon^u_* = m^u c^2$ calculated in two delocalization volumes $V^u_{max}$ and $V^u_{min}$ physically different and both allowed for $m^u$. Having replaced in eqs 4,57 the size $(\chi^u \lambda^u) l^u_o l^u_*$ of $V^u_*$ with $(c/\nu^u_*)^3$, i.e. as a function of the corresponding frequency $\nu^u_*$, the volumes defining $\eta^u_{min}$ and $\eta^u_{max}$ read now $V^u_{max} = \zeta^u (c/\nu_{oc})^3$ and $V^u_{min} = \zeta^u (c/\nu_c)^3$, whereas $\varepsilon^u_*$ is replaced by $h\nu_{oc}$. Of course it is unessential that $h\nu_{oc}$ does not coincide numerically with $m^u c^2$, being instead essential that $\varepsilon^u_* / V^u_* = h\nu_{oc} / (\zeta^u (c/\nu^u_*)^3)$; in other words, eqs 4,57 rewrite the respective eqs 4,47 and 4,48 through a multiplicative scale factor at numerator and denominator. From a physical point of view, however, these equations are perfectly equivalent despite their different notations; in particular, eq 4,47 requires

$$\eta^u_{max} = \frac{c^3}{\zeta^u G \Delta\tau^u_o \alpha \lambda^u} = \frac{h\nu_{oc}}{\zeta^u} \left(\frac{\nu_c}{c}\right)^3 \qquad 4,58$$

This result allows to calculate $\Delta\tau^u_o$ as a function of $\nu_{oc}$ and $\nu_c$. Note that $\eta^u_{max}/\eta^u_{min}$, and thus the ratio $V^u_{max}/V^u_{min} = N^3 = (\nu_c/\nu_{oc})^3$, does not depend on $\zeta^u$; if the the idea of space-time quantization is correct, it is sensible to regard $V^u_{max}$ as a cluster of $N^3$ elementary boxes of volume $V^u_{min}$. Of course $N$ must be integer, which entails thus $\nu_c = N\nu_{oc}$; it means regarding reasonably $\nu_{oc}$ as fundamental frequency and $\nu_c$ as an integer multiple of $\nu_{oc}$. In summary

$$\frac{\eta^u_{max}}{\eta^u_{min}} = N^3 \qquad \nu_c = N\nu_{oc} \qquad V^u_{min} = \zeta^u \left(\frac{c}{N\nu_{oc}}\right)^3 \qquad V^u_{max} = \zeta^u \left(\frac{c}{\nu_{oc}}\right)^3 \qquad 4,59$$

The importance of these conclusions will appear later in eqs 4,65 and 4,69 that concern the Casimir effect. Let us exploit eqs 4,42 to calculate relevant cosmological data of the vacuum defining appropriately $\nu_G$, $m_G$ and $V_G$. Specifying $\nu_G = c/\Lambda^u_{un}$, i.e. regarding in the absence of matter $\nu_G$ as the frequency of an electromagnetic wave, and $d_G = \eta^u_{min}/c^2$ in $V_G = m_G G \nu^{-2}_G$ as before, one finds $\Lambda^u_{un} = c^2 (G\eta^u_{min})^{-1/2}$. Moreover the average density $d^u_{un}$ of virtual matter in the volume $\zeta^u (\Lambda^u_{un})^3$ is estimated putting in eq 4,42 $m_G V^{-1}_G = d^u_{un}$ and $\nu_G = c/\Lambda^u_{un}$, so that $d^u_{un} = (c/\Lambda^u_{un})^2 G^{-1}$. In summary, we have obtained the following equations

$$\Lambda^u_{un} = \frac{c^2}{\sqrt{G\eta^u_{min}}} \qquad d^u_{un} = \frac{(c/\Lambda^u_{un})^2}{G} \qquad 4,60$$

These equations depend on a unique quantity, the ground value of vacuum energy density $\eta^u_{min}$, which is actually function of $\nu_{oc}$ only. Of course, the proton mass does not longer appear in eqs 4,57 to 4,60. As $\nu_{oc}$ is not known nor calculable because it depends on $\Delta\tau^u_o$ basically unknowable, eq 4,58 suggests exploiting just the fact that $\eta^u_{min}$ and thus $\nu_{oc}$ are functions of $\Delta\tau^u_o = t - t^u_{beg}$; this highlights that also the quantities of eqs 4,60 are functions of time themselves. Even without defining $t^u_{beg}$, these equations can be calculated regarding $\nu_{oc}$ as arbitrary time parameter; as done in eq 4,53, introducing input values of $\nu_{oc}$ means fixing a particular value of the time at which the resulting values possibly agree with observable results. If so, then it should be possible to identify this particular value as that where all of the quantities fit the corresponding experimental values today observable. This procedure is significant as it is based on the time profile of several relevant quantities that must be simultaneously correct to validate the present model and provide a comprehensive view of our universe. Before considering this aspect of the problem, let us show that the equations above are sensible indeed. Whatever the actual time dependence of $v_{oc}$ upon the time might be, at the particular time, say today, where

$$v_{oc} = 1.56 \cdot 10^{12} \text{ s}^{-1} \qquad 4,61$$

eqs 4,58 and 4,60 yield then

$$\eta^u_{\min} = 1.5 \cdot 10^{-10} \text{ J m}^{-3} \qquad \Lambda^u_{un} = 9.1 \cdot 10^{26} \text{ m} \qquad V^u_{\min} = 3.2 \cdot 10^{-83} \text{ m}^3$$
$$V^u_{\max} = 7.1 \cdot 10^{-12} \text{ m}^3 \qquad d^u_{un} = 1.6 \cdot 10^{-27} \text{ Kg m}^{-3} \qquad \Delta\tau^u_o = 7.9 \cdot 10^{17} \text{ s} \qquad 4,62$$

Before calculating $\eta^u_{\max}$ these results deserve some comments. The time dependence of the value of $v_{oc}$ corresponds to that of the position $V_{\min} \gtrsim V_\lambda$ leading to eq 4,53. The value of $\Lambda^u_{un}$, formally introduced as a wavelength, is surprisingly similar to $\Lambda_{un}$ of eqs 4,56 despite the different ways to infer them; hence also $\Lambda^u_{un}$ is the wavelength of a giant standing wave crossing all the universe. The value of $\eta^u_{\min}$ is reasonable: the average vacuum density $d^u_{un}$ in the universe is of the order of one proton mass per unit volume, hence it is realistic that at local level also $\eta^u_{\min}$ is very close to $m_p c^2 \simeq 1.5 \cdot 10^{-10}$ J per unit volume, although calculated in a different way. This consideration confirms the idea previously introduced about $\Delta\tau_o$ to infer eq 4,54, i.e. quantum local scale and cosmological global scale are somehow correlated; this explains why in eq 4,49 $m \approx m_p$, whatever $m_p$ might actually represent. It also appears that $\Delta\tau_o \approx \Delta\tau^u_o$. The similarity of these time ranges, previously regarded as lifetimes of $m^u$ within $V^u_*$ and $m$ later identified with the proton mass in $V_*$, could not be accidental; rather, just this similarity could justify eq 4,61. Moreover $\Delta\tau_o$ and $\Delta\tau^u_o$ pose in fact the question about whether the former is preceding or following the latter. Although the link between $m^u$ and $m$, eq 4,49, does not suggest itself any hint to decide either time sequence, one could guess three chances: (i) a universe initially empty and then filled with matter created by its own vacuum energy fluctuation, which should therefore entail matter annihilation after the lifetime of the fluctuation; (ii) an empty universe with vacuum energy resulting from annihilation of matter previously present, without excluding even (iii) a cyclic energy-matter sequence of vacuum states. This latter chance seems consistent with the previous remark that the vacuum energy per unit volume, $\eta^u_{\min}$, is consistent with $m_p c^2$ per unit volume. This stimulates to estimate that the total vacuum energy $\eta^u_{\min} \zeta^u \Lambda^{u\,3}_{un}$ in the universe is equivalent to the mass $M^u_p = \eta^u_{\min} \zeta^u \Lambda^{u\,3}_{un} c^2$ and number $n^u_p = \eta^u_{\min} \zeta^u \Lambda^{u\,3}_{un} / m_p c^2$ of protons at rest respectively of the order of $1.2 \cdot 10^{54}$ Kg and $7.3 \cdot 10^{81}$, i.e. greater than the corresponding numbers previously estimated for real matter. So eqs 4,56 estimate

$$\frac{M^u_p}{M^u_p + M^{tot}_p} = 0.77$$

It is worth stressing at this point that the results above do not imply determining $\Delta\tau_o$ and $\Delta\tau^u_o$; rather, although an exact knowledge of $\Delta\tau_o$ and $\Delta\tau^u_o$ is forbidden by the uncertainty, hold also now the considerations carried out in section 2 about the hydrogenlike atoms and harmonic oscillators. So the universe evolves as a function of time simply because of the time uncertainty. The elementary calculation of the energy levels has evidenced the propensity of nature to fulfil a general criterion, the minimum energy, in principle not required by the fundamental positions 1,1; so, some among all the possible space ranges appeared effective to obtain results in fact observable. Now this idea seems still valid and regards the character necessarily transient of any energy fluctuation, which agrees in effect with the expected time dependence of eqs 4,61 and 4,62; if so the leading choice of the nature should be that of allowing a sort of time symmetry between creation and annihilation of matter, to which corresponds compatibly with eq 4,49 an analogous time symmetry between creation and annihilation of antimatter. Thus matter and antimatter flow forwards and backwards on time. Anyhow, at this point of the exposition the attention moves to the problem of the quantum fluctuation that should necessarily regard just $\eta^u_{min}$; considering this latter as ground energy density subjected to time dependent fluctuation seems a reasonable advancement to complete the picture of universe hitherto exposed, in particular to explain in a natural way the time dependence of eqs 4,60. Let $\eta_{\Delta r_o} - \eta^u_{min}$ be the quantum fluctuation in an arbitrary region of the vacuum having space-time volume $V_0 = \zeta^u \Delta r_o^3$; here $\eta_{\Delta r_o}$ is the maximum energy density triggered by the fluctuation at $t = t_o$ in $V_o$, so that the vacuum energy in this volume at this time increases to $\varepsilon^u_o = \zeta^u \Delta r_o^3 \eta_{\Delta r_o}$. The energy density gap initially confined within $V_o$ is described by the Heaviside stepwise function $(\eta_{\Delta r} - \eta^u_{min})H(\Delta r_o - \Delta r)$. In agreement with the dependence of $\Delta r_o$ on space and time coordinates, one expects that during an arbitrary time range $\Delta t = t - t_o$ the local fluctuation energy spreads around $V_o$ and propagates to a larger volume of space-time $V_r = \zeta^u \Delta r^3$; the initial energy gap $(\eta_{\Delta r_o} - \eta^u_{min})\zeta^u \Delta r_o^3$ in $V_o$ at $t = t_o$ becomes then $(\eta_{\Delta r} - \eta^u_{min})\zeta^u \Delta r^3$ in $V_r$ at $t_o + \Delta t$. Since $(\eta_{\Delta r} - \eta^u_{min})\zeta^u \Delta r^3 = (\eta_{\Delta r_o} - \eta^u_{min})\zeta^u \Delta r_o^3$ is required by the energy conservation, the $\Delta r$-profile of the initial energy fluctuation results to be a decreasing function of $\Delta r$ in $V_r$. The link between $\eta_{\Delta r_o}$ and $\eta_{\Delta r}$ at any $t$ reads therefore

$$\eta_{\Delta r} = \eta_o \left(\frac{\Delta r_o}{\Delta r}\right)^3 + \eta^u_{min} \qquad \Delta r \geq \Delta r_o \qquad \eta_o = \eta_{\Delta r_o} - \eta^u_{min} \qquad 4,63$$

In conclusion, we have three key energies in the present problem: the unperturbed energy of the vacuum $h\nu_{oc}$ corresponding to $\eta^u_{min}$, see eq 4,57, the energy gap $\varepsilon_o = \zeta^u \Delta r_o^3 \eta_o$ and the initial peak energy $\varepsilon^u_o$. Note that $\eta_{\Delta r}$ and $\eta_{\Delta r_o}$ do not depend on the particular geometry of volume elements of the perturbed and unperturbed energy densities if, as assumed here, $\zeta^u$ does not depend on time. Also, $\eta_{\Delta r} \to \eta^u_{min}$ for $\Delta r \gg \Delta r_o$; this is reasonable, as the fluctuation energy density tends to the ground limit $\eta^u_{min}$ if propagated in space volumes $\zeta^u \Delta r^3$ very large with respect to that where it was early generated. From the dimensional point of view the energy density defines in general a force per unit surface, so that any change of $\eta_{\Delta r}$ generates a net resulting local force. To highlight this point and check the validity of eq 4,63, let us calculate the change of $\eta_{\Delta r}$ when increasing $\Delta r$ by $\delta r$ along an arbitrary direction; at the first order, the energy density between $\Delta r$ and $\Delta r + \delta r$ is

$$\delta \eta_{\Delta r} = -3\eta_o \frac{\Delta r_o^3}{\Delta r^4} \delta r \qquad 4,64$$

The minus sign is obvious consequence of the $\eta_{\Delta r}$ profile as a function of $\Delta r$; so, if $\eta_o$ is positive, $\delta \eta_r > 0$ for $\delta r < 0$; this means that the force per unit surface corresponding to $\delta \eta_{\Delta r}$ and acting within $\delta r$ is attractive, i.e. it tends to shrink the volume where is defined $\delta \eta_{\Delta r}$. Writing by dimensional reasons $3\eta_o \Delta r_o^3 \delta r = k\hbar c$, i.e. collecting $\eta_o$, $\Delta r_o$ and $\delta r$ into the unique proportionality constant $k$, eq 4,64 reads

$$\delta \eta_{\Delta r} = \frac{F}{A} = -k\frac{\hbar c}{\Delta r^4} \qquad k = \frac{3\varepsilon_o \delta r}{\zeta^u \hbar c} \qquad \varepsilon_o = \zeta^u \Delta r_o^3 \eta_o \qquad 4,65$$

Regard one side $\Delta r$ of $V_r$ as separation gap between two arbitrary surfaces of area $A = \zeta'^u \Delta r^2$; with a proper value of the constant $k$ eq 4,65 is nothing else but the Casimir force, which therefore depends on the specific geometry linking $\Delta r$ to $V_r$. Particular physical interest has the position $\delta r = \Delta r_o / 3$, through which eqs 4,64 and 4,65 yield

$$\delta \eta_{\Delta r} = -\eta_o \left( \frac{\Delta r_o}{\Delta r} \right)^4 \qquad \eta_o = k \frac{\hbar c}{\Delta r_o^4} \qquad \Delta r_o = \zeta^u k \frac{\hbar c}{\varepsilon_o} \qquad 4,66$$

The fact that $\delta \eta_{\Delta r_o} = -\eta_o$ at $\Delta r = \Delta r_o$ agrees with eq 4,63: so the energy density transferred from $V_o$ to $V_r$ is just that lost at the boundary of $V_o$, whereas the force $F_o$ acting at $t = t_o$ on the area $A$ of this boundary is related just to the local energy density gap $\eta_o$ across $\Delta r_o$. The third equation suggests to express $\hbar c / \varepsilon_o$ in Compton length units of a particle, still to be identified, having mass $m_o$ through a parameter $q$ such that

$$\Delta r_o = q \frac{h}{m_o c} \qquad \varepsilon_o = \frac{\zeta^u k}{2\pi q} m_o c^2 \qquad \eta_o = \frac{k}{(2\pi q)^4} \frac{(m_o c^2)^4}{(\hbar c)^3} \qquad 4,67$$

At $t = t_o$ the increment $\eta_o$ of energy density exceeding $\eta_{\min}^u$ is proportional to $m_o^4$. If so however also $\varepsilon_o^u$ can be regarded itself as the energy $m_o^u c^2$ in the volume $\zeta^u \Delta r_o^3$; this position defines thus the further mass $m_o^u$ originated by the vacuum quantum fluctuation. Moreover, regarding in the same way even $\eta_{\min}^u$ itself, it is possible to define $\zeta^u \Delta r_o^3 \eta_{\min}^u = m_{oc} c^2$. In conclusion the fluctuation equation introduces three particles of masses $m_o$, $m_o^u$ and $m_{oc}$ corresponding to the aforesaid three energies that fulfil the condition

$$m_o^u = \frac{\zeta^u k}{2\pi q} m_o + m_{oc} \qquad m_{oc} c^2 = h \nu_{oc} \qquad 4,68$$

that is direct consequence of eqs 4,63. Owing to the finite lifetime of the quantum fluctuations, the masses $m_o$, $m_o^u$ and $m_{oc}$ should randomly change depending on the maximum energy $\varepsilon_{\Delta r_o}$ and energy gap $\varepsilon_o$ of each fluctuation. Suppose however that it is possible to define a statistical average of a great number of fluctuations randomly occurring with frequencies $\varepsilon_o / nh$ and consistent with a well defined value $\overline{\varepsilon_{\Delta r_o}}$; if so the results above are determined by the respective average values of $\overline{m_o}$, $\overline{m_o^u}$, $\overline{m_{oc}}$. Alternatively, it is also likely to expect that appropriate numbers of such particles are generated during the vacuum quantum fluctuations. Exploit now the fact that any space-time range requires a corresponding momentum range, defined in the present case through appropriate wavelengths inside $\delta r$, i.e. at distances $\Delta r$ and $\Delta r + \delta r$ from $V_o$; these latter suggest in turn the existence of vacuum momentum wavelengths also outside the thickness $\delta r$. Denoting these wavelengths as $\lambda_i$ and $\lambda_o$ respectively, the equation $(\lambda_i^{-1} - \lambda_o^{-1}) h \delta r = n\hbar$ yields

$$n \frac{3}{2\pi} \frac{\lambda_o \lambda_i}{\lambda_o - \lambda_i} = \frac{h}{m_o c} \qquad 4,69$$

According to eqs 2,5 and B4 of appendix B, the number $n$ of vacuum states is related to $n$ as $\delta \eta_{\Delta r} / \eta_\delta = \pm \sqrt{n}$, where $\eta_\delta$ is any random value of energy density within $\delta \eta_{\Delta r}$. So

$$\eta_\delta = \pm \frac{k\hbar c}{\sqrt{n} \Delta r^4} \qquad \eta_\delta^o = \pm \frac{k\hbar c}{\sqrt{n_o} \Delta r_o^4} = \pm \frac{\eta_o}{\sqrt{n_o}} \qquad 4,70$$

The first equation is defined regarding the increment $\delta r$ in an arbitrary point at distance $\Delta r$ from $V_o$, the second one just at $\Delta r_o$; so $n$ and $n_o$ are the respective numbers of states allowed to the particle of mass $m_o$ defining $\eta_o$. It is not surprising that, factors $\pm \sqrt{n}$ and $\pm \sqrt{n_o}$ apart, the form of $\eta_\delta$ and $\eta_\delta^o$ is the same as that of $\delta\eta_{\Delta r}$ and $\eta_o$; in effect the third equation links $\eta_\delta$ and $\eta_o$ previously introduced. As $n$ is mere notation indicating arbitrary integers rather than any specific value, it follows that $n$ has a unique physical meaning in eqs 4,68, 4,69 and 4,70 and thus links these equations. Eqs 4,65 describe the force per unit surface resulting from the change of energy density across $\delta r$, thus without emphasizing the vacuum states inside $\delta r$; this means that the constant $k$ should be determined summing up or averaging the various vacuum wavelengths $\lambda_i$, as in effect it is known. The first eq 4,70 shows that any local value of energy density $\eta_\delta$ within $\delta r$ depends on $n^{-1/2}$; in effect it does not provide detailed information about this random energy just because of the arbitrariness of $n$. This situation inside $\delta r$ also suggests a corresponding number of states outside $\delta r$; analogous reasoning holds of course for $\varepsilon_{\delta_o}$ at $t = t_o$. Moreover, eqs 4,70 show that $\eta_o$ can in principle take also negative values, which prospects also the possibility of a repulsive Casimir force and negative masses of the three particles introduced above; these latter three admit therefore the existence of three corresponding antiparticles as well. The next considerations exploit $v_{oc}$ to estimate the value of the constant $\zeta^u k$ that defines the Casimir force in eq 4,65 and to propose a reasonable conclusion about the nature of the three particles and respective antiparticles just introduced in eq 4,68. Exploit to this purpose the first eq 4,57 and second eq 4,66 that yield $\eta_o / \eta_{\min}^u = (2\pi)^{-1} \zeta^u k (c/\Delta r_o v_{oc})^4$. As $v_{oc}$ is a function of time, it is reasonable to write in general $c/\Delta r_o v_{oc} = 1 + g$ with $g = g(t)$ whatever $\Delta r_o$ might be; this suggests that $\Delta r_o$ is anyway related to the wavelength corresponding to $v_{oc}$. Also, noting that in eqs 4,47 and 4,48 both $\eta_{\min}^u$ and $\eta_{\max}^u$ are proportional to $\alpha^{-1}$, assume here $\eta_{\min}^u = \eta_o \alpha^{-1}$ in order that the reciprocal fine structure constant still appears in this expression of vacuum energy density. So

$$\eta_{\min}^u = \eta_o \alpha^{-1} \qquad \zeta^u k = (1+g)^{-4} 2\pi\alpha \qquad g = g(t) \qquad 2\pi\alpha = 0.046 \qquad 4,71$$

Let us explicate $\varepsilon_o$ noting that $\eta_{\min}^u = \eta_o \alpha^{-1}$ has the form $h v_{oc} (\zeta^u)^{-1} (c/v_{oc})^{-3} = \varepsilon_o (\alpha \zeta^u)^{-1} \Delta r_o^{-3}$; trivial manipulations with the help of the first and second eqs 4,67 yield then

$$\Delta r_o = \left(\frac{\varepsilon_o}{\alpha h v_{oc}}\right)^{1/3} \frac{c}{v_{oc}} \qquad m_o c^2 = \left(\frac{2\pi\alpha}{\zeta^u k}\right)^{1/4} q h v_{oc} \qquad m_o^u c^2 = \left((\zeta^u k / 2\pi\alpha)^{3/4} \alpha + 1\right) h v_{oc} \qquad 4,72$$

The last equation has been inferred from eq 4,68. To clarify the physical meaning of eqs 4,71 and 4,72, consider a further way to calculate $\zeta^u k$ based on the fact that any value $\eta_*^u$ between $\eta_{\min}^u$ and $\eta_{\max}^u$ of eqs 4,57 is physically allowed and given by $\eta_*^u = \eta_{\max}^u V_* / V_{\max}^u$ in the vacuum volume $V_{\max}^u \geq V_* \geq V_{\min}^u$; thus $\eta_{\max}^u V_* = \varepsilon_* = n_* h v_{oc}$ with $1 \leq n_* \leq N^3$ represents the amount of energy existing in a region $V_*$ of space-time defined by the number $n_* = V_* / V_{\min}^u$ of elementary volume elements. This reasoning is in fact nothing else but that previously exposed to introduce the vacuum energy fluctuation: the energy density gradient responsible of the Casimir effect is just the ground energy $h v_{oc}$ excited at $t = t_o$ to a higher value $n h v_{oc}$ in an initial cluster of $n_*$ elementary volumes $V_{\min}^u$ with total size $V_o$ and then spreading towards an increasing number of neighbouring space-time elementary volumes at $t > t_o$, whence $\varepsilon_o^u = n h v_{oc}$. Therefore eqs 4,71 and 4,72 describe through $g$ and $v_{oc}$ the space-time profile evolution of the fluctuation energy density, thus showing the link between the Casimir effect and the vacuum energy density of eqs 4,57. At $t = t_o$ one expects $g(t_o) = 0$, being the fluctuation energy confined within $V_o$, and thus $g = g(t - t_o)$. So at any $t$ such that $g \ll 1$ one finds $\zeta^u k \approx 0.046$; in effect this value agrees significantly with the known value $\pi^2 / 240 = 0.041$ of the Casimir effect coefficient putting once again $\zeta^u = 1$, as it is natural in the case of parallel plates $\Delta r$ apart in the vacuum. To confirm this result put in eqs 4,67 $\Delta r_o = (1+g)^{-1} c / v_{oc} = w' \lambda_o^0$ and $\varepsilon_o = w \eta_{max}^u V_*$, which is possible through appropriate proportionality factors $w$ and $w'$. So

$$m_o c^2 = (1+g) q h v_{oc} \qquad m_o c = (q/w') p_o^0 \qquad \zeta^u k = 2\pi w w' \eta_{max}^u V_* / p_o^0 c \qquad 4,73$$

If $2\pi w' w = 1$, then the third equation is particularly interesting: the vacuum energy $\eta_{max}^u V_*$ results proportional to the field energy $p_o^0 c$, the proportionality factor being just the coefficient of the Casimir effect $\zeta^u k$. Calculate $V_*$ specifying purposely $V_G$ of eq 4,42 with the position $v_G = v_{oc}$ suggested by the fact that $\eta_{max}^u$ is defined by $v_{oc}$. As concerns $m_G$, it must agree with the idea that the matter perturbs the local properties of the vacuum in between the plates; also, $V_*$ must account for the presence of matter with a value of $m_G$ such that $\zeta^u k$ results comparable with $2\pi\alpha$. Putting $m_G = m_p$ in eq 4,42, so that

$$V_* = \frac{G m_p}{v_{oc}^2}$$

eqs 4,57 and 4,59 yield

$$\zeta^u k = \frac{\eta_{max}^u V_*}{p_o^0 c} = \frac{N^3 h v_{oc}^2 G m_p}{p_o^0 c^4} \qquad 4,74$$

To calculate this equation it is necessary to guess the value of $N$; if

$$N = 6.02214199 \cdot 10^{23} \qquad 4,75$$

then the results provided by eq 4,74 are

$$\zeta^u k = 0.034 \qquad \eta_{max}^u = 3.2 \cdot 10^{61} \text{ J m}^{-3} \qquad \eta_{max}^u V_* = 2.1 \text{ J} \qquad V_* = 6.6 \cdot 10^{-62} \text{ m}^3 \qquad 4,76$$

As expected, just the proton mass expresses the perturbation of the vacuum state. The fact that $\zeta^u k$ calculated in eq 4,74 is slightly different from $2\pi\alpha$ is not surprising in principle: the former is the current value at the time that defines $v_{oc}$, the latter is a reference fixed value for $\zeta^u k$, likewise as $V_\lambda$ of eq 4,52 was a reference fixed value for the time dependent $V_{min}$ quoted in eq 4,54. For this reason the value of $v_{oc}$ of eq 4,61 has been introduced without requiring that $\zeta^u k$ matches $2\pi\alpha$ in eq 4,74, although the minor correction sufficient to this purpose would not have modified appreciably the reasonableness of eqs 4,56 and 4,62; rather, it appears significant that at the time where $v_{oc}$ is given by eq 4,61 hold both $V_{min} \approx V_\lambda$ and $\zeta^u k \approx 2\pi\alpha$. Note that combining eqs 4,74 and 4,6 one finds

$$p_o^0 = \sqrt{\frac{\eta_{max}^u m_p \hbar c}{v_{oc}^2 m^u \zeta^u k}} = 1.2 \cdot 10^{-7} \text{ Kg m s}^{-1} \qquad G = \sqrt{\frac{\hbar c^3 v_{oc}^2 \zeta^u k}{m^u \eta_{max}^u m_p}} = 7.7 \cdot 10^{-11} \text{ m}^3 \text{ Kg}^{-1} \text{ s}^{-2}$$

These equations calculate therefore $G$ and $p_o^0$ through the Casimir coefficient $\zeta^u k$, for simplicity put approximately equal to $2\pi\alpha$; the present results compare quite well with the values of eqs 4,40 although calculated in a completely different way. With the positions above and taking into account that in fact the physical properties of the universe change as a function of time, the energy corresponding to $\eta_{min}^u$ yields

$$m_{oc} c^2 = h v_{oc} = 1.1 \cdot 10^{-21} \text{ J} = 6.4 \cdot 10^{-3} \text{ eV} \qquad 4,77$$

A preliminary inspection of eqs 4,72 reveals that $m_{oc}$ and $m_o^u$ are very similar: the latter differs from the former by a multiplicative factor of the order of $1+\alpha$, i.e. $m_o^u c^2 = 6.5 \cdot 10^{-3}$ eV. In fact this estimate is true without approximation at $t = t_o$, at which is in effect defined $\eta_o$; at this time $\zeta^u k = 2\pi\alpha$ and $\Delta r_o = c / v_{oc}$, whereas eqs 4,72 read

$$m_o c^2 = qh\nu_{oc} \qquad m_{oc} c^2 = h\nu_{oc} \qquad m_o^u c^2 = (\alpha+1) h\nu_{oc} \qquad t = t_o \qquad 4,78$$

These equations describe what happens as soon as the quantum fluctuation triggers the energy gap in $V_o$. Note however in this respect that: (i) the mass $m_o$ is related to the vacuum energy and to the field momentum $p_o^0$ of eq 4,40, not to its fluctuation; (ii) the masses $m_{oc}$ and $m_o^u$ are defined in a reference system $R$ where they are at rest, as at the left hand sides of the second and third equation appear indeed their rest energies; (iii) the same holds for $m_o$ in eq 4,68; (iv) the first equation 4,78 differs from the others because of the factor $q$, absent in eq 4,68 and still to be defined. These remarks suggest that eq 4,68 and eqs 4,78 are defined in two different reference systems. In effect, being $m_o$ related to the field energy $p_o^0 c$, it is presumably referred to the reference system $R^u$ where is at rest $m^u$ defining $p_o^0$; instead $m_{oc}$ and $m_o^u$ are defined in $R$ displacing with respect to $R^u$ at rate v. So $m_{oc}$ and $m_o^u$ at rest in $R$ move in general in $R^u$ with velocity v assumed for simplicity constant and equal for both of them. The fact that $q$ appears in the first equation only is understood as the correction factor taking into account the reciprocal motion of $R$ and $R^u$: if $q = (1 - v^2/c^2)^{1/2}$, then $m_o$ moves in $R$ with kinetic energy $m_o' c^2 = m_o c^2 (1 - v^2/c^2)^{-1/2}$, so that $m_o c^2 = qh\nu_{oc}$ must be actually regarded as $m_o' c^2 = h\nu_{oc}$; owing to its small mass in eq 4,77, $m_o$ shoud move at velocity $v \approx c$, which means a relevant relativistic correction. In $R^u$ the particles $m_{oc}$ and $m_o^u$ have kinetic energies $m_{oc}' c^2 = m_{oc} c^2 (1 - v^2/c^2)^{-1/2}$ and $m_o'^u c^2 = m_o^u c^2 (1 - v^2/c^2)^{-1/2}$, as if their rest masses would be $m_{oc}'$ and $m_o'^u$; therefore for an observer in $R^u$

$$m_o c^2 = h\nu_{oc} \qquad m_{oc}' c^2 = h\nu_{oc}' \qquad m_o'^u c^2 = (\alpha+1) h\nu_{oc}' \qquad \nu_{oc}' = \frac{\nu_{oc}}{\sqrt{1 - v^2/c^2}}$$

$$t = t_o$$

For purposes of comparison with experimental data, it is interesting to estimate the masses $m_o$, $m_{oc}'$ and $m_o'^u$ in $R^u$ through differences between couples of masses squared; with $v \approx 0.994c$ one finds

$$q^{-1} = 9 \qquad \begin{cases} (m_{oc}' c^2)^2 - (m_o c^2)^2 = 3.3 \cdot 10^{-3} & eV^2 \\ (m_o'^u c^2)^2 - (m_{oc}' c^2)^2 = 4.9 \cdot 10^{-5} & eV^2 \\ (m_o'^u c^2)^2 - (m_o c^2)^2 = 3.4 \cdot 10^{-3} & eV^2 \end{cases}$$

Experimental values of differences between squared masses of electron, muon and tauon neutrinos are reported in literature: $\Delta m_{21}^2 = 7.9 \cdot 10^{-5}$ $eV^2$ [20] and $|\Delta m_{32}^2| = 2.7 \cdot 10^{-3}$ $eV^2$ [21]. Considering the difficulty of collecting experimental data on neutrino masses, the calculated values appear quite satisfactory to validate the present results. Moreover, it is significant the possibility of explaining why experimental data on the third difference are not explicitly quoted in literature: the third difference is in fact undistinguishable from the first one. The last problem, so far left behind, concerns the charge $e$ also inherent the definition of $G$. After having introduced the proton mass, the electroneutrality balance of the total virtual charge per unit volume of the vacuum suggests considering also the electrons (let us remark however that the present statement would also hold for anti-protons and positrons as well). In practice, from the mass point of view, the results previously inferred considering protons only do not change; with equal number of protons and electrons, the mass of these latter can be neglected in the previous order of magnitude estimates. Is it however possible to acknowledge any signature of the electrons in the universe? Considering that the evolution of the temperature in the early stages of life of the universe allowed to form hydrogen atoms plus protons and electrons plus progressively increasing amounts of heavy elements, it is difficult to guess what kind of physical mechanism specifically stimulates the cosmic background radiation. It is however possible to bypass complicated speculations about such a mechanism, whatever it might be, starting from the vacuum energy density of eq 4,57 with the help of eqs 4,59. The fact that $v_c = N v_{oc}$ with $N$ integer suggests that the quantized vacuum vibration frequencies should admit also their zero point energy $h v_{oc}/2$, which of course is in the volume $\zeta^u (c/v_{oc})^3$ of the vacuum defined by the propagation rate $c$ and frequencies $v_{oc}$. If in the unit volume of the vacuum there is on average one virtual proton and one virtual electron, then one could think that even in $\zeta^u (c/v_{oc})^3$ there should be an equal amount of both particles: indeed $m_e c^2 = 8.2 \cdot 10^{-14}$ J per unit volume compares with $\eta^u_{min}/1836.15 = 9.3 \cdot 10^{-14}$ J m$^{-3}$ quite well; both values would match even better increasing $m_e c^2$, i.e. admitting that also the electron, like the proton, is not at rest but has kinetic energy as well. In any case, even without complicating the present discussion, consider $\eta^u_{min} m_e / (m_p + m_e)$ and $\eta^u_{min} m_p / (m_p + m_e)$ for the respective energy density related outcomes and define therefore the following quantity concerning in particular the zero point energy density of virtual electron oscillators

$$\frac{1}{2} \frac{\eta^u_{min} m_e}{m_p + m_e}$$

This expression is the key to evidence the presence of virtual electrons together with that of virtual protons. After having described the properties of the universe through the proton mass only, it is possible to ask if this residual energy is also measurable. It is known that the spectral energy density of a black body is obtained integrating $8\pi c^{-3} h \int (\exp(hv/k_b T) - 1)^{-1} v^3 dv$ over all the frequencies. Having found the consequences of one proton per unit volume of the vacuum, let us consider therefore also the presence of one electron per unit volume; so one concludes

$$\frac{\eta^u_{min}/2}{1836.15} = \frac{4\sigma}{c} T^4 \qquad \sigma = 5.67 \cdot 10^{-8} \text{ W m}^{-2} \text{ T}^{-4}$$

In effect the calculation yields $T = 2.69$ K; this shows that electrons in the zero point state radiate the energy measured as a background cosmic radiation. Note that is crucial the value of $v_{oc}$ defining $\eta^u_{min}$ to find this agreement with the temperature, which results therefore strongly linked just through $v_{oc}$ to all of the properties of the universe hitherto described; this could have been indeed a good reason to introduce just the value of eq 4,61. Also, the result just obtained suggests a possible hypothesis that highlights further the physical essence of the invariance of $c$. Whether regarded as Einstein's postulate or corollary of the uncertainty in the present conceptual frame, this invariance must be accepted as such without need of further explanation. Besides the mathematical worth of velocity sum rule, however, a physical hint comes in this respect just from the existence of the virtual electron oscillators as a property of the vacuum, whose zero point energy density appears to us as background cosmic radiation. If so, however, not only the zero point frequency $v_o^u$ but also the excited energy levels of these virtual oscillators should be effective as concerns the radiation emission. Imagine now a free atom travelling at velocity v in the reference system $R^u$ excited to a higher energy level at time $t_o$ when it was in the arbitrary point $x_o$ of the vacuum. After a proper time range $t - t_o$, the atom decays to the ground energy level when it is at the point $x$ defined by $x - x_o = (t - t_o)v$. The lack of any information about $x$ and $x_o$ excludes the deterministic knowledge about $t_o$ and $t$, as it is known. One expects therefore that at the time $t$, whatever it might be, the decay process generates one photon of frequency $v_{\Delta \varepsilon}$. In addition to this usual point of view, however, the previous considerations also suggest the chance that the atom actually does not emit any photon but interacts with the virtual electron oscillators of the vacuum, which are therefore excited to a higher virtual energy level and return afterwards to their ground virtual state; if so the vacuum, not the atom, emits the photon that of course propagates with velocity $c$ from the point $x$ where occurred the energy exchange. This emission mechanism does not need any addition rule between $c$ and $v$ since, as assumed here, the photon has nothing to do with the flight velocity of the decaying atom; rather, the velocity addition rule takes its appropriate role of implementing such a physical idea. This hypothesis does not conflict with the fact that in general $v_o^u \neq v_{\Delta\varepsilon}$. Indeed $v_o^u$ is a property of the vacuum in the absence of matter, whereas $v_{\Delta\varepsilon}$ is a local property of the vacuum in the presence of matter; as emphasized before, the latter affects the properties of the former, for instance turning $\zeta^u$ into $\zeta$ and $\lambda^u$ into $\lambda$ and so on. Likewise, the effect of replacing $m^u$ with $m$ could be just that of shifting the characteristic frequency $v_o^u$ of the vacuum to that $v_{\Delta\varepsilon}$ characteristic of the matter. This radiation mechanism entails some time delay between electron decay and photon emission: could be such a delay somehow experimentally detectable, for instance comparing the light emission rate inside and outside the plates enabling the Casimir effect? A closing remark concerns the equations 4,50 to 4,56 that deal with the presence of matter in the universe. Consider again these equations but replacing $m$ with $-m$, as allowed by eq 4,49. Eqs 4,53, 4,54 and 4,55 remain unchanged if $\Delta\tau_o$ is replaced by $-\Delta\tau_o$; so, an universe of antimatter would expand like that of matter. While the signs of $m\Delta\tau_o$ and $\Delta\tau_o/m$ remain unchanged, that of $V_{min}$ of eq 4,51 and $V_\lambda$ of eq 4,52 changes because of $\Delta\tau_o$ and $m$ respectively; the same holds for $\Lambda_{un}$ of eq 4,55. Hence space and time coordinates must be regarded with different sign in an universe of antimatter, whereas the densities $d_{max}$ and $d_\lambda$ of eqs 4,51 and 4,52 maintain the sign unchanged; the same holds for the energy densities as well, see for example $\eta_{max}$ or $\eta_{max}^\lambda$ of eqs 4,51 and 4,52. The four eqs 4,56 are consistent with these remarks; the sign change of $M_p^{tot}$, now to be identified with the total mass of antiprotons, agrees with the sign change of $\Lambda_{un}^3$, whereas the total number of antiprotons $n_p^{tot}$ maintains of course its positive value like $d_{un}$ and $\eta_{min}$. In conclusion, both signs of $m$ and $e$ in eqs 4,49 and 4,43 seem admissible without problems of logical consistency in describing the respective universes. Strictly speaking, different signs of $\Delta\tau_o$ and $\Lambda_{un}$ do not raise any conceptual problem; also the latter is indeed a whole spacetime range size where hold the physical properties so far described. Even the ordinary concept of volume, whether we think to the quantum elementary box $V_{min}$ or to the whole universe volume $\Lambda_{un}^3$, is related to nothing else but the product of three range sizes; for the moment we discard here possible extra dimensions, which would require regarding separately the chances of even or odd numbers of these extra dimensions. Nothing compels in principle to accept an uncertainty range size with positive sign only: any consideration about the random, unknown and unpredictable coordinate $x$ within $\Delta x = x - x_o$ would identically hold regarding this local coordinate in $\Delta x = x_o - x$, i.e. even changing sign to $\Delta x$ the delocalization range is exactly the same. Exchanging the role of $x$ and $x_o$ is surely possible because both are arbitrary, unknown and unpredictable by fundamental assumption and no physical property discriminates either range boundary; also, as shown in section 2, no eigenvalue depends explicitly on how are defined the signs of the uncertainty ranges and in particular their boundaries. Therefore any reasoning about $\Lambda_{un}$ has in principle the same worth and physical implication as that about $-\Lambda_{un}$, in agreement with the considerations already emphasized for $\Delta\tau_o$. This is also evident recalling the basic equations of the uncertainty 1,2 and 2,10 on which rests entirely the present paper: changing sign of $\Delta x$ requires changing that of $\Delta p$ and thus that of $m$, whereas changing sign of $\Delta t$ requires changing that of $\Delta\varepsilon$ in agreement with the physical dimensions of the energy $ml^2t^{-2}$. So it seems that the existence of an universe of matter or antimatter is simply related to the sign reversal of the time range within which evolves the time arrow: the definition $\Delta\tau_o = t_1 - t_o$ means that $t$ evolves from $t_o$ towards $t_1$; exchanging the role of $t_1$ and $t_o$ means that $t$ evolves from $t_1$ towards $t_o$. Of course it is unphysical the attempt to speculate which time arrow describes the universe and which the antiuniverse; it is only possible to say that either of them is related to either way of defining the signs of the space and time ranges. Of course this conclusion does not prevent the existence of limited islands of antimatter in a universe of matter, in agreement not only with the chance of local energy fluctuations but also with the lack of deterministic information about local coordinates and related phenomena therein allowed to occur.

A closing remarks about the values of eqs 4,55 and 4,56 deserves attention. Replacing $M$ of eq 4,31 with $M_p^{tot}$ one finds $r_{Schw}^{M_p^{tot}} = 5.5 \cdot 10^{26}$ m, i.e. just the limit size $\Lambda_{un}$ of the universe. At first sight this numerical coincidence seems merely accidental, although a physical explanation of this result is easily found; the mass $M_p^{tot}$ links indeed the light confinement within $\Lambda_{un}$ and within $r_{Schw}^{M_p^{tot}}$. The former has been introduced as the limit distance travelled by a light beam originated anywhere and propagating elsewhere in the universe, so that by definition no light escapes beyond this size; the latter is the distance that also prevents photons moving inside it to cross the event horizon and escape outwards. The only way to discriminate between numerical fortuity and physical link is to examine a further consequence of this explanation. Imagine a light beam that starts from any point of the universe, assumed here spherical for simplicity, and points towards its boundary. The impossibility of the beam to escape outside $\Lambda_{un}$ does not prevent however its propagation on the boundary surface through a mechanism similar to that described in subsection *4.5*; accordingly, after having assumed that the light defines the radial size of the universe, this statement simply emphasizes that the boundary layer is outlined itself by the curved path of a light wave trapped by the total mass $M_p^{tot}$. Define thus the circular frequency of this light wave as $\omega_{un} = c/(2\pi \Lambda_{un}/2)$; if so the boundary layer of the universe vibrates with the frequency $\omega_{un}$ of the electromagnetic field surrounding the mass $M_p^{tot}$. Consider now the zero point energy $\varepsilon_{un}^{zp} = \hbar \omega_{un}/2 = \hbar c/(2\pi \Lambda_{un})$ of this vibration; having found that $\Lambda_{un} = r_{Schw}^{M_p^{tot}}$, eq 4,31 yields $\varepsilon_{un}^{zp} = hc^3/\left((2\pi)^2 2 M_p^{tot} G\right)$. Moreover it is also natural to introduce the total energy $\varepsilon_{un}^{M_p^{tot}} = M_p^{tot} c^2$ of the universe due to the whole amount of matter present in it. Try now to relate $\varepsilon_{un}^{M_p^{tot}}$ and $\varepsilon_{un}^{zp}$ through a dimensionless proportionality factor $\sigma$, i.e. $\varepsilon_{un}^{M_p^{tot}} = \sigma \varepsilon_{un}^{zp}$. Eventually, recalling the Plank length $l_P$ introduced in eq 4,7, this position yields immediately

$$\sigma = \frac{\varepsilon_{un}^{M_p^{tot}}}{\varepsilon_{un}^{zp}} = \frac{A}{4 l_P^2} \qquad l_P = \sqrt{\frac{\hbar G}{c^3}} \qquad A = 4\pi \left(\frac{2 G M_p^{tot}}{c^2}\right)^2$$

By definition $\sigma$ is the ratio of two energies, a bulk mass energy and a boundary zero point oscillation. To explain this result consider first an harmonic oscillator and note that the ratio between its $n$-th vibrational energy and its zero point energy is $2n$. Since in the present approach the vibrational quantum number $n$ is actually the number of allowed vibrational states, it follows that even the ratio $\varepsilon^{vib}/\varepsilon^{zp}$ should be somehow related itself to the number of allowed states of oscillation. This is enough to regard by analogy also the ratio $\varepsilon_{un}^{M_p^{tot}}/\varepsilon_{un}^{zp}$ in a similar way, which should be related to an appropriate number of quantum states allowed for the vibration of the boundary layer of the universe. So this ratio must be somehow related to a surface entropy. In effect, $\sigma$ coincides just with the Hawking surface entropy in Boltzmann's units. The big bang energy balances the tendency of matter to collapse on the centre, as it was at the beginning of time.

*4.10 The MOND approach.*

The previous subsection has exploited the concept of unit mass $m^u$ introduced in eq 4,5. It is instructive to examine further on the physical meaning of $m^u$ through simple non-relativistic considerations. Let us replace one of the masses of eq 4,6, say $m_2$, with $m^u$. The particular case of Newtonian interaction of the real mass $m_1$ with the vacuum virtual mass $m^u$ yields

$$F_N^u = -\frac{Gm^*}{\Delta x_o^2} M^* \qquad M^* = M^u \left(\frac{\Delta x_o}{\Delta x}\right)^2 \qquad M^u = m_1 + m^u \qquad m^* = m_1 m^u / (m_1 + m^u) \qquad 4,79$$

The physical meaning of $\Delta x_o$ is highlighted considering that this interaction relates $F_N^u$ to the function $\mu$ defined as follows

$$F_N^u = -a^u \mu M^* \qquad \mu = \mu(m_1 / m^u) = \frac{m_1 / m^u}{m_1 / m^u + 1} \qquad a^u = \frac{Gm^u}{\Delta x_o^2} \qquad 4,80$$

Thus $\Delta x_o$ is the range including any distance around the virtual particle $m^u$ at which is calculated the local acceleration $a^u$. The limit values of $F_N^u$ calculated at $m_1 \gg m^u$ and $m_1 \ll m^u$ are

$$-F_N^u = \begin{cases} a^u M^* & \text{for } m_1 \gg m^u \\ a^u M^* m_1 / m^u & \text{for } m_1 \ll m^u \end{cases} \qquad 4,81$$

Consider also now what happens replacing $m^u$ with a real mass $m$: clearly $m_1$ interacts with $m$ while $a^u$ is replaced by $a = (m/m^u) a^u$, with $a$ calculated at the same distance $\Delta x_o$ from $m_1$. So

$$\mu(m_1 / m^u) \to \mu(a / a^u) = \frac{(m_1 / m) a / a^u}{(m_1 / m) a / a^u + 1} \qquad a = \frac{Gm}{\Delta x_o^2}$$

whereas eq 4,81 turns into

$$-F_N' = \begin{cases} M' a & \text{for } (m_1 / m) a / a^u \gg 1 \qquad M' = m_1 + m \\ M'' a^2 / a^u & \text{for } (m_1 / m) a / a^u \ll 1 \qquad M'' = M'(m_1 / m) \end{cases}$$

Being both $m_1$ and $m$ arbitrary, in general $M'$ and $M''$ are arbitrary as well. One recognizes in this result the basis of the so called MOND (MOdified Newtonian Dynamics) approach, introduced since 1959 to explain the galaxy rotation problem [22]. The implications of the given inequalities are well known in literature and for brevity will not be further commented here. The order of magnitude of $a^u$ is estimated noting that must hold for the velocity of $m^u$, whatever its initial value might be, the obvious condition $a^u \delta t < c$ after any time range $\delta t$. This inequality must be true even for a time range equal to the lifetime of the universe, to ensure that the light speed limit is never violated. Putting therefore $\delta t \equiv \Delta \tau_0$, one finds $a^u < 5 \cdot 10^{-10}$ m/s$^2$ i.e. $\Delta x_o > 0.35$ m. Note that the last eq 4,80 reads $m^u a^u \Delta x_o = Gm^{u2} / \Delta x_o$; the energy at left hand side corresponds to the displacement of $m^u$ by a length $\Delta x_o$ with constant acceleration $a^u$, the right hand side shows that this energy can be nothing else but that of the vacuum. So one expects that $Gm^{u2} / \Delta x_o = \eta_{\min}^u \Delta x_o^3$ and then $\Delta x_o = (Gm^{u2} / \eta_{\min}^u)^{1/4}$; this yields $\Delta x_o = 0.8$ m, in agreement with the inequality above, and thus $a^u = 1.1 \cdot 10^{-10}$ m/s$^2$. It is instructive to confirm this result comparing $m^u a^u \Delta x_o$ with the energy $m_p c^2$ of a proton at rest; this link defines through $\Delta x_o$ the energy balance necessary to replace the vacuum virtual mass $m^u$ with the proton mass. This way of thinking replicates here the ideas of the previous subsection that involve the proton mass according to the Dirac intuition. As the last eq 4,80 is one-dimensional by definition, let us write more appropriately this energy balance as $m_p c^2 = m^u a^u \sqrt{\Delta x_o^2 + \Delta y_o^2 + \Delta z_o^2}$; putting then for simplicity $\Delta x_o^2 \approx \Delta y_o^2 \approx \Delta z_o^2$, one finds $\sqrt{3} m^u a^u \Delta x_o \approx m_p c^2$. So, eliminating $\Delta x_o$ from the last eq 4,80 with the help of this last result, one finds $a^u = (m_p c^2)^2 (3Gm^{u3})^{-1}$ and then $\Delta x_o = 0.81$ m and $a^u = 1.1 \cdot 10^{-10}$ m/s$^2$. The link between these ways to calculate $a^u$ rests on the fact that $\eta_{min}$ correspond to the energy of one proton per unit volume of the space-time. The present value of $a^u$ agrees with the experimental result $1.2 \cdot 10^{-10}$ m/s$^2$ reported in literature for the constant acceleration $a_0$ of the MOND approach [23,24]. Note that this experimental value is very simply inferred considering that $G\eta_{min}$ has physical dimensions of squared acceleration; in effect, the last eq 4,56 yields $\sqrt{G\eta_{min}} = 1.2 \cdot 10^{-10}$ m s$^{-2}$. Thus one could calculate $G$ as a function of $a^u$ exploiting once again the frequency $v_{oc}$ of eq 4,61. Is significant the derivation of the cosmological quantity $a^u$ in the frame of a unique theoretical model including both quantum mechanics and relativity. Let us show a further consequence of the interaction of a real particle of mass $m$ with the virtual mass $m^u$. In agreement with the first eq 4,79, rewrite eq 4,6 as $F_N^u/m^u = Gm/\Delta x_{12}^2$ and consider this equation at the Plank scale of mass and length; the right hand side reads thus $c^4(G\hbar c)^{-1/2}$. The left hand side has physical dimensions of an acceleration, so it can be written as $\phi\lambda^u v^{u2}$ being $\phi$ a proportionality constant and $v^u$ a frequency; $\lambda^u$ is the Compton length of $m^u$ already introduced in eq 4,45. As eq 4,6 defines $G$ through $m^u$ and $\omega_o^0$ only, the physical properties of the vacuum should be the key ingredients enough to calculate $G$; in effect eq 4,6 has been rewritten here just according to this idea, which suggests therefore $2\pi v^u = \omega_o^0$ as well. Hence $\phi\lambda^u v^{u2} = c^4/\sqrt{\hbar c G}$ yields $G = c(\hbar c G)^{1/4}(\phi\lambda^u)^{1/2}(2\pi m^u)^{-1}$. This equation calculates $G$ if $\phi$ is appropriately defined, plausibly through a suitable fundamental constant of nature. The fact that the sought constant must be dimensionless suggests $\phi = N$ or $\phi = \alpha$: the latter chance seems however unreasonable, because in eq 4,53 $G$ was found proportional to $\alpha^{-1}$. Putting $\phi = N$ one finds $G = 6.62 \cdot 10^{-11}$ m$^3$ Kg$^{-1}$ s$^{-2}$.

5. The spinless free particle.
This topic is well known and widely reported in literature; it is concerned here to better outline and complete the theoretical quantum frame within which have been found the relativistic results of sections 3 and 4. The appendixes A and B ensure the conceptual link between the present approach, based uniquely on the uncertainty positions 1,1 through eqs 1,2 and 2,10, and the usual quantum approach based on operator formalism of wave mechanics. Therefore it is reasonable to expect that even the weird aspects of the quantum world, e.g. the wave/particle duality, can be included in the unique very general context where are also rooted the relativistic appearances of nature, e.g. the light beam bending. This is the aim of the present section: the concepts exposed hereafter are already known; yet the non-trivial fact deserving attention is that they are inferred through the same conceptual scheme so far followed without distinction between macro or quantum scale of the physical phenomena. Let us exploit to this purpose, with the help of eqs A7 and B2, the wave equation corresponding to eqs 2,16. The following discussion concerns in particular the spinless free particle for the reasons already stated in section 4. With the position

$$\psi = \psi(x,t) = W\sqrt{\Pi_x}\sqrt{\Pi_t} \qquad 5,1$$

where $W$ is an arbitrary constant, the resulting equation is

$$\frac{\partial^2 \psi}{\partial x^2} - \frac{1}{c^2}\frac{\partial^2 \psi}{\partial t^2} - \left(\frac{mc}{\hbar}\right)^2 \psi = 0 \qquad 5,2$$

Eq 5,2 is the 1D Lorentz covariant Klein-Gordon equation. Let the solution $\psi$ of eq 5,2 be defined in order to fulfil eqs A11 and B4 too; then

$$\psi = a(x,t)\exp[iS(x,t)/\hbar \pm i\omega mc^2 t/\hbar\omega_o] \qquad 5,3$$

where $S(x,t)$ is real, $a(x,t)$ is complex, $\omega$ and $\omega_o$ are real constants. Eqs B6 and B7 require

$$S(x,t) = g_\varepsilon \varepsilon \delta t + g_p p \delta x \qquad 5,4$$

The coefficients $g_\varepsilon$ and $g_p$ must be determined. Replacing $\psi$ into eq 5,2 yields

$$\pm\frac{i}{\hbar}\left\{\left[\frac{\partial^2 S}{\partial x^2}-\frac{1}{c^2}\frac{\partial^2 S}{\partial t^2}\right]a+2\left[\frac{\partial a}{\partial x}\frac{\partial S}{\partial x}-\left(\frac{m\omega}{\omega_o}+\frac{1}{c^2}\frac{\partial S}{\partial t}\right)\frac{\partial a}{\partial t}\right]\right\}+ \quad\quad 5,5$$

$$+\frac{\partial^2 a}{\partial x^2}-\frac{1}{c^2}\frac{\partial^2 a}{\partial t^2}+\left[2\frac{m\omega}{\omega_o}\frac{\partial S}{\partial t}+\frac{1}{c^2}\left(\frac{\partial S}{\partial t}\right)^2-\left(\frac{\partial S}{\partial x}\right)^2+\left(\frac{\omega^2}{\omega_o^2}-1\right)m^2c^2\right]a=0$$

Let us split now this equation as follows for reasons that will be clear soon

$$-a\frac{\partial^2 S}{\partial x^2}-2\frac{\partial a}{\partial x}\frac{\partial S}{\partial x}+\left(a\frac{\partial^2 S}{\partial t^2}+2\frac{\partial a}{\partial t}\frac{\partial S}{\partial t}\right)\frac{1}{c^2}\pm 2a\frac{im\omega}{\hbar\omega_o}\frac{\partial S}{\partial t}=0 \quad\quad 5,6a$$

$$\frac{\partial^2 a}{\partial x^2}\mp\frac{2mi\omega}{\hbar\omega_o}\frac{\partial a}{\partial t}-\frac{1}{c^2}\frac{\partial^2 a}{\partial t^2}+\left[\left(\frac{mc}{\hbar}\right)^2\left(\frac{\omega^2}{\omega_o^2}-1\right)-\left(\frac{1}{\hbar}\frac{\partial S}{\partial x}\right)^2+\left(\frac{1}{\hbar c}\frac{\partial S}{\partial t}\right)^2\right]a=0 \quad\quad 5,6b$$

Multiply eq 5,6a by $[a(x,t)]^*$ and its conjugate by $a(x,t)$; summing the resulting equations, one finds the following condition for $a^*a$

$$-\frac{\partial}{\partial x}\left(aa^*\frac{\partial S}{\partial x}\right)+\frac{1}{c^2}\frac{\partial}{\partial t}\left(aa^*\frac{\partial S}{\partial t}\right)=0 \quad\quad 5,7$$

With the help of eqs 5,4, eq 5,7 reads $-g_p p\partial(aa^*)/\partial x+g_\varepsilon\varepsilon/c^2\partial(aa^*)/\partial t=0$ and, owing to eq 2,14, yields $-g_p v\partial(aa^*)/\partial x+g_\varepsilon\partial(aa^*)/\partial t=0$. Put now $g_\varepsilon=-g_p$ for reasons that will be clear later, although an intuitive consideration is possible even now recalling that v of a free particle is constant: the positions $v\partial_x=\partial_\tau$ and $\partial_t=v^\S\partial_{x^\S}$ yield $-g_p\partial(aa^*)/\partial\tau+g_\varepsilon v^\S\partial(aa^*)/\partial x^\S=0$ that reads with the given conditions on the coefficients $g_\varepsilon\partial(aa^*)/\partial\tau-g_p v^\S\partial(aa^*)/\partial x^\S=0$; being $t$, v and $x$ completely arbitrary like $\tau$ and $v^\S$, this result is identical to the initial equation, which is then symmetrical with respect to the exchange of $t$ with $x/v$ in agreement with the way to infer eq 2,10 from eq 1,2. With these conditions, $aa^*$ fulfils the equation $v\partial(aa^*)/\partial x+\partial(aa^*)/\partial t=0$ and

$$S(x,t)=\pm g(\varepsilon\delta t-p\delta x) \quad p=\pm\frac{\partial S}{\partial x} \quad -\varepsilon=\pm\frac{\partial S}{\partial t} \quad g=|g_p|=|g_\varepsilon| \quad g_p=-g_\varepsilon \quad 5,8$$

Replacing eqs 5,8 into eq 5,6b yields $(\partial S/\partial x)^2-c^{-2}(\partial S/\partial t)^2+(mc)^2=0$, while eq 5,6b reduces to

$$\frac{\partial^2 a}{\partial x^2}\mp\frac{2mi\omega}{\hbar\omega_o}\frac{\partial a}{\partial t}-\frac{1}{c^2}\frac{\partial^2 a}{\partial t^2}+\left(\frac{mc}{\hbar}\right)^2\frac{\omega^2}{\omega_o^2}a=0 \quad\quad 5,9$$

Considering the three space coordinates, the first result reads

$$\left(\frac{\partial S}{\partial x}\right)^2+\left(\frac{\partial S}{\partial y}\right)^2+\left(\frac{\partial S}{\partial z}\right)^2-\left(\frac{\partial S}{\partial ct}\right)^2+(mc)^2=0 \quad\quad 5,10$$

So, eq 5,10 is the relativistic Hamilton-Jacobi equation and $S(x,t)$ is the action of the free particle. This result suggests in turn that $S$ of eq 5,4 should be invariant with the conditions 5,8. Let $S$ be defined in a reference system $R$ where $S=\pm g\delta t(\varepsilon-pV)$ with $V=\delta x/\delta t$. Replace eq 2,14 in eq 5,8 putting $v=V$, so that $S'=\pm g\varepsilon'\delta t'(1-V^2/c^2)$: this position introduces a new reference system $R'$ where the particle is at rest and $\varepsilon'$ is its energy defined by $\partial S'/\partial t'$, whence the notation. If $S$ is invariant, then it should be true that $S=S'$. Since $\delta t'=\delta t/\sqrt{1-V^2/c^2}$, as shown in section 3, the invariance of $S$ requires $\varepsilon'\sqrt{1-V^2/c^2}=\varepsilon-pV$, as in effect it is well known. This result supports eqs 5,8 that, replaced in eq 5,6a together with eq 2,14, yields

$$v\frac{\partial a}{\partial x} + \frac{\partial a}{\partial t} \pm a\frac{imc^2\omega}{\hbar\omega_o} = 0 \qquad 5,11$$

The solution of this equation is

$$a(x,t) = \left[f_1(\xi) + f_2(-\xi)\right]\exp(\mp i\chi x) \qquad \xi = \frac{x-vt}{v\tau} \qquad \chi = \frac{\omega}{v\omega_o}\frac{mc^2}{\hbar} \qquad 5,12$$

where $f_1$ and $f_2$ are two arbitrary functions of the argument and $\tau$ an arbitrary constant. Note that in this way $a(x,t)a^*(x,t)$, i.e. $\psi\psi^*$ according to eq 5,3, has the physical meaning of standing wave moving with velocity $\pm v$. This is not surprising even if $m \neq 0$: let us derive in fact eq 5,11 once with respect to $x$ and once with respect to $t$; subtracting the latter from the former multiplied by $v$ yields $v^2\partial^2 a/\partial x^2 - \partial^2 a/\partial t^2 \pm (v\partial a/\partial x - \partial a/\partial t)imc^2\omega/\hbar\omega_o = 0$, i.e. thanks to eq 5,11 itself

$$v^2\frac{\partial^2 a}{\partial x^2} - \frac{\partial^2 a}{\partial t^2} \mp \frac{2imc^2\omega}{\hbar\omega_o}\frac{\partial a}{\partial t} + \left(\frac{mc^2\omega}{\hbar\omega_o}\right)^2 a = 0 \qquad 5,13$$

This equation is actually equal to eq 5,9 rewritten as a function of $v$ instead of $c$, the rest energy $mc^2$ of the particle being however the same; it is still fulfilled by eq 5,12 and yields for $m=0$ the known D'Alembert wave equation

$$\frac{\partial^2 a}{\partial x^2} - \frac{1}{v^2}\frac{\partial^2 a}{\partial t^2} = 0 \qquad 5,14$$

Note that the same equation is also obtained putting $\omega_o \to \infty$ in eq 5,13 even though $m \neq 0$. This explains the wave nature of particles, regardless of whether $m \neq 0$ or $m = 0$. Further information on this concept is obtained from eq 5,6b, which reads now with the help of eqs 5,8 and 2,14

$$\frac{\partial^2 a}{\partial x^2} \mp \frac{2mi\omega}{\hbar\omega_o}\frac{\partial a}{\partial t} - \frac{1}{c^2}\frac{\partial^2 a}{\partial t^2} + \left[\left(\frac{mc}{\hbar}\right)^2\left(\frac{\omega^2}{\omega_o^2} - 1\right) + \left(\frac{\varepsilon}{\hbar c}\right)^2\left(1 - \frac{v^2}{c^2}\right)\right]a = 0 \qquad 5,15$$

Eq 5,12 introduces $a(x,t)$ as solution of eq 5,11, in turn inferred from eq 5,6a. Let us replace eq 5,12 into eq 5,15 to show the consistency of eq 5,13 with eq 5,15, i.e. the self consistency of eqs 5,6

$$\left(\frac{1}{v^2} - \frac{1}{c^2}\right)\frac{\partial^2 f}{\partial \xi^2} \pm 2i\frac{m\omega\tau}{\hbar\omega_o}\left(\frac{c^2}{v^2} - 1\right)\frac{\partial f}{\partial \xi} + \gamma\tau^2 f = 0 \qquad 5,16$$

where

$$f = f(\xi) = f_1(\xi) + f_2(-\xi) \qquad \gamma = \left(1 - \frac{v^2}{c^2}\right)\left(\frac{\varepsilon}{c\hbar}\right)^2 + \left(\frac{mc}{\hbar}\frac{\omega}{\omega_o}\right)^2\left(1 - \frac{c^2}{v^2}\right) - \left(\frac{mc}{\hbar}\right)^2$$

This equation determines the analytical form of $f$ and defines the solution $a(x,t)$ of the initial equation 5,5. Before considering this point, note that if $v = c$ in eq 5,16 then necessarily $\gamma = 0$, so that $m = 0$ as well in agreement with eqs 2,16. Yet the reverse is not true: $m = 0$ does not necessarily entail $v = c$, rather eq 5,16 reduces to $\partial^2 f/\partial\xi^2 + (v\varepsilon/c\hbar)^2 f\tau^2 = 0$. To clarify this point, let us consider eq 5,16 without any assumption on $m$ and regard in general $v \neq c$. Define

$$f = A\exp(i\varphi) \qquad A = A(\xi) \qquad \varphi = \varphi(\xi) \qquad r_i = c/v \qquad 5,17$$

where $A(\xi)$ and $\varphi(\xi)$ are functions to be determined. Replacing eq 5,17 into eq 5,16 and putting the real and imaginary parts of the resulting equation separately equal to zero, one finds

$$\frac{\partial^2 A}{\partial\xi^2} - \left[\left(\frac{\partial\Phi}{\partial\xi}\right)^2 - \Omega\right]A = 0 \qquad\qquad A\frac{\partial^2\Phi}{\partial\xi^2} + 2\frac{\partial\Phi}{\partial\xi}\frac{\partial A}{\partial\xi} = 0 \qquad 5,18$$

where

$$\Omega = \left(\frac{\varepsilon\tau}{r_i\hbar}\right)^2 - \frac{(mc^2)^2}{r_i^2-1}\left(\frac{\tau}{\hbar}\right)^2 \qquad \frac{\partial\Phi}{\partial\xi} = \frac{\partial\varphi}{\partial\xi} \pm \frac{mc^2\tau}{\hbar\omega_o} \qquad 5,19$$

Replacing $\varepsilon$ in $\Omega$ with $mc^2/\sqrt{1-v^2/c^2}$ only, the result is $\Omega=0$. If so, eqs 2,16 do not yield eqs 5,18 a relevant physical interest; moreover eqs 5,18 have a different form and physical meaning depending on whether $m=0$ or $m\neq 0$, since $\varepsilon\neq 0$ in the latter case only. To overcome this discrepancy, let us consider first $m=0$; with the following positions, suggested by eqs 2,10 and 2,11,

$$\varepsilon_{wave} = \frac{\hbar}{\tau_0} \qquad \Omega_0 = \frac{1}{r_i^2} \qquad 5,20$$

$\Omega_0 = \Omega(m=0)$ depends on $r_i$ only, whereas eq 2,14 yields with analogous notations

$$p_{wave} = \frac{\hbar}{\lambda_0} \qquad \lambda_0 = \frac{c^2\tau_0}{v} \qquad 5,21$$

Now eqs 5,18 are better understood: rewritten as explicit functions of $x$ and $v$ keeping the time constant, they read with the help of eqs 5,12 and 5,17

$$\frac{\partial^2 A}{\partial x^2} - A\left(\frac{\partial\varphi}{\partial x}\right)^2 + Ak_0^2\left(\frac{c}{v}\right)^2 = 0 \qquad 2\frac{\partial A}{\partial x}\frac{\partial\varphi}{\partial x} + A\frac{\partial^2\varphi}{\partial x^2} = 0 \qquad k_0 = \frac{1}{\lambda_0} \qquad m=0 \quad 5,22$$

It is evident that in a 4D approach, where $A = A(x,y,z,t)$ and $\varphi = k_0\varphi'(x,y,z,t)$, eqs 5,22 read $2\nabla A \cdot \nabla(k_0\varphi') + A\nabla^2(k_0\varphi') = 0$ and $\nabla^2 A - A[\nabla(k_0\varphi')]^2 + (c/v)^2 k_0^2 A = 0$. Eqs A11 and B4 together with $\psi$ of eq 5,3 enable thus the eikonal equation and the well known solenoidal vector $A^2\nabla(k_0\varphi')$ to be inferred; it follows by consequence that light propagates according to the laws of wave or geometrical optics depending on $\lambda_0$, with $r_i$ of eq 5,18 having the meaning of refraction index. The momentum of eq 2,14 turns into the De Broglie wave; with reference to eq 2,15, eqs 5,21 yield now

$$\lim_{v\to 0}\frac{p_{wave}}{v} = \frac{\hbar/\tau_0}{c^2}$$

i.e. the rest mass $m$ is replaced by the wave energy $\hbar/\tau_0$ times $c^{-2}$. Let us impose now $\Omega\neq 0$ even for $m\neq 0$, as suggested by the considerations on eq 5,16 and for convenience define its value as $\Omega = 1/(qr_i)^2$, without loss of generality owing to the unknown parameter $q$. Likewise put in general $\varepsilon^{wave} = \hbar/\tau$ with $\tau\neq\tau_0$. So eqs 5,18 hold again even though $m\neq 0$ simply considering $\Phi$ instead of $\varphi$, i.e. the conclusions for a light beam described by $\varphi$ hold for a beam of massive particles described by $\Phi$; as expected both cases resulting from eqs 5,13 and 5,14, i.e. $m=0$ and $m\neq 0$ in the limit $\omega_o\to\infty$, are consistent with the eikonal equations of light and matter, in agreement with the fact that neither $\omega_o$ nor $\tau$ appear in eqs 5,18 and 5,19. It means that even a beam of particles propagates like a light beam at velocity $v$, thus explaining why even massive particles exhibit diffraction phenomena. So eq 2,16 would yield $\Omega=0$ if directly put into eq 5,19 since it considers the corpuscular aspect of matter only, without taking into account the wave behaviour as well. It is easy to show that this latter is hidden in $\Omega$. With the previous positions one finds

$$\Omega = \frac{\Omega_0}{q^2} \qquad \varepsilon_{wave}^2 - \varepsilon_{corp}^2 = \left(\frac{\hbar}{q\tau}\right)^2 \qquad \varepsilon_{corp}^2 = \frac{(mc^2)^2}{1-1/r^2} \qquad 5,23$$

Clearly $\varepsilon_{wave}$ is conceptually equal to that of eq 5,20 and arbitrary because of $\tau$; the second addend, also arbitrary because of $v$, is the squared corpuscular kinetic energy of the particle already concerned in section 2. According to eqs 2,10 and 2,11, $\varepsilon_{wave}$ is a random value within $\Delta\varepsilon_{wave}$, whereas $\varepsilon_{corp}$ is a random value within its own $\Delta\varepsilon_{corp}$. The ratio $(\hbar/q\tau)^2$, defined as difference of the local terms, reads thus $(\Delta\eta)^2$ in agreement with the lack of hypotheses about the value of $\tau$; this position makes physically irrelevant the factor $q$: whatever this latter might be, $\tau$ or $q\tau$ express the arbitrary character of any uncertainty range. Regardless of the specific value of $v$, the form of eq 5,23 $\varepsilon_{wave}^2 - \varepsilon_{corp}^2 = \Delta\eta^2$ shows that the random and unknown values of the variable $\eta^2$ range between $\varepsilon_{wave}^2$ and $\varepsilon_{corp}^2$ within $\Delta\eta^2$ defined by these energies: the former is related to the wave behaviour of the particle, the latter to its corpuscular behaviour, both consistent and coexisting in $\Omega$ through the uncertainty. This result shows the dual particle/wave nature of matter and once again confirms that the energy uncertainty contains inherently both aspects; it is explicitly evident also in eqs 2,16, 5,13 and 5,14, in agreement with eqs A10 and B3 concurrently found in appendixes A and B: as already emphasized therein, eqs A10a and B3a describe the particle, for which can be defined at least in principle the probability of being somewhere within $\Delta x$ according to the random, unknown and unpredictable position of $\delta x$, whereas eqs A10b and B3b replace the particle with the abstract concept of probability density wave. Massless and massive particles are both consistent in principle with this dual behaviour confirmed by the eikonal equations and differ only because of the value of $\lambdabar$ according to eq 5,21. This conclusion shows the mutual consistency of eqs 5,6 resulting from the initial eq 5,5. The non-relativistic limit of eq 5,2 is obtained from eq 5,15 for $v \ll c$. By definition of $\xi$ in eq 5,12, the inequality $(1/c^2)(\partial^2 a/\partial t^2) \ll \partial^2 a/\partial x^2$ reads $(1/\tau^2 c^2)(\partial^2 a/\partial \xi^2) \ll (1/\tau^2 v^2)\partial^2 a/\partial \xi^2$; owing to eq 2,14, one finds

$$\frac{\partial^2 a_{nr}}{\partial x^2} \mp \frac{2mi\omega}{\hbar\omega_o}\frac{\partial a_{nr}}{\partial t} + \left(\frac{mc}{\hbar}\frac{\omega}{\omega_o}\right)^2 a_{nr} = 0 \qquad v \ll c \qquad 5,24$$

where the subscript "nr" denotes the non-relativistic limit of $a$. If also $\omega \to \omega_o$ for $v \ll c$, then eq 5,24 is nothing else but the time dependent Schrodinger equation of the particle in presence of a constant potential $-mc^2/2$. As expected, eq 5,24 coincides with that obtained directly from eqs B7 and B8 in the case where eq 2,14 reduces to $p \approx mv$ and $\varepsilon \approx mc^2 + p^2/2m$. The fact of having found the non-relativistic wave equation uniquely as the limit of the relativistic equation 5,2 through eq 5,16 is a significant check of the procedure followed in the present section.

6. Discussion.
The strategy followed in the present paper was to show first that the quantum point of view introduced in section 1 and highlighted in section 2 is conceptually consistent with the known results of quantum mechanics and with the principles of special relativity as well; before carrying out further calculations, it has been shown in section 3 that effectively the invariance of interval and Lorentz's transformation are inherent the concept of space-time uncertainty. Only thereafter have been approached the problems of general relativity, formulated and carried out as in section 2 with trust in the quantum nature of general concepts like the space-time curvature. Revealing evidences were in effect the corollary of Einstein's equivalence principle and the immediate connection between force $\Delta\dot{P}$ and deformation rate $\Delta\dot{x}$ of the space-time uncertainty ranges $\Delta x$, inferred in agreement with the Newton law simply deriving the uncertainty equations 4,1a with respect to time under the condition of their conceptual equivalence with eqs 4,1b; in fact the Newton law was found in eq 4,6 as approximation of eq 4,21 for $p_o \approx p_o^0$ legitimated by the weak dependence of $p_o$ on $\Delta x$. In turn, the quantum origin of gravity force supported the idea that also relevant relativistic effects should be explained in the theoretical frame previously outlined; in fact red shift and light beam bending were immediately acknowledged in subsection *4.1*, effects (a) and (b), even without the specific discussion carried out in the following subsections *4.3* and *4.5*. This surprisingly simple evidence suggests that the deformation rate of phase space ranges effectively includes also the local mass driven space-time deformation itself: for instance, the subsection *4.5* has shown that the curvature of the space-time uncertainty ranges $\Delta x$ is simply a particular case of the more general concept of deformation rate $\Delta \dot{x}$. There is a conceptual analogy between the ways to find quantum and relativistic results. In wave mechanics the eigenvalues are inferred solving the appropriate wave function; yet, even without solving any wave equation, the real observables verifiable by the experiment are found in section 2 simply introducing the concept of total uncertainty into the pertinent classical problem and counting the number of allowed states. The same reasoning holds also for the relativistic results of section 4: instead of solving Einstein's field equations through functions of generalized coordinates in curvilinear reference systems, which would require the formalism of tensor calculus, the present approach introduces since the beginning the space-time quantum uncertainty into the classical formulation of the various problems of section 4, thus finding in effect the same results of general realtivity verifiable by the experiment. As a consequence of this strategy, the 4-vector and tensor formalisms of have been surrogated by that of section 2 simply merging time and space uncertainty ranges into a unique concept of generalized uncertainty; it allowed to plug both special and general relativity into the realm of quantum mechanics, governed by its severe limits about the knowledge of the physical world. In the theoretical frame so far exposed the uncertainty plays the double role of unsurmountable boundary within which we can formulate our considerations, think for instance to the missing information about the space components of angular momentum, and of stimulation to formulate a correct description of the reality; even so, however, unambiguous agreement with the experimental evidence is eventually feasible, despite the initial agnostic randomness on which is based the present approach. Consider for instance the light beam deflection in the gravitational field: the general relativity describes in detail the bending effect through the local space-time curvature, whereas the actual observation concerns merely the overall deflection of star light in the field of sun. This final outcome, which is in fact the result experimentally available, is however just that calculated in eq 4,18 even without concerning the local details of bending dynamics of light path. As concerns the conceptual basis o the present model, the necessary assumption is that the uncertainty be a basic principle of our universe, even more fundamental than the geometry of the curved space-time itself. For this reason the formulae obtained in the various cases of section 4 are significant despite they correspond to the respective approximate solutions of Einstein's field equations. Even concerning weak fields, the physical meaning of the present results is essentially heuristic; once having shown that the gravity is rooted into the concept of uncertainty, is legitimate the idea that higher order terms could be also found to further improve the present results within the same conceptual frame. The main task in this respect is to demonstrate that the present theoretical model includes correctly the relativity without hypotheses *ad hoc*, but simply extending ideas and formalism of quantum mechanics to gravitation problems. Are remarkable to this purpose the initial considerations of section 3 about the preminent importance of the numbers of allowed quantum states, rather than the local random values of position and momentum, time and energy: the essential physical idea is to calculate appropriately these numbers through simple algebraic manipulations of classical equations thanks to eqs 1,2 and 2,10. Just these results, obtained introducing first the classical formalism and then conveying this latter into the quantum world through the concept of particle delocalization inherent the positions 1,1, confirm once again that only the ranges of dynamical variables have essential physical meaning. This primary strategy, formerly aimed merely to solve specific quantum problems, soon revealed a wider horizon; in addition to the results of special relativity, examined in section 3, the section 4 showed that effectively the concept uncertainty spreads outside the realm of quantum mechanics and becomes the "added value" to the mere Newtonian gravity enabling the most relevant outcomes of general relativity to be found. Strictly speaking, all of the considerations of section 4 concern quantum particles subjected to gravitational interaction; yet, the subsection *4.6* has shown that the behaviour of a quantum particle orbiting in the gravity field of another particle is analogous to that of an orbiting planet. The Kepler problem and the other cases quoted in section 4 (light deflection, red shift and time dilation as well) suggest that, at least at the present order of approximation, simply a scale factor discriminates quantum problems and cosmological problems; also, the behaviour of light and matter interacting with gravity field is substantially the same since the fields are with good approximation additive. This conclusion is actually possible because the gravity is deeply rooted into the quantum concept of uncertainty, so that in effect there is no reason to expect a different behaviour for single particles or aggregates of an arbitrary number of particles. Moreover, there is no evidence in the present model that the energy of the particles affects their gravitational behaviour; in other words, no approximation has been introduced that could suggest a different behaviour of high energy or low energy particles in the gravity field, which would have accordingly affected the conclusions of section 4. For these reasons the subsection *4.1* is the most important one of the present paper. The reasoning holds also for the gravitational waves, which could be evidenced even in an elementary particle experiment, simpli allowing to form in a beam of particles local orbiting systems emitting gravitational waves that wiggle the paths of neighbouring particles: this is a prediction of the present model according to eqs 4,37. The general relativity is therefore nothing else but the quantum aspect of the Newton mechanics: for this reason the positions 1,1 and eqs 1,2 and 2,10 are enough to infer the former from the latter. The arbitrariness of the ranges is not mere restriction of information but source itself of information; it is indeed remarkable the fact that none of the concepts typical of general relativity has been postulated in the present approach, e.g. invariance of light speed and equivalence of inertial and gravitational mass. Since the mass is unambiguously defined by eq 2,15, it is not surprising that the inertia principle is itself a corollary of eq 3,1. This principle, inferred for the gravity force in section 4 through the equation $F_i / m_i = \dot{v}_i$, holds in general for any kind of force. Owing to the essential concept of delocalization, basic hypotheses of relativity like the inertia principle or equivalence principle of inertial and gravitational mass are also corollaries in the present theoretical frame. In fact, the curvature of the space-time uncertainty ranges has been introduced as a natural consequence of eq 3,1, even discarding the local coordinates as a function of which operates the tensor formalism of general relativity. Also the known properties of the gravity force, e.g. conservation of momentum and angular momentum shown first in subsection *2.1* and inferred again in subsection *4.2*, are straightforward consequences of quantum uncertainty, being obtained without integrating any equation of motion based on a postulated interaction law. It is not surprising the result of having found the conservation laws before introducing the gravity field; indeed this latter is consequence of the same principle, the uncertainty, that underlies the angular momentum conservation as well. This confirms the profound link between gravity field and quantum properties of matter. A heuristic aspect of the present conceptual frame concerns $|\mathbf{v}|$ of eq 3,1. Suppose that $\mathbf{v}$ has a finite number of components $v_j$ along specified directions of space with respect to an arbitrary reference system; if so, being $\mathbf{v}$ related to a corresponding momentum $\mathbf{P}_v$, are by consequence defined for each $v_j$ local momentum components $P_{v\,j}$. These latter are in turn defined within ranges $\Delta P_j$, clearly conjugate to the scalars $x_j - x_{oj}$ of eq 3,1. From a physical point of view, therefore, if a finite number $j_{max}$ would exist such that really $1 \leq j \leq j_{max}$, this reasoning would introduce in fact a $j_{max}$-dimensional space; each dimension identified by its own $j$ would be in fact legitimated in the present theoretical frame by an uncertainty equation conceptually identical to eqs 1,2 and related 2,10. Certainly the compelling role of the uncertainty on the understanding of the reality makes the existence of further dimensions hidden by eq 3,1 an open point of the present model. However any speculation about $\mathbf{v}$ would require a valid physical reason, for instance some form of space-time anisotropy, to justify existence and physical consequences of the components $v_j$. According to the previous reasoning such an anisotropy should concern $\Delta x_{xt}$ only, not the space and time ranges separately; also, even the anisotropy of uncertainty should not affect $c$ for the reasons sketched in sections 2 and 3, and could be evidenced in a relativistic experiment distinctive of the space-time properties of uncertainty. In principle nothing hinders thinking so as concerns the consistency of the results of section 4, which seem however enough to explain a wide amount of experimental evidences as a function of a unique fundamental assumption. Thus it does not seem really legitimate to introduce anything without experimental evidences and without a compelling necessity to clarify unexplained effects; the fact that nothing in principle hinders the existence of these extra dimensions, which would support the string theory, is not enough however to justify speculations on their effective physical reality. In any case, nothing in the present model prevents or contradicts the existence of extra-dimensions. Finally, note that the present physical model does not exclude the infinities; so, the dynamical variables can take in principle even infinite values because the respective uncertainty ranges defining them are completely arbitrary. In effect the eigenvalues of section 2 have been calculated just postulating the random and unpredictable character of the local dynamical variables on the one side and the arbitrary sizes of the respective uncertainty ranges on the other side. So infinite ranges peacefully coexist with well defined results without divergence problems because not necessarily the local values of the dynamical variables must be infinite themselves. In general a range is unrelated to the local properties of its own variable, whereas two conjugate ranges are necessary to define the numbers of quantum states, i.e. the eigenvalues of the particle. Is challenging the idea that the finite size and, presumably, time length of our universe are explained from the microscopic quantum scale to the macroscopic relativistic scale by physical variables conceptually described by indefinable ranges. May be, the key of this intriguing paradox stems on the fact that the nature admits in principle infinite ranges allowed to its physical parameters, but in practice does not need them. As a first example, it has been emphasized in section *4.3* that the gravitational interaction between light and matter removes the infinite values of frequency shift $\Delta\omega$ and $\Delta P_2$ of the photon, in principle allowed to the ranges, by merging eqs 4,10 and 4,11 into eq 4,12; in this case is just the interaction the way to eliminate the infinities. The examples of section 2 have evidenced another aspect of this problem: among the range sizes in principle possible, particular values exist that fulfil some appropriate selection condition not excluding or contradicting however the total randomness of eqs 1,2 and 2,10. The harmonic oscillator and hydrogenlike atoms reveal propensity of nature to fulfil the condition of minimum energy; with this preferential condition, which is proven effective in general even though not explicitly required by any fundamental physical law, the eigenvalues of eqs 2,2 and 2,9 are definite even being in principle consequence of total arbitrariness of the uncertainty ranges. Is the condition of minimum energy the other way to pass over the infinities in principle possible? It would seem so, despite the lack of a compelling reason. In effect, notwithstanding the unambiguous agreement with the experimental results, it appears that the outcomes of present theoretical model are more flexible than that of the wave formalism; one reason of it rests just on the condition of minimum energy. Once writing the quantum Hamiltonian of the harmonic oscillator, for instance, the solution of the wave equation yields uniquely eq 2,9 without any other chance. This equation, as obtained in section 2, is instead compatible with other results easily calculated simply admitting that the minimum condition is not fulfilled. The most intuitive physical concept evoked by this conclusion is that of non-equilibrium state of matter. A lattice oscillator could be described through the non-equilibrium parameter $\alpha_{ne} = \Delta p_x / \Delta p_x^{(\min)} \neq 1$ by $\varepsilon_{ne} = (\alpha_{ne}^4 + 1)(2\alpha_{ne}^2)^{-1} n\hbar\omega + \alpha_{ne}^2 \hbar\omega/2$. This idea is actually more general and concerns even more fundamental laws of nature. Recall for instance the way to infer in section 3 the Lorentz transformations and the invariancy rule of interval through the ranges $c\Delta t = x_s - x_o + \delta X$ and $c\Delta t' = x'_s - x'_o + \delta X'$ reciprocally sliding at constant rate $V = (x'_o - x_o)/\Delta t$; if $(x_s - x_o)^2 = (x'_s - x'_o)^2$ by definition, one finds $c^2\Delta t^2 - \Delta X^2 = c^2\Delta t'^2 - \Delta X'^2$. This result was immediately considered conclusive, being in agreement with the special relativity and with the experience. There is no reason however to exclude in principle even further positions, like for instance $c^4\Delta t^4 = (x_s - x_o)^4 + \Delta X^4$ and $c^4\Delta t'^4 = (x'_s - x'_o)^4 + \Delta X'^4$; if so $(x_s - x_o)^4 = (x'_s - x'_o)^4$ would be now consistent with the hypothetical invariant $\delta s^{(4)} = \sqrt[4]{c^4\Delta t^4 - \Delta X^4}$, once more with $\Delta t' \neq \Delta t$ and $\delta X' \neq \delta X$. Of course such an invariant interval has no physical interest, at least as far as we know, although it is still in agreement with invariant $c$ and introduces the factor $(1 - v^4/c^4)^{-1/4}$ with Galilean limit for $v \ll c$ and with the consequences expected for $v \to c$. So the question "should be physically excluded *a priori* an invariant interval like $\sqrt{c^2 \Delta t^2 - \Delta x^2} + a\sqrt[4]{c^4 \Delta t^4 - \Delta X^4}$ with $|a| \ll 1$?" has surely speculative character, being not required by any experimental evidence; yet it merely emphasizes that the squared interval describing correctly the reality is not the only one conceptually definable, rather it is the simplest one and sensibly the right one among the many in principle possible. As in the case of the minimum energy, also now the actual physical laws appear to be the result of a preferential choice of nature, not excluding however other possible choices and laws provided that in agreement with the fundamental idea of uncertainty. The conclusion is that a subtle wire links apparently different concepts like steady eigenvalues and non-equilibrium states of matter, gravitational interaction and weirdness of quantum world, time dilation and wave-corpuscle dualism; actually, the underlying and unifying concept to all these different aspects of realty are the uncertainty and its necessary infinities that ensure maximum global randomness and minimum local information; these assumptions must be accepted thoroughly, without transgressions or exceptions. Yet remains evident the paradox of infinities conceptually allowed in a finite universe. A possible hint to solve this apparent irrationality comes from the indication that the physical laws are the result of two opposite instances: eq 1,1 introduces first the total arbitrariness into the description of a physical system through the total delocalization of its constituent particles, as it reasonable for a quantum approach; yet fine tuning of this arbitrariness on particular values, the eigenvalues, is provided by eqs 1,2 and 2,10, for which also holds the correspondence principle. In other words: the positions 1,1 are the basic conceptual condition, eqs 1,2 and 2,10 allow in fact the information essential for the existence of rational life in the universe. This circumstance, which connects quantum results and macroscopic world, e.g. thermodynamics and conservation of momentum and angular momentum, has been highlighted in the case of hydrogenlike atoms: the available physical information is the same regardless of considering probability density or total ignorance about conjugate dynamical variables. If so it is easy to understand why the total uncertainty does not prevent the existence of observables; think for instance that a particle, free or bound, is in fact also a wave propagating or a solid corpuscle moving from minus infinity to infinity. This dual behaviour clearly appeared in section 5 through the eikonal eqs 5,18 and in appendixes A and B through the weird link between real probability $\Pi$ on the one side and complex wave function $\psi(x,t)$ to it related on the other side. Indeed the couples of equations A10 and B3 were inferred contextually, i.e. without hierarchical priority for either behaviour of the particles. Yet, just with the uncertainty premises, the indefinable boundaries of the ranges are in fact non-elusive even at the infinite limit: the reference systems are arbitrary themselves, so an infinite coordinate could be regarded likewise any other finite coordinate with respect to a proper reference system at infinity itself. For this reason the infinity is not a failure of the model but a possible chance for a quantum particle likewise any other finite coordinate. Moreover even the infinity does not entail any "spooky action at distance": once having disregarded the local variables, is in fact missing the definition itself of distance. For instance the Newton law resulted expressed as a function of ranges enclosing any possible distances between two masses; the Coulomb law to calculate the eigenvalues of hydrogenlike atoms has been expressed through a range of possible distances between electron and nucleus. Strictly speaking, it is impossible to know how close or how remote are actually the respective particles, so it would be more appropriate to assert "action at spooky distance". In effect, it is physically meaningless to inquire if the particles are able or not to inform each other about their own status: even in an infinite range they could identically be infinitely apart or maybe infinitely close (despite the existence of physical observables, nobody will ever know it) without changing their physical state defined by the conservation laws. This behaviour, which anyway depends upon couples of conjugate ranges, is typically non-local and clear consequence of the uncertainty; yet it does not contradict for instance the conservation of angular momentum, which is independent consequence itself of the uncertainty as shown in section 2. Is then the total uncertainty a concept really more agnostic than the probabilistic knowledge provided by the wave mechanics? A possible answer could be that the question is physically meaningless and badly posed and that only the results justify a positive or negative reply. Yet, a better answer is probably that the history of our universe does not depend on its own physical limits. Would then an endless universe be the same as the actual one we are trying to describe? Would then an ever lasting universe evolve as the actual one we are trying to foresee?

FIGURE CAPTIONS

Fig 1:The Regge-Chew-Frautschi diagrams of experimental meson and baryon masses ($\alpha$, $\beta$, $\gamma$, $\delta$ trajectories) [15]. The plot collects the data of 47 particles.

Fig 2: Diagram of the Regge-Chew-Frautschi data of Fig 1 expressed as a function of $(m_H - m_o)^2$. The data merge into a unique trajectory. The regression parameters are shown in figure.

# APPENDIX A

Most of the results of this appendix and next appendix B are well known; they have signed the birth of the new quantum mechanics after Bohr's early hypothesis. Yet the non-trivial reason to infer these results here is to show their derivation from and consistency with the positions 1,1 and eqs 1,2 and 2,10, on which are rooted the relativistic results of sections 3 and 4 too. Both appendixes concur thus to highlight the connection between relativity and quantum mechanics through their common root in the fundamental concept of uncertainty. Consider the uncertainty equation 1,2 for a free particle; let $x_1, p_1$ and $x_2, p_2$ be two arbitrary couples of dynamical variables such that

$$\Delta p \Delta x = n\hbar \qquad \Delta x = x_2 - x_1 \qquad \Delta p = p_2 - p_1 \qquad \text{A1}$$

No information is allowed about current position and momentum of the particle in the respective ranges of the phase space. Yet the total uncertainty does not prevent to define in principle the probability $\Pi$ that the particle be in an arbitrary subrange $\delta x$ inside the total allowed range $\Delta x$

$$\frac{\delta x}{\Delta x} = \Pi \qquad \delta x = x - x_o \qquad \delta x \leq \Delta x \qquad \text{A2}$$

considering thus $x_o$ and $x$ arbitrary coordinates within $\Delta x$ unknown likewise as $x_1$ and $x_2$ themselves, i.e. without possibility to define width or location of $\delta x$ inside the total allowed range $\Delta x$ and without possibility to distinguish $\delta x$ with respect to any other possible sub-range. No hypothesis is necessary about $\delta x$ and $\Delta x$. In general $\Pi$ is expected to depend on coordinate and time. Yet the time dependence will be considered in the next appendix B; here we consider explicitly the dependence of $\Pi$ on the space coordinates only, i.e. $t$ is regarded as fixed parameter in correspondence to which are examined the properties of $\Pi$ as a function of the random and unknown coordinate $x$. Regard the width of $\delta x$ variable, with $x$ current coordinate and constant $x_o$; also, the couples of coordinates $x_1, x_2$ and momenta $p_1, p_2$ are considered fixed. So eqs A2 yield

$$\frac{1}{\Delta x} = \frac{\partial \Pi}{\partial x} \qquad \Pi = \Pi(x,t) \qquad \text{A3}$$

Let us introduce the probability $\Pi$ into eq A1 considering both possibilities that the particle be or not within $\delta x$. Moreover, let $n_+$ and $n_-$ be two arbitrary numbers of states consistent with the respective probabilities $\Pi$ and $1-\Pi$. Putting then

$$\delta x \Delta p = n_+ \hbar \qquad (\Delta x - \delta x)\Delta p = n_- \hbar \qquad n_+ + n_- = n \qquad \text{A4}$$

it appears that effectively $n_+/n + n_-/n = 1$; moreover eq A4 yields the identity

$$(1-\Pi)\Pi \Delta p^2 = n_- n_+ \hbar^2 \left(\frac{\partial \Pi}{\partial x}\right)^2 \qquad \text{A5}$$

With the position $n' + n'' = n_+ n_-$, where $n'$ and $n''$ are further arbitrary numbers, eq A5 splits as

$$\Pi \Delta p^2 = n' \hbar^2 \left(\frac{\partial \Pi}{\partial x}\right)^2 \qquad \text{A6a}$$

$$\Pi^2 \Delta p^2 = -n'' \hbar^2 \left(\frac{\partial \Pi}{\partial x}\right)^2 \qquad \text{A6b}$$

Being $n_+$ and $n_-$ by definition positive, at least one among $n'$ and $n''$ or both must be necessarily positive. Eqs A6 are now discussed considering separately the possible signs of $n'$ and $n''$.

(i) $n' > 0$ and $n'' < 0$. Eq A6a and b read also $\delta x \Delta p = (n'/n)\hbar$ and $\delta x^2 \Delta p^2 = |n''|\hbar^2$ respectively thanks to eqs A2 and A3: hence $(n'/n)^2 = |n''|$ and $\Pi = |n''|/n'$. Both results are possible for any $n$ because $n'$ and $n''$ are arbitrary. Eqs A6 are formally analogous to the initial eq A1, from which

they differ because of the widths of the uncertainty ranges only: multiplying both sides by $n^\S n/n'$, with $n^\S$ arbitrary integer, one finds $\Delta x^\S \Delta p^\S = n^\S \hbar$, where $\Delta x^\S$ and $\Delta p^\S$ are any ranges related to the initial ones $\delta x$ and $\Delta p$ through the condition $\Delta x^\S \Delta p^\S = \delta x \Delta p (n^\S n/n')$. Clearly the physical interest of eq A6a rests on the possibility of being expressed through the integer $n^\S$, not on the new sizes of the uncertainty ranges appearing in its mathematical form; in fact, however, the physical meaning of $\Delta x^\S \Delta p^\S$ is the same as that of eq A1. Of course the same holds for eq A6b, identical to eq A6a with the given sign of $n''$. In conclusion, nothing conceptually new with respect to eq A1 is inferred from this combination of signs of $n'$ and $n''$. This possibility has no physical interest.

(ii) $n' < 0$ and $n'' > 0$. The right hand sides of both eqs A6 have now negative sign, so that neither of them can have the same physical meaning of the initial eq A1; they read $\Pi = -|n'|/n^2$ and $\Pi^2 = -n''/n^2$ because of eq A3. Yet the result $\Pi = n''/|n'| = -|n'|/n^2$ is clearly absurd; also this combination of signs is to be excluded.

(iii) $n' > 0$ and $n'' > 0$. Now eqs A6a and b are physically different because of the signs: their ratio would yield $\Pi$ negative. The reasonable conclusion is that these equations cannot be combined together because they provide different ways to describe the particle delocalized in its uncertainty range. Thus let us consider them separately. Eq A6a is conceptually analogous to eq A1 and can be worked out with the help of eq A3; eq A6b excludes instead eq A3 and admits the solution $\Pi = A' \exp(\pm i x \Delta p / \hbar \sqrt{n''})$, where $A'$ is the integration constant. Rewritten more expressively as $\Pi = A \exp(\pm i \varphi \delta x / \Delta x)$ with $\varphi = n/\sqrt{n''}$, this solution significantly differs from $\Pi$ of eq A6a: despite the same notation the latter coincides conceptually with the probability introduced in eq A2, conceivable and definable even if random and unknown, the former is instead complex function of this probability. Thus eq A6b still retains the essential concept of delocalization within an arbitrary uncertainty range; yet it does not longer concern through $\Pi$ the ability of the particle to be in some specific point of $\Delta x$. The only possibility to regard both eqs A6 together is to admit their different physical meaning, i.e. their different way to describe the particle dynamics inside $\Delta x$. This dual outcome compels in fact the impossibility to regard the particle simply as a corpuscle delocalized somewhere in its uncertainty range, as done by eq A6a through its own $\delta x$ and related probability $\delta x / \Delta x$; so one concludes that eq A6b is incompatible with the simple corpuscle-like behaviour of eq A6a, despite the particle must anyway be randomly moving in $\Delta x$. Moreover, a further difficulty to regard together eqs A6a and A6b comes from the fact that $\Pi$ defined by this latter is not real, as instead $\Pi^* \Pi = |const|^2$ does. Yet just this property suggests a possible way out from this difficulty, simply supposing that eq A6b requires wave-like propagation of the particle: so $\Pi^* \Pi$ could stand for particle wave amplitude whereas $A$, in fact regarded as $A_0 A(t)$ without contradicting any previous step, could define frequency and phase of the particle wave. This idea is better elucidated repeating here the steps used to infer eq 2,10 from eq 1,2: rewrite the exponential $x \Delta p$ of $\Pi$ as $t \Delta \varepsilon$ dividing and multiplying by v in order that $\pm i x \Delta p / \hbar \sqrt{n''}$ becomes $\pm i t \Delta \varepsilon / \hbar \sqrt{n''}$. So one expects that $A(t)$ is defined just by this requirement, i.e. $\Pi = A_0 \exp[\pm i (c_x x \Delta p + c_t t \Delta \varepsilon) / \hbar \sqrt{n''}]$ where $c_x$ and $c_t$ are arbitrary coefficients of the linear combination expressing the most general way to combine the space and time functions. Calculate now $\partial^2 \Pi / \partial x^2 = -(c_x \Delta p)^2 \Pi$ to extract the real quantity $c_x \Delta p$ from $\Pi$, and then also $\partial^2 \Pi / \partial t^2 = -(c_t \Delta \varepsilon)^2 \Pi$ by analogy; eliminating $\Pi$ between these equations and noting that by dimensional reasons $(c_x \Delta p / c_t \Delta \varepsilon)^2 = v^{-2}$, one finds the result $\partial^2 \Pi / \partial x^2 - v^{-2} \partial^2 \Pi / \partial t^2 = 0$ that, whatever v might be, confirms the wave-like character of particle delocalization provided by eq A6b. Of course the equation of waves can be in no way inferred from eq A6a. In this manner however the physical properties of the wave are related to $\Pi^*\Pi$, whereas eq A6a suggests their dependence upon $\Pi$ or $\Pi^*$ because the physical properties of the particle are related to the probability $\Pi$ rather than to its square. Let us introduce thus the complex function $\sqrt{\Pi}$ in place of $\Pi$ and rewrite eq A6b as a function of the former instead of the latter; dividing both sides by $\Pi$, eq A6b reads

$$\left(\pm\hbar\frac{\partial\sqrt{\Pi}}{\partial x}\right)^2 = -\left(p^\S\sqrt{\Pi}\right)^2 \qquad p^\S = \pm\frac{\Delta p}{2\sqrt{n''}} \qquad \text{A7}$$

The notation emphasizes that $p^\S$ does not depend on $x$ and is not a range; being uniquely defined by the solution of the differential equation A7 it has a fixed value not longer related to $\Delta p$, i.e. it is an eigenvalue of $\sqrt{\Pi}$. This is possible because $n''$ is arbitrary like $\Delta p$; so the ratio $\Delta p/2\sqrt{n''}$ is a well determined quantity, summarized just by $p^\S$, whose value and signs correspond to either component of momentum along the direction where are defined $\delta x$ and $\Delta x$. Thus eq A7 becomes

$$\frac{\hbar}{i}\frac{\partial\sqrt{\Pi}}{\partial x} = p^\S\sqrt{\Pi} \qquad \sqrt{\Pi} = \sqrt{A}\sqrt{\exp(\pm i\varphi\delta x/\Delta x)} \qquad \text{A8}$$

Now $\sqrt{\Pi}\sqrt{\Pi^*}$ expresses the probability to find the particle somewhere in $\Delta x$. Write

$$\sqrt{\Pi}\sqrt{\Pi^*} = \frac{\hbar}{ip^\S}\frac{\sqrt{\Pi^*}\partial\sqrt{\Pi}}{\partial x}$$

The right hand side is real and reads $\sqrt{\Pi}\sqrt{\Pi^*} = \delta x_0/\Delta x = A_0^2$, being $\delta x_0 = A_0^2\hbar\varphi/2p^\S$. Since a proper value of $A_0^2$ surely exists such that $\delta x_0 \leq \Delta x$, it follows that $\sqrt{\Pi}\sqrt{\Pi^*}$ is still consistent with the concept of probability similarly to the early $\delta x/\Delta x$ of eq A2; yet this latter is replaced in the last equation by a constant value, which entails thus equal probability to find the particle in any subrange $\delta x_0$, regardless of its size and position in $\Delta x$. To understand the physical meaning of this result, let us integrate both sides of eq A8 with respect to $x$ in the subrange $\delta x_0$; one finds

$$p^\S = \left(\int_{x_{01}}^{x_{02}}\sqrt{\Pi}\sqrt{\Pi^*}dx\right)^{-1}\int_{x_{01}}^{x_{02}}\left(\sqrt{\Pi^*}\frac{\hbar}{i}\frac{\partial}{\partial x}\sqrt{\Pi}\right)dx \qquad \delta x_0 = x_{02} - x_{01} \qquad \text{A9}$$

The average value of momentum equal to the eigenvalue, expected for the steady motion of a free particle, suggests regarding $\delta x_0/\Delta x$ as average probability that the particle is in the subrange $\delta x_0$. It is clearly convenient therefore to define $A_0$ in order that $\delta x_0 = \Delta x$ through $\int\sqrt{\Pi}\sqrt{\Pi^*}dx = 1$, so that the momentum eigenvalue concerns the certainty that the particle is delocalized in the total range $\Delta x$. Since this latter is arbitrary, it allows to consider in general the particle from $-\infty$ to $\infty$. The physical information provided by eq A6b is thus really different from that of eq A6a, although the consistency of eqs A8 and A9 with the initial eq A1 is unquestionable; despite their different formulation, repeating backwards the same steps just shown eq A8 leads to eq A6b, originated together with eq A6a from the unique uncertainty equation A1. For this reason it is not surprising that the uncertainty is still inherent $\sqrt{\Pi}$ and consistent with the existence of the eigenvalue $p^\S$. To extrapolate these results to the classical formulation of quantum mechanics, it is enough to regard in general the wave functions in analogous way, e.g. as it will be shown in appendix B for the energy eigenfunction. So, one exploits $\psi = const\sqrt{\Pi}$ and $\psi^* = const\sqrt{\Pi^*}$ to normalize $\psi\psi^*$ and define the probability density of the particle within $\Delta x\Delta y\Delta z$; being the uncertainty ranges arbitrary, this probability density concerns actually the whole space allowed to the particle. The constant of normalization is not essential for the purposes of the present paper and not explicitly concerned hereafter and in appendix B. The result of interest is that, after having introduced the probability $\Pi$ of eq A1, one finds two distinct equations concurrently inferred from the respective eqs A6

$$\Delta p^\S \Delta x^\S = n^\S \hbar \qquad \text{A10a}$$

$$\frac{\hbar}{i}\frac{\partial \sqrt{\Pi}}{\partial x} = p^\S \sqrt{\Pi} \qquad \text{A10b}$$

An interesting property of $p^\S$ is inferred from its definition in eq A7 noting that $(p^\S)^2 = (\Delta p^\S)^2 / 4n'' = (n/4n'')\hbar \Delta p^\S / \Delta x^\S$. Including the numerical factor $n/4n''$ into $p^\S$ and omitting the superscripts, once again because all of the range sizes are arbitrary, one finds in general

$$p^2 = \hbar \frac{\Delta p}{\Delta x} \qquad \text{A11}$$

Summarizing the results so far exposed, the particle is described by:
(i) eq A10a, which differs trivially from the initial eq A1 merely because of the width of the uncertainty ranges and related number of states;
(ii) a differential equation defining the momentum through the probability that the particle be in a given point of its allowed range $\Delta x^\S$.

The point of view of eq A10a does not consider explicitly the particle, but only its random location somewhere inside $\Delta x^\S$; the same holds also for the momentum, which does not appear explicitly because it is replaced by its uncertainty range too. The only information available through this equation concerns therefore the number of states $n^\S$ consistent with the delocalisation ranges $\Delta x^\S$ and $\Delta p^\S$; nothing can be inferred about the dynamical variables themselves. However, the results of section 2 show that even renouncing "*ab initio*" to any information about these latter, the quantum properties of the particle are correctly described. The point of view of eq A10b is different. This equation considers explicitly the subrange $\delta x$ through $\sqrt{\Pi}$ and, even without hypothesizing anything about its size and position within $\Delta x^\S$, concerns directly the particle itself through its properties $\Pi$ and $p^\S$, both explicitly calculated solving the differential equation. Yet the common derivation of both eqs A10 from the initial eq A1 shows that actually the respective ways to describe the particle must be consistent and conceptually equivalent: this fact justifies why the same results are expected through both points of view. This coincidence evidences the conceptual link between properties of the particles and phase space; it explains why the quantum energy levels and angular momentum do not depend on the current values of the dynamical variables of the particles, even when calculated solving the differential equation A10b. Initially $\Pi$ has been introduced in eq A2 as mere function of uncertainty ranges of the phase space; thereafter, however, it has also taken through the steps from eqs A3 to A10 the physical meaning of wave function $\sqrt{\Pi}$ of the particle defining the momentum eigenvalue $p^\S$. Eq A10a considers uniquely the phase space, whereas eq A10b concerns explicitly the particle and introduces the operator formalism of wave mechanics. Since no hypothesis is made on the physical nature of the particle, this conclusion has general validity. Eq A7 is consistent with $\partial(\pm\sqrt{\Pi})/(\pm\partial x)$ and $\partial(\mp\sqrt{\Pi})/(\pm\partial x)$; both signs agree of course with the initial equations A6. Exchanging $dx$ with $-dx$ means replacing $x$ with $-x$, i.e. moving the particle from the positive side of the coordinate axis to the negative side. Clearly $\Pi(-x) = \Pi(x)$, since the ratio $\delta x$ to $\Delta x$ does not change; it only requires that in a given reference system $\Delta x$ extends arbitrarily with respect to the origin from the negative side of the coordinate axis to the positive side, where are respectively located $-\delta x$ and $\delta x$. Even so, however, $\sqrt{\Pi(-x)} = \pm\sqrt{\Pi(x)}$; hence the symmetric and anti-symmetric character of the wave function ($\Delta x$ is the same in both cases) seems a physical property rather than a mathematical result. The idea of parity has also a further implication since in a different reference system, shifted with respect to the former, $\Delta x$ can be entirely located on the positive side of coordinates, which means that the particle is delocalized in either arbitrary subrange $\delta x_1$ or $\delta x_2$ having equal sizes. In both cases these considerations have little interest for one particle only, because no new physical information is to be expected from a particle delocalized in $\delta x$ rather than $-\delta x$; if so, one should accept $\sqrt{\Pi(x_1)} = \sqrt{\Pi(x_2)}$ only. Yet once having introduced these subranges appears more interesting the case of two particles delocalized in $\Delta x$, for simplicity assumed non-interacting. Let the first be in $\delta x_1$ and the second in $\delta x_2$; so are defined the respective $\sqrt{\Pi_1}$ and $\sqrt{\Pi_2}$, each one with its own eigenvalue $p_1^\S$ and $p_2^\S$. Eq A10b concerns thus either particle in either subrange, eq A10a skips a priori such an information because it concerns the total range only, regardless of their local coordinates and momenta, unknown and then ignored. So the concept of indistinguishability, already sketched in subsection 2.2 and further emphasized here, is basically inherent eq A10a, being conceptually impossible to identify particles whose properties are in fact unspecified; eq A10b requires instead a specific rule, to be introduced as a postulate, to get the same conclusion. In other words, eq A10a entails the corollary of indistinguishability of identical particles; without paying attention to the steps from eq A1 to eqs A9, the operator formalism of eq A10b needs introducing "ad hoc" this requirement to ensure its physical consistency with eq A10a. The approach starting directly from eq 1,2 has therefore more general character than that utilizing the operator formalism of wave mechanics, which starts just postulating eq A10b: the basic reason is that eq A10a contains less information than eq A10b. These equations can be now regarded together in agreement with their connected derivation from eq A1. On the one side eqs A10 introduce the corpuscle/wave dual nature of particles: eq A10a admits that the particle is somewhere in $\Delta x$, even though renouncing to know exactly where because of the delocalization; eq A10b instead regards the particle as a wave propagating within $\Delta x$ thus still delocalized but excluding in principle even the idea of unknown position of a material corpuscle. These ideas will be more clearly detailed in section 5. On the other side eqs A10 confirm that properties of particles and properties of phase space must not be regarded separately, rather they are intrinsically correlated: in effect the results of section 2 show that the numbers of quantum states (properties of the phase space) coincide with the quantum numbers that define the eigenvalues (properties of the wave function of the particle). If $p_1^\S \neq p_2^\S$ then $\sqrt{\Pi_1} \neq \sqrt{\Pi_2}$ while, from the point of view of eq A10a, different quantum states are allowed to the respective particles; only if these latter are in the same quantum state, identical quantum numbers are extracted from the respective wave functions. Just this crucial coincidence suggests the reasonable link between symmetric or anti-symmetric character of many particle wave functions and ability of the respective particles to be or not in the same quantum state. If so, the wave function of the whole system $\sqrt{\Pi_{tot}}$ is affected or not by the interchange of particles depending on either behaviour of these latter: whatever its analytical form might be, it must be such that $\sqrt{\Pi_{tot}(x_1, x_2)} = \sqrt{\Pi_{tot}(x_2, x_1)}$ if both particles are allowed to be in the same quantum state, whereas $\sqrt{\Pi_{tot}(x_1, x_2)} \neq \sqrt{\Pi_{tot}(x_2, x_1)}$ if the particles are in quantum states necessarily different; the latter chance is thus consistent with $\sqrt{\Pi_{tot}(x_1, x_2)} = -\sqrt{\Pi_{tot}(x_2, x_1)}$. This general conclusion is inferred here in principle without need of considering in detail the specific form of the many particle $\sqrt{\Pi_{tot}}$, i.e. the known expression $\sqrt{\Pi_1(x_1)}\sqrt{\Pi_2(x_2)} \pm \sqrt{\Pi_1(x_2)}\sqrt{\Pi_2(x_1)}$; this result will be furthermore concerned in section 3 in connection with the exclusion principle. We note here that the form of this wave function is justified by the arbitrariness of the coefficients $c_x$ and $c_t$ in $\Pi = A_0 \exp[\pm i(c_x x \Delta p + c_t t \Delta \varepsilon)/\hbar\sqrt{n''}]$, which is actually given by the more general form $\Pi = \sum_j A_{0j} \exp[\pm i(c_{xj} x \Delta p_j + c_{tj} t \Delta \varepsilon_j)/\hbar\sqrt{n_j''}]$. We conclude that the anti-symmetric or symmetric behaviour of $\sqrt{\Pi}$ is in fact a physical property, not a mere mathematical consequence of eqs A6. The growth of these concepts till today's physics is well known and of course does not need to be further concerned in the present paper.

## APPENDIX B

The steps to find the energy operator are conceptually identical to those reported in appendix A, yet now the probability that the particle be in the range $\delta x$ is regarded as a function of time; $\Pi$ is now defined as ratio between the time range $\delta t = t - t_o$ spent by the particle within fixed $\delta x$ and the total time range $\Delta t = t_2 - t_1$ spent within the total range $\Delta x$. Let us write then $\Pi = \delta t / \Delta t$, without contradicting anyone of the steps of appendix A but simply regarding $\Pi$ as a function of time at fixed coordinate $x$; eqs A3 and A5 read now $\Delta t^{-1} = \partial \Pi / \partial t$ and $(1-\Pi)\Pi \Delta \varepsilon^2 = n_- n_+ \hbar^2 (\partial \Pi / \partial t)^2$. Replacing position and momentum with time and energy in eq A1, eq A7 reads

$$\left(\pm \hbar \frac{\partial \sqrt{\Pi}}{\partial t}\right)^2 = -\left(\varepsilon^\S \sqrt{\Pi}\right)^2 \qquad \varepsilon^\S = \pm \frac{\Delta \varepsilon}{2\sqrt{n''}} \qquad \text{B1}$$

whereas eq A8 reads

$$-\frac{\hbar}{i}\frac{\partial \sqrt{\Pi}}{\partial t} = \pm \varepsilon^\S \sqrt{\Pi} \qquad \text{B2}$$

The upper sign at right hand side of eq B2 makes the classical Hamiltonian written with the help of eq A8 consistent with the result $\varepsilon^\S = p^{\S 2}/2m$ in the particular case of a free particle having mass $m$ and momentum $p^\S$. Yet the lower sign, also allowed as a consequence of eq B1, shows the possibility of states with negative energy as well. The couple of equations A10 turns into

$$\Delta t^\S \Delta \varepsilon^\S = n^\S \hbar \qquad \text{B3a}$$

$$-\frac{\hbar}{i}\frac{\partial \sqrt{\Pi}}{\partial t} = \pm \varepsilon^\S \sqrt{\Pi} \qquad \text{B3b}$$

The considerations carried out for $p^\S$ can be repeated also for $\varepsilon^\S$, which is indeed the eigenvalue of eq B3b. Holds therefore the following equation, conceptually analogous to eq A11,

$$\varepsilon^2 = \hbar \frac{\Delta \varepsilon}{\Delta t} \qquad \text{B4}$$

The comparison between eqs A10 and B3 is interesting because it shows the strict analogy between time and space coordinates in defining the complex wave function $\psi(x,t) \equiv \sqrt{\Pi(x,t)}$. These conclusions have mere formal valence in non-relativistic physics; yet their actual meaning will appear in the sections 3 and 4 dedicated to the special and general relativity. The reasoning so far exposed can be further extended handling in eq 1,2 $\Delta p$ exactly in the same way as $\Delta x$. In other words, one could define a probability $\Pi_p$ in analogy with eqs A1 as

$$\frac{\delta p}{\Delta p} = \Pi_p \qquad \delta p = p - p_o \qquad \Pi_p = \Pi_p(p) \qquad \delta p \leq \Delta p \qquad \text{B5}$$

considering again $p_o$ and $p$ unknown in principle, as mentioned in appendix A for $\delta x$ in $\Delta x$. The notation in eq B5 emphasizes the variable concerned by the probability that the momentum of the particle be just in the range $\delta p$; hence the property $1/\Delta p = \partial \Pi_p / \partial p$ is completely analogous to eq A3 and the same elementary steps as from eqs A4 to A7, yield again the phase space equation $\Delta p^\S \Delta x^\S = n^\S \hbar$ in the form of eq A10a. Yet one finds now the following equation

$$\pm \frac{\hbar}{i}\frac{\partial \sqrt{\Pi_p}}{\partial p} = x\sqrt{\Pi_p} \qquad \text{B6}$$

Having merely exchanged the roles of $\delta p$ and $\delta x$, now $x$ is defined by an equation analogous to A7. As expected, eqs A7 and B6 show that either $p$ or $x$, not both simultaneously, can be inferred solving the respective differential equations; clearly the unique eq 1,2 cannot be utilized twice, depending on whether $\Pi_x$ or $\Pi_p$ is concerned, to obtain information about two conjugate dynamical variables within their own uncertainty ranges. This conclusion is immediately inferred as straightforward consequence of eq 1,2, without need of considering the commutation rule of the coordinate and momentum operators. In conclusion, extending this reasoning to the energy and to the time as well, one finds the following set of equations, whose notations emphasize the dynamical variables that define the respective probabilities

$$\pm \frac{\hbar}{i} \frac{\partial \sqrt{\Pi_\varepsilon}}{\partial \varepsilon} = t\sqrt{\Pi_\varepsilon} \qquad \pm \frac{\hbar}{i} \frac{\partial \sqrt{\Pi_t}}{\partial t} = \varepsilon \sqrt{\Pi_t} \qquad \text{B7}$$

$$\pm \frac{\hbar}{i} \frac{\partial \sqrt{\Pi_p}}{\partial p} = x\sqrt{\Pi_p} \qquad \pm \frac{\hbar}{i} \frac{\partial \sqrt{\Pi_x}}{\partial x} = p\sqrt{\Pi_x} \qquad \text{B8}$$

## APPENDIX C

The Lorentz transformations are consequence of the interval invariance rule, already found in section 3 in the frame of the present theoretical approach. In this appendix we assume therefore known the Lorentz transformations of momentum $P'_x = (P_x - \varepsilon V_x/c^2)/\beta_x$ and coordinate $x' = (x - tV_x)/\beta_x$ between two inertial reference systems $R$ and $R'$ moving with relative velocity $V_x$ along the x-axis and the consequent basic equations of special relativity; $\beta_x = \sqrt{1-(V_x/c)^2}$, time $t$ and energy $\varepsilon$ of the particle are defined in $R$. To find more general transformation vector formulae, put $P_x = \mathbf{P} \cdot \mathbf{V}/|\mathbf{V}|$ and $P'_x = \mathbf{P}' \cdot \mathbf{V}/|\mathbf{V}|$; since the components of momentum normal to the drift velocity $\mathbf{V}$ of the reference systems are unchanged, $\mathbf{P}' - (\mathbf{P}' \cdot \mathbf{V})\mathbf{V}/V^2 = \mathbf{P} - (\mathbf{P} \cdot \mathbf{V})\mathbf{V}/V^2$, for an observer in $R'$

$$\mathbf{P}' = \mathbf{P} - \mu \mathbf{V} \qquad \mu = \mathbf{P} \cdot \mathbf{V}/V^2 - \left(\mathbf{P} \cdot \mathbf{V}/V^2 - \varepsilon/c^2\right)/\beta \qquad \beta = \sqrt{1-V^2/c^2} \qquad \text{C1}$$

Analogous considerations hold of course also for the coordinates, thus obtaining

$$\mathbf{r}' = \mathbf{r} - \sigma^* \mathbf{V} \qquad \sigma^* = \mathbf{r} \cdot \mathbf{V}/V^2 - \left(\mathbf{r} \cdot \mathbf{V}/V^2 - t\right)/\beta \qquad \text{C2}$$

Let us modify the approach of subsection *2.1* to take into account these requirements of relativity. The angular momentum of a system of $j$ particles is an anti-symmetric 4-tensor built of two 3-vectors: $\mathbf{M} = \sum_j (\mathbf{r} \times \mathbf{P})$ and $\mathbf{M}_4 = ic \sum_j (t\mathbf{P} - \varepsilon \mathbf{r}/c^2)$; $\mathbf{M}_4$ is defined by the centre of inertia of the system of particles. Consider one free quantum particle whose rest mass, velocity, linear momentum and proper distance from the origin are $m$, $\mathbf{v}$, $\mathbf{P}$ and $\mathbf{r}$ respectively in $R$. Eqs C1 and C2 enable $\mathbf{M}' = \mathbf{r}' \times \mathbf{P}'$ in $R'$ to be calculated as a function of $\mathbf{M} = \mathbf{r} \times \mathbf{P}$ in $R$

$$\mathbf{M}' = \mathbf{M} - \left[\frac{\mathbf{r} \cdot \mathbf{V}}{V^2} - \frac{\mathbf{r} \cdot \mathbf{V}/V^2 - t}{\beta}\right]\mathbf{V} \times \mathbf{P} - \left[\frac{\mathbf{P} \cdot \mathbf{V}}{V^2} - \frac{\mathbf{P} \cdot \mathbf{V}/V^2 - \varepsilon/c^2}{\beta}\right]\mathbf{r} \times \mathbf{V} \qquad \text{C3}$$

that is equal to that obtained directly from the general theory of Lorentz transformations of 4-tensors

$$\mathbf{M}' = \frac{1}{\beta}\left[\mathbf{M} + \frac{\mathbf{V}}{V^2}(\mathbf{V} \cdot \mathbf{M})(\beta - 1) - \mathbf{V} \times \left(t\mathbf{P} - \frac{\varepsilon}{c^2}\mathbf{r}\right)\right] \qquad \text{C4}$$

Indeed, collecting with respect to $\beta$ the terms of eq C3 and comparing with eq C4, one obtains the result $(\mathbf{V} \cdot \mathbf{M})\mathbf{V} - \mathbf{M}V^2 = [(\mathbf{r} \cdot \mathbf{V})\mathbf{P} - (\mathbf{P} \cdot \mathbf{V})\mathbf{r}] \times \mathbf{V}$. Since $\mathbf{V} \times (\mathbf{V} \times \mathbf{M}) = (\mathbf{V} \cdot \mathbf{M})\mathbf{V} - V^2\mathbf{M}$, this equation reads $[\mathbf{V}\times\mathbf{M}+(\mathbf{r}\cdot\mathbf{V})\mathbf{P}-(\mathbf{P}\cdot\mathbf{V})\mathbf{r}]\times\mathbf{V}=0$, which is an identity because $\mathbf{V}\times\mathbf{M}=(\mathbf{V}\cdot\mathbf{P})\mathbf{r}-(\mathbf{V}\cdot\mathbf{r})\mathbf{P}$. Let us assume now without loss of generality that the origins of $R$ and $R'$ coincide at a given time $t$ and are displaced by a distance $\mathbf{V}\delta t$ after a time range $\delta t$, with $\delta t$ defined in $R$. Summing and subtracting $\mathbf{V}\delta t$ at right hand side of the first eq C2, one obtains

$$\mathbf{r}' = \mathbf{r}_G - \sigma\mathbf{V} \qquad \mathbf{r}_G = \mathbf{r} - \mathbf{V}\delta t \qquad \sigma = \mathbf{r}\cdot\mathbf{V}/V^2 - \delta t - \left(\mathbf{r}\cdot\mathbf{V}/V^2 - \delta t\right)/\beta \qquad \text{C5}$$

Here $\mathbf{r}_G$ is a Galilean transformation of $\mathbf{r}$, whereas $\sigma\mathbf{V}$ is the relativistic contraction; in effect $\sigma \to 0$ for $c \to \infty$. The moduli of $\mathbf{r}$ and $\mathbf{r}_G$ must be equal; $|\mathbf{r}_G|=|\mathbf{r}|$ yields then $2\mathbf{r}\cdot\mathbf{V}\delta t = (\mathbf{V}\delta t)^2$, i.e.

$$\frac{\mathbf{r}\cdot\mathbf{V}}{V^2} = \frac{1}{2}\delta t \qquad \text{C6}$$

Rewriting $\mathbf{M}' = (\mathbf{r}_G - \sigma\mathbf{V})\times(\mathbf{P}-\mu\mathbf{V})$ as a function of $\mathbf{r}_G$ with the help of eqs C5 and C1, the result is

$$\mathbf{M}' = \mathbf{L}+\mathbf{S} \qquad \mathbf{L} = \mathbf{r}_G\times(\mathbf{P}-\mu\mathbf{V}) \qquad \mathbf{S} = -\sigma\mathbf{V}\times\mathbf{P} \qquad \text{C7}$$

Replacing eqs C5, C6 and $\mathbf{P} = m\mathbf{v}/\sqrt{1-v^2/c^2}$ of the particle into eq C7, $\mathbf{S}$ reads

$$\mathbf{S} = \frac{1}{2}[\mathbf{V}\delta t - \mathbf{V}\delta t/\beta]\times m\mathbf{v}\left(1-v^2/c^2\right)^{-1/2} \qquad \text{C8}$$

If in eq C7 $c$ is put equal to infinity the component $\mathbf{L}$ of $\mathbf{M}'$ takes the classical form $\mathbf{L}_{cl} = \mathbf{r}_G\times m\mathbf{v}$ because $\mu$ vanishes according to eq C1, whereas the component $\mathbf{S}$ vanishes because of eq C5. The vector $\mathbf{S}$ is therefore a relativistic correction to $\mathbf{L}$. Let us recall now that $\delta t$ is a time range in $R$; thus the corresponding time range $\delta t'$ for an observer in $R'$ is $\delta t = \delta t'\beta$. Then eq C8 yields

$$\mathbf{S} = \frac{1}{2}(\delta t - \delta t')\mathbf{V}\times m\mathbf{v}\left(1-v^2/c^2\right)^{-1/2} \qquad \text{C9}$$

Exchanging the vectors $\mathbf{v}/\sqrt{1-v^2/c^2}$ and $\mathbf{V}$ in the cross product, eq C9 reads

$$\mathbf{S} = \frac{1}{2}\frac{\boldsymbol{\lambda}}{\sqrt{1-v^2/c^2}}\times\mathbf{P}_V \qquad \boldsymbol{\lambda} = (\delta t' - \delta t)\mathbf{v} \qquad \mathbf{P}_V = m\mathbf{V} \qquad \text{C10}$$

The step from eq C9 to eq C10 is not merely formal; relevant physical information is introduced in eq C10 through the vector $\boldsymbol{\lambda}_v$ defined as follows

$$\boldsymbol{\lambda}_v = \frac{\boldsymbol{\lambda}}{\sqrt{1-v^2/c^2}} \qquad \text{C11}$$

Being $\mathbf{v}$ of a free particle constant, $\boldsymbol{\lambda}$ can be regarded as the Lorentz contraction of the proper length $\boldsymbol{\lambda}_v$ defined in a reference system solidal with the particle itself. Then $\mathbf{S}$ of eq C10 reads

$$\mathbf{S} = \frac{1}{2}\boldsymbol{\lambda}_v\times\mathbf{P}_V \qquad \text{C12}$$

To summarize: according to eq C7, $\mathbf{M}'$ in $R'$ is expressed as a function of $\mathbf{M}$ in $R$ through the component $\mathbf{L}$ plus the relativistic component $\mathbf{S}$, given by eq C12 and formerly introduced in eq C5 because the simple Galilean transformations of coordinate and momenta have been replaced by the corresponding Lorentz transformations. Moreover the step from eq C9 to eq C10 shows that the angular momentum $\mathbf{S}$ of eq C12 does not depend on the state of motion of the particle, rather it is an intrinsic property of the particle itself being function of $\mathbf{P}_V$ and $\boldsymbol{\lambda}_v$ only: the latter is an internal degree of freedom of the particle and not a vector related to its state of motion, the former defines unambiguously Lorentz's transformations between $R$ and $R'$ through $\mathbf{V}$. Thus $\mathbf{S}$ is related uniquely to the arbitrary translation speed of two inertial reference systems, regardless of the kinetic properties $\varepsilon$ and $\mathbf{v}$ of the particle. As concerns $\mathbf{M}_4$, in general the centre of mass of a system of particles $\mathbf{R}_{cm} = \sum_j \varepsilon \mathbf{r} / \sum_j \varepsilon$ moves in $R$ at rate $\mathbf{V}_{cm} = c^2 \sum_j \mathbf{P} / \sum_j \varepsilon$. Since $\mathbf{R}_{cm} \equiv \mathbf{r}$ and $\mathbf{V}_{cm} \equiv \mathbf{v}$ for one free particle only, $\mathbf{P} = \varepsilon \mathbf{v}/c^2$ yields $\mathbf{M}_4 = (\mathbf{v}t - \mathbf{r}) i \varepsilon / c$, which is constant because of the conservation law of angular momentum; being $\varepsilon$ constant too, $\mathbf{v}t - \mathbf{r}$ is a constant vector, i.e. $\mathbf{r}$ moves thus in $R$ with velocity $\mathbf{v}$. These results require only that $c$ is finite, being instead irrelevant in principle the particular value of v with respect to $c$; only $c = \infty$ entails in any case $\mathbf{S} = 0$ and $\mathbf{L} = \mathbf{r}_G \times m\mathbf{v}$, thus obtaining the mere non-relativistic component $l\hbar$ of the kinetic angular momentum. The reasoning above holds then in general for any free particle. At this point we assume that the basic ideas previously introduced in subsection *2.1* to infer the non-relativistic quantum angular momentum still hold to handle the cross products $\mathbf{r}_G \times (\mathbf{P} - \mu \mathbf{V})$ of eq C7 and $\boldsymbol{\lambda}_v \times \mathbf{P}_V$ of eq C12 in the same way as sketched for the non-relativistic vectors; in other words, the relativistic dynamical variables are treated as quantities whose allowed values fall within arbitrary ranges having physical meaning of quantum uncertainties. The only additional condition imposed by relativity is that the transformation properties of coordinate and momentum ranges are the same as those of the respective dynamical variables. This condition does not modify conceptually the reasoning of subsection *2.1* because the uncertainty ranges are arbitrary; in effect the approach followed for $\Delta \rho$ and $\Delta p$ holds identically for $\Delta \rho'$ and $\Delta p'$, regardless of the fact that the latter are actually Lorentz transformations of the former. The component of $\mathbf{r}_G \times (\mathbf{P} - \mu \mathbf{V})$ along an arbitrary direction defined by the unit vector $\mathbf{w}$ yields then $\pm l\hbar$, whereas $\boldsymbol{\lambda}_v \times \mathbf{P}_V$ yields $\pm l'\hbar$; the component of total angular momentum of the particle requires

$$\mathbf{M} = \mathbf{L} + \mathbf{S} \qquad M_w = L_w + S_w = \pm l\hbar \pm \frac{1}{2} l'\hbar \qquad l, l' \text{ integers including zero} \qquad \text{C13}$$

The properties of the relativistic component $\mathbf{S}$ of angular momentum, e.g. the impossibility to know simultaneously its x, y and z components, are inferred repeating exactly the reasoning shown in subsection *2.1*; moreover $S^2 = \hbar^2 (L'/2 + 1) L'/2$ is again obtained summing the square average components $\langle S_x^2 \rangle$, $\langle S_y^2 \rangle$, $\langle S_z^2 \rangle$, each term being defined now as $\langle S_i^2 \rangle = \hbar^2 (L' + 1)^{-1} \sum_{l'=-L'/2}^{l'=L'/2} l'^2$. The positions 1,1 and the uncertainty equation 1,2 plus the Lorentz transformations, which however follow themselves from the former, are conditions necessary and enough to show the existence of an angular momentum number $l'$ of states additional to $l$. Eq C13 further emphasizes why $l'/2$ and then both $S_w$ and $S^2$ are properties of the particle regardless of its state of motion: $l$ and $l'$ are independent numbers of states inferred from two independent uncertainty equations so that in general $l' \neq 0$ even though $l = 0$. In conclusion $\mathbf{S}$ can be nothing else but what we call spin of quantum particles. In this respect it is interesting the fact that the analysis of states in the relativistic phase space allows to infer also a form of angular momentum that, strictly speaking, is an intrinsic property of the particle rather than a true kinematical property. It is essential in this respect that: (i) also the relativistic dynamical variables are regarded as randomly changing within the respective quantum uncertainties and (ii) the number of allowed states in the relativistic phase space is calculated through quantum uncertainties that fulfil Lorentz's transformations. The generalization of the non-relativistic approach to the special relativity is legitimated by the conceptual consistency of the results of section 3 with the foundations of relativity. A closing remark concerns the constant vector $\mathbf{v}t - \mathbf{r}$ and the fact that the reference system $R'$ where the free particle is at rest requires $\mathbf{V} = \mathbf{v}$; so, the third term of eq C4 reads $\alpha = \mathbf{v} \times (\mathbf{v}t - \mathbf{r}) \varepsilon / c^2$ because $\mathbf{p} = \varepsilon \mathbf{v}/c^2$. Being this term constant, because $\mathbf{v}$ and $\varepsilon$ are constants, the boundary condition $\alpha \to 0$ at values of $t$ such that $\mathbf{v}t \gg \mathbf{r}$ requires $\alpha$ null at any time. Hence eq C4 reads $\mathbf{M}' = \left[ \mathbf{M} + (\mathbf{V} \cdot \mathbf{M})(\beta - 1) \mathbf{V}/V^2 \right]/\beta$ if $\mathbf{M}'$ is defined $R'$. Put without loss of generality $\mathbf{V} = \mathbf{w}V$, since $\mathbf{w}$ is arbitrary; this equation reads then $M'^2 = M^2 - (M^2 - M_w^2)(1 - \beta^{-2})$ and agrees with eq 2,1 $M'^2 = M^2 - (M^2 - M_w^2)(1 - \Delta r'^2 / \Delta r^2)$ if $\Delta r^2 / \Delta r'^2 = \beta^2$, as in effect it is true.

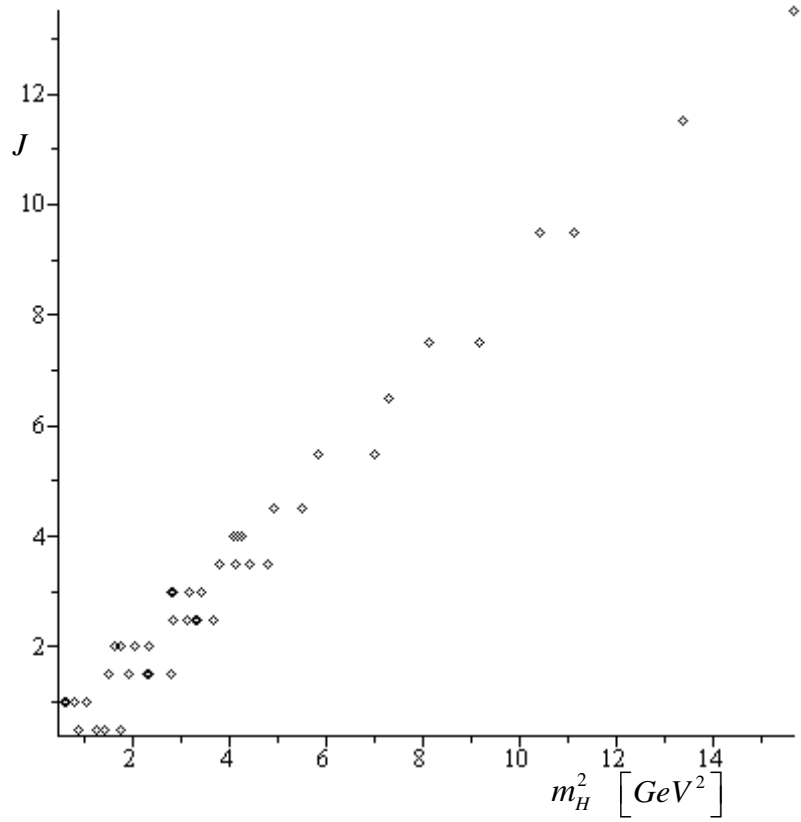

Fig 1

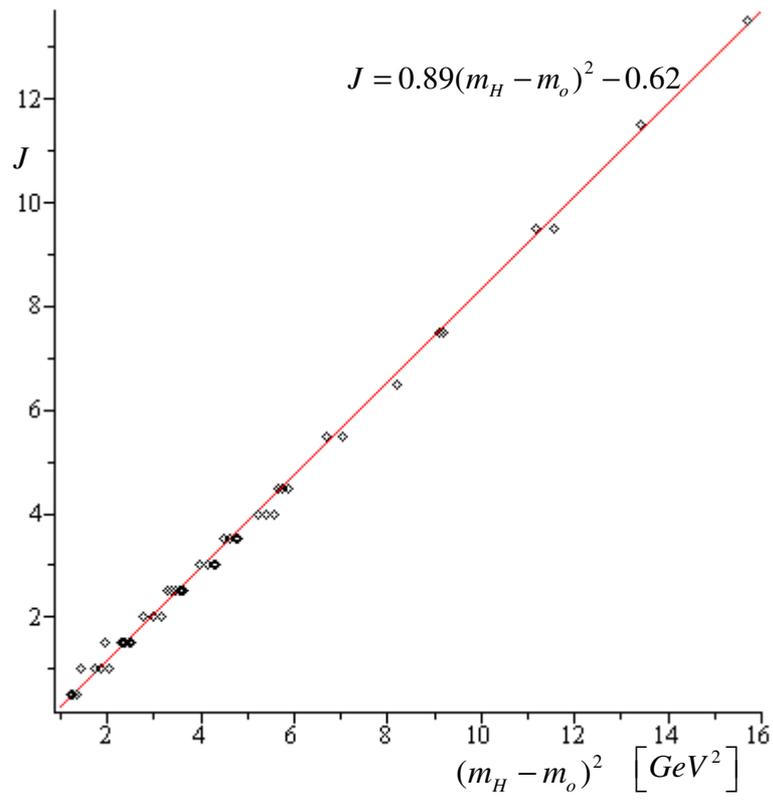

$J = 0.89(m_H - m_o)^2 - 0.62$

Fig 2